\documentclass[10pt, conference, compsocconf]{IEEEtran}

\usepackage{url}        
\usepackage{hyperref}
\usepackage{amsmath}
\usepackage{amssymb}
\usepackage{graphicx}
\usepackage{algorithm}
\usepackage[noend]{algorithmic}
\usepackage{MnSymbol}
\usepackage{subfig}
\usepackage{stmaryrd}
\usepackage{changepage}
\usepackage{amsthm}
\usepackage{float}

\newcommand{\CASE}[1]{\STATE \textbf{case} (#1)\textbf{:} \begin{ALC@g}}
\newcommand{\ENDCASE}{\end{ALC@g}}

\newcommand{\IFLINE}[2]{\STATE \textbf{if} #1 \textbf{then} #2}

\newcommand{\NENDFOR}{\end{ALC@g}\algorithmicendfor}
\newcommand{\NFOR}{\STATE \begin{ALC@g}}

\newcommand{\fn}[2]{\mathrm{\mathbf{#1}}(#2)}
\newcommand{\fnn}[1]{\mathrm{\mathbf{#1}}}

\newcommand{\dvm}[1]{\ensuremath{\text{DEX}_\mathcal{#1}}}
\newcommand{\kobs}{\ensuremath{k_{\mathrm{obs}}}}
\newcommand{\state}[1]{\langle i#1, \rho#1, h#1 \rangle}
\newcommand{\cstate}[1]{\langle #1 \rangle}

\newcommand{\RTVal}{(\mathcal{R} \rightharpoonup \mathcal{V})}
\newcommand{\RTSpec}{(\mathcal{R} \rightarrow \mathcal{S})}

\newcommand{\norm}{\ensuremath{\mathrm{Norm}}}
\newcommand{\np}{\ensuremath{\mathrm{np}}}

\newcommand{\sext}{\ensuremath{\mathcal{S}^{\mathrm{ext}}}}
\newcommand{\sqcupext}{\ensuremath{\sqcup^{\mathrm{ext}}}}
\newcommand{\nleqext}{\ensuremath{\nleq^{\mathrm{ext}}}}
\newcommand{\leqext}{\ensuremath{\leq^{\mathrm{ext}}}}

\newcommand{\colfrac}[3]{\multicolumn{#1}{c}{\dfrac{#2}{#3}}}

\newcommand{\ins}[1]{\ensuremath{\mathrm{\mathbf{#1}}}}
\newcommand{\dexins}[2]{\mathrm{\mathbf{#1}}(#2)}
\newcommand{\object}[1]{\ensuremath{\mathrm{\mathbf{#1}}}}
\newcommand{\field}[1]{\ensuremath{\mathrm{\mathbf{#1}}}}
\newcommand{\compile}[1]{\ensuremath{\llbracket #1 \rrbracket}}
\newcommand{\ordout}[1]{\ensuremath{\llceil #1 \rrceil}}
\newcommand{\translate}[1]{\ensuremath{\llfloor #1 \rrfloor}}

\def\rightsquigarrow{\leadsto}

\newtheorem{theorem}{Theorem}[section]
\newtheorem{lemma}{Lemma}[section]
\newtheorem*{lem}{Lemma}
\newtheorem{definition}{Definition}[section]
\theoremstyle{definition}
\newtheorem{example}{Example}[section]
\theoremstyle{Corollary}
\newtheorem{property}{Property}[section]

\begin{document}

\author{
\IEEEauthorblockN{Alwen Tiu}
	\IEEEauthorblockA{ School of Computer Engineering\\ 
	Nanyang Technological University\\
	Email: atiu@ntu.edu.sg}, 
\and
\IEEEauthorblockN{Hendra Gunadi}
	\IEEEauthorblockA{Research School of Computer Science\\ 
	The Australian National University\\
	Email: hendra.gunadi@anu.edu.au} 
\and 
\IEEEauthorblockN{Rajeev Gore}
  \IEEEauthorblockA{Research School of Computer Science\\ 
  The Australian National University\\
  Email: rajeev.gore@anu.edu.au} }

\title{Formal Certification of Android Bytecode}

\maketitle
\begin{abstract}
Android is an operating system that has been used in a majority of
mobile devices. Each application in Android runs in an instance of the
Dalvik virtual machine, which is a register-based virtual machine
(VM).  Most applications for Android are developed using Java,
compiled to Java bytecode and then translated to DEX bytecode using
the dx tool in the Android SDK.  In this work, we aim to develop a
type-based method for certifying non-interference properties of DEX
bytecode, following a methodology that has been developed for Java
bytecode certification by Barthe et al.  To this end, we develop a
formal operational semantics of the Dalvik VM, a type system for DEX
bytecode, and prove the soundness of the type system with respect to a
notion of non-interference.  We then study the translation process
from Java bytecode to DEX bytecode, as implemented in the dx tool in
the Android SDK.  We show that an abstracted version of the
translation from Java bytecode to DEX bytecode preserves the
non-interference property. More precisely, we show that if the Java
bytecode is typable in Barthe et al's type system (which guarantees
non-interference) then its translation is typable in our type system.
This result opens up the possibility to leverage existing bytecode
verifiers for Java to certify non-interference properties of Android
bytecode.
\end{abstract}

\section{Introduction}

Android is an operating system that has been used in many mobile devices. 
According to \cite{StatCounter}, Android has the largest market share for mobile 
devices, making it an attractive target for malwares, so verification of the security
properties of Android apps is crucial. 
To install an application, users can download applications from Google Play or third-party app stores 
in the form of an Android Application Package (APK). Each of these 
applications runs in an instance of a Dalvik virtual machine (VM) on top of the Linux 
operating system. Contained in each of these APKs is a DEX file containing specific 
instructions \cite{DEXBytecode} to be executed by the Dalvik VM, so from here 
on we will refer to these bytecode instructions as DEX instructions. The Dalvik VM is a 
register-based VM, unlike the Java Virtual Machine (JVM) which is a stack-based VM. Dalvik is now superseded
by a new runtime framework called ART, but this does not affect our analysis
since both Dalvik and ART use the same DEX instructions.

We aim at providing a framework for constructing trustworthy apps, where developers
of apps can provide guarantees that the (sensitive) information the apps 
use is not leaked outside the device without the user's consent.
The framework should also provide a mean for the end user to
verify that apps constructed using the framework adhere to their
advertised security policies. This is, of course, not a new concept,
and it is essentially a rehash of the (foundational) 
proof carrying code (PCC)~\cite{Necula97POPL,Appel01LICS}, 
applied to the Android setting.
We follow a type-based approach for restricting 
information flow~\cite{sabelfeld2003language} 
in Android apps. Semantically, information flow properties of apps are specified 
via a notion of non-interference~\cite{goguen1982noninterference}. In this setting, 
typeable programs are guaranteed to be non-interferrent, with respect to
a given policy, and typing derivations serve as certificates of non-interference. 
Our eventual goal is to produce a compiler tool chain that can
help developers to develop Android applications that complies
with a given policy, and automate the process of generating
the final non-interference certificates for DEX bytecode.

An Android application is typically written in Java and
compiled to Java classes (JVM bytecode). Then using tools provided in the
Android Software Development Kit (SDK), these Java classes are further
compiled into an Android application in the form of an APK. One
important tool in this compilation chain is the dx tool, which will
aggregate the Java classes and produce a DEX file to be
bundled together with other resource files in the APK. 
Non-interference type systems exist for Java source code~\cite{Banerjee05JFP},
JVM~\cite{barthe13mscs} and (abstracted) 
DEX bytecode without exception handling mechanism~\cite{cassandra}. 
To build a framework that allows end-to-end certificate production,
one needs to study certificate translation between these different type systems.
The connection between Java and JVM type systems for non-interference has
been studied in \cite{barthe2006deriving}. In this work, we fill the gap by showing
that the connection between JVM and DEX type systems.
Our contributions are the following:
\begin{itemize}
\item We give a formal account of the compilation process 
from JVM bytecode to DEX bytecode as implemented in 
the official dx tool in Android SDK. Section~\ref{section:Translation} 
details some of the translation processes.

\item We provide a proof that the translation from JVM to DEX preserves
typeability. That is, JVM programs typeable in the non-interference type system
for JVM translates into typeable programs in the non-interference
type system for DEX.

\end{itemize}

The development of the operational semantics and the type systems for 
DEX bytecode follows closely the framework set up in \cite{barthe13mscs}. 
Although Dalvik is a registered-based
machine and JVM is a stack-based machine, the translation from one instruction 
set to the other is for most part quite straightforward. The adaptation of the 
type system for JVM to its DEX counterpart is complicated
slightly by the need to simulate JVM stacks in DEX registered-based instructions.
The non-trivial parts are when we want to capture
both direct (via operand stacks) and indirect information flow 
(e.g., those resulting from branching on high value).
In \cite{barthe13mscs}, to deal with both direct and indirect flow, 
several techniques are used, among others, the introduction of
operand stack types (each stack element carries a type which is a security label),
a notion of safe control dependence region (CDR), which keeps track of the 
regions of the bytecode executing under a 'high' security level, and the notion 
of security environment, which attaches security levels to points in programs. 
Since Dalvik is a register-based machine, when translating from JVM to DEX, 
the dx tool simulates the operand stack using DEX registers.
As the type system for JVM is parameterised by
a safe CDR and a security environment, we also need to define how these are affected by the translation,
e.g., whether one can construct a safe CDR for DEX given a safe CDR for JVM. 
This was complicated by the fact that the translation by dx in general is organized along blocks of sequential
(non-branching) codes, so one needs to relate blocks of codes in the image of the translation back to
the original codes (see Section~\ref{section:Translation}) .

The rest of the paper is organized as follows : in the next section we describe 
some work done for Java bytecode security and the work on static analysis for Android 
bytecode. This is important as what we are doing is bridging the relationship between 
these two security measures. Then we review the work of Barthe et.al. on a 
non-interferent type system for JVM bytecode. In the remainder of the paper we will 
describe our work, namely providing the type system for DEX and proving the translation 
of typability. We also give examples to demonstrate how our methodology is able to 
detect interference by failure of typability. Before concluding, we provide 
our design of implementation for the proof of concept.

\section{Related Work}

As we already mentioned, our work is heavily influenced by the work of Barthe et. al. \cite{barthe2006proof,barthe13mscs}
on enforcing non-interference via type systems. We discuss other related work in the following.

The cloest to our work is the Cassandra project~\cite{cassandra,cassandratr}, 
that aims at developing certified app stores, 
where apps can be certified, using an information-flow type system
similar to ours, for absence of specific information flow. 
Specifically, the authors of \cite{cassandra,cassandratr} 
have developed an abstract Dalvik language (ADL), similar to Dalvik bytecode, and
a type system for enforcing non-interference properties for ADL. 
Our type system for Dalvik has many similarities with that of Cassandra, but one
main difference is that we consider a larger fragment of Dalvik, which includes
exception handling, something that is not present in Cassandra.
We choose to deal directly
with Dalvik rather than ADL since we aim to eventually integrate our
certificate compilation into existing compiler tool chains for Android apps,
without having to modify those tool chains. 

Bian et. al. \cite{bian2007java} targets the JVM bytecode to check 
whether a program has the non-interference property. Differently from Barthe et. al. 
their approach uses the idea of the compilation technique where they analyse a variable 
in the bytecode for its definition and usage. Using this dependence analysis, their 
tool can detect whether a program leaks confidential information. This is an interesting
technique in itself and it is possible to adopt their approach to analyze DEX bytecode.
Nevertheless, we are more interested in the transferability of properties instead of 
the technique in itself, i.e., if we were to use their approach instead of a type system,
the question we are trying to answer would become ``if the JVM bytecode is non-interferent
according to their approach, is the compiled DEX bytecode also non-interferent?''.  

In the case of preservation of properties itself the idea that a non-optimizing 
compiler preserves a property is not something new. The work by Barthe et. al. 
\cite{barthe2006proof} shows that with a non-optimizing compiler, the proof obligation 
from a source language to a simple stack based language will be the same, thus allowing 
the reuse of the proof for the proof obligation in the source language. In showing
the preservation of a property, they introduce the source imperative language and 
target language for a stack-based abstract machine.
This is the main difference with our work where 
we are analyzing the actual dx tool from Android
which compiles the bytecode language for stack-based virtual machine (JVM bytecode) to 
the actual language for register-based machine (DEX bytecode).
There are also works that address this non-interference preservation from 
Java source code to JVM bytecode \cite{barthe2006deriving}. Our work can then be seen
as a complement to their work in that we are extending the type preservation
to include the compilation from JVM bytecode to DEX bytecode. 

To deal with information flow properties in Android, there are several works addressing
the problem \cite{fuchs2009scandroid, bugliesi2013lintent, enck2014taintdroid, 
zhao2012trustdroid, kim2012scandal, android:esorics12, android:esorics13, 
IPCInspection, enck2009lightweight, enck2009understanding}
although some of them are geared towards the privilege escalation problem. These works
base their context of Android security studied in \cite{enck2009understanding}.
The tool in the study, which is called Kirin, is also of great interest for us since 
they deal with the certification of Android applications. Kirin is a lightweight tool 
which certifies an Android application at install time based on permissions requested.
Some of these works are similar to ours in a sense working on static analysis for 
Android. The closest one to mention is ScanDroid \cite{fuchs2009scandroid}, with the 
underlying type system and static analysis tool for security specification in Android 
\cite{chaudhuri2009language}. 
Then along the line of type system there is also work by Bugliesi et. al. 
called Lintent that tries to address non-interference on the source code level
\cite{bugliesi2013lintent}. The main difference with what we do lies 
in that  the analysis itself is relying on the existence of 
the source (the JVM bytecode for ScanDroid and Java source code for Lintent)
from which the DEX program is translated. 

There are some other static analysis tools for Android which do not stem from the idea 
of type system, e.g. TrustDroid \cite{zhao2012trustdroid} and ScanDal 
\cite{kim2012scandal}. TrustDroid is another static analysis tool on Android bytecode, 
trying to prevent information leaking. 
TrustDroid is more interested in doing taint analysis on the program, although different
from TaintDroid \cite{enck2014taintdroid} in that TrustDroid is doing taint analysis 
statically from decompiled DEX bytecode whereas TaintDroid is enforcing run time taint 
analysis. ScanDal is also a static analysis for Android applications targetting the 
DEX instructions directly, aggregating the instructions in a language they call Dalvik 
Core. They enumerate all possible states and note when any value from any predefined 
information source is flowing through a predefined information sink. 
Their work assumed that predefined sources and sinks are given, whereas 
we are more interested in a flexible policy to define them.

Since the property that we are interested in is non-interference, it is also worth
mentioning Sorbet, a run time enforcement of the property by modifying the Android 
operating system \cite{android:esorics12, android:esorics13}. Their approach is 
different from our ultimate goal which motivates this work in that 
we are aiming for no modification in the Android operating system.

\section {Type System for JVM}\label{javaTypeSystem}

In this section, we give an overview of Barthe et. al's type system for JVM.
Due to space constraints, some details are omitted and the reader is referred
to \cite{barthe13mscs} for a more detailed explanation and intuitions behind
the design of the type system. Readers who are already familiar with the work
of Barthe et. al may skip this section.

A program $P$ is given by its list of instructions given in 
Figure~\ref{figure:javaInstructionList}. The set 
$\mathcal{X}$ is the set of local variables, 
$\mathcal{V} = \mathbb{Z} \bigcup \mathcal{L} \bigcup \{null\}$ is the set of values, 
where $\mathcal{L}$ is an (infinite) set of locations and $null$ denotes the null 
pointer, and $\mathcal{PP}$ is the set of program points. 
We use the notation $^*$ to mean that for any set $X$, 
$X^*$ is a stack of elements of $X$. 
Programs are also implicitly 
parameterized by a set $\mathcal{C}$ of class names, a set $\mathcal{F}$ of field 
identifiers, a set $\mathcal{M}$ of method names, and a set of Java types 
$\mathcal{T}_J$. The instructions listing can be seen in 
Figure~\ref{figure:javaInstructionList}.
\begin{figure}
\[\boxed{
\begin{array}{l c l}
	\ins{binop}\ op & : & \text{binary operation on stack} \\
	\ins{push}\ c & : & \text{push value on top of a stack} \\
	\ins{pop} & : & \text{pop value from top of a stack} \\
	\ins{swap} & : & \text{swap top two operand stack values} \\
	\ins{load}\ x & : & \text{load value of } x \text{ on stack} \\
	\ins{store}\ x & : & \text{store top of stack in variable }x \\
	\ins{ifeq}\ j & : & \text{conditional jump} \\
	\ins{goto}\ j & : & \text{unconditional jump} \\
	\ins{return} & : & \text{return the top value of the stack}\\
	\ins{new }\ C & : & \text{create new object in the heap} \\
	\ins{getfield }\ f & : & \text{load value of field } f \text{on stack} \\
	\ins{putfield }\ f & : & \text{store top of stack in field } f \\
	\ins{newarray }\ t & : & \text{create new array of type } t \text{ in the heap}\\
	\ins{arraylength} & : & \text{get the length of an array} \\	
	\ins{arrayload} & : & \text{load value from an array} \\
	\ins{arraystore} & : & \text{store value in array} \\
	\ins{invoke}\ m_\mathrm{ID}& : & \text{Invoke method indicated by } m_\mathrm{ID}\\
	   && \text{with arguments on top of the stack} \\	
	\ins{throw} & : & \text{Throw exception at the top of a stack}\\
	\hline \\
	\multicolumn{3}{l}{\text{where }op \in \{+, -, \times, /\}, c \in \mathbb{Z}, x \in 
	\mathcal{X}, j \in \mathcal{PP}, C \in \mathcal{C}\text{, } }\\
 	\multicolumn{3}{l}{f \in \mathcal{F}, t \in \mathcal{T}_J, \text{ and } 
 	m_\mathrm{ID} \in \mathcal{M}.}
\end{array}
}\]
\caption{JVM Instruction List}
\label{figure:javaInstructionList}
\end{figure}

\noindent \textbf{Operational Semantics} The operational semantics is given as a relation
$\rightsquigarrow_{m, \tau} \subseteq \mathrm{State} \times 
(\mathrm{State} + (\mathcal{V}, \object{heap}))$ 
where $m$ indicates the method under which the relation is considered 
and $\tau$ indicates whether the instruction is executing normally (indicated by
$\norm$) or throwing an exception.
(sometimes we omit $m$ whenever it is clear which $m$ we are referring to,
we may also remove $\tau$ when it is clear from the context whether the instruction
is executing normally or not
). 
State here represents a set of JVM states, which is a tuple 
$\langle i, \rho, os, h \rangle$ where i $\in \mathcal{PP}$ is the program counter that 
points to the next instruction to be executed; $\rho \in \mathcal{X} \rightharpoonup 
\mathcal{V}$ is a partial function from local variables to values, $os\in\mathcal{V}^*$ 
is an operand stack, and  $h \in \object{heap}$ is the heap for that particular state. 
Heaps are modeled as partial functions 
$\fnn{h} : \mathcal{L} \rightharpoonup \mathcal{O} + \mathcal{A}$, where the set 
$\mathcal{O}$ of objects is modeled as $\mathcal{C} \times (\mathcal{F} \rightharpoonup 
\mathcal{V})$, i.e. each object $o \in \mathcal{O}$ possess a class $\fn{class}{o}$ 
and a partial function to access field values, which is denoted by 
$o.f$ to access the value of field $f$ of object $o$. $\mathcal{A}$ is the set of 
arrays modeled as $\mathbb{N} \times (\mathbb{N} \rightharpoonup \mathcal{V}) \times 
\mathcal{PP}$ i.e. each array has a length, partial function from index to value, and a 
creation point. The creation point will be used to define the notion of array 
indistinguishability. \object{Heap} is the set of heaps.

The program also comes equipped with a partial function 
$\fnn{Handler_m}: \mathcal{PP} \times \mathcal{C} \rightharpoonup \mathcal{PP}$.
We write $\fn{Handler_m}{i, C} = t$ for an exception of class $C \in \mathcal{C}$
thrown at program point $i$, which will be caught by a handler with its
starting program point $t$. In the case where the exception is uncaught, we write
$\fn{Handler_m}{i, C} \uparrow$ instead.
The final states will be $(\mathcal{V} + \mathcal{L}) \times 
\object{Heap}$ to differentiate between normal termination $(v, h) \in \mathcal{V} 
\times \object{Heap}$, and an uncaught exception $(\langle l \rangle, h) \in 
\mathcal{L} + \object{Heap}$ which contains the location $l$ for the exception in the 
heap $h$.

\underline{op} denotes here the standard interpretation of arithmetic operation 
of $op$ in the domain of values $\mathcal{V}$ (although there is no arithmetic 
operation on locations). 

\begin{figure*}
\[
\boxed{
\begin{array}{c}
\begin{array}{c c c c} \hspace{0.25\textwidth} & \hspace{0.22\textwidth} & 
  \hspace{0.2\textwidth} & \hspace{0.2\textwidth}\\
	
	\hfill \dfrac{P_m[i] = \ins{push }\ n}
	  {\cstate{i, \rho, os} \rightsquigarrow \cstate{i + 1, \rho, n :: os}} \hfill &
	\hfill \dfrac{P_m[i] = \ins{pop}}
	  {\cstate{i, \rho, v :: os} \rightsquigarrow \cstate{i + 1, \rho, os}} \hfill &
	\hfill \dfrac{P_m[i] = \ins{return}}
	  {\cstate{i, \rho, v :: os} \rightsquigarrow v, h} \hfill &
	\hfill \dfrac{P_m[i] = \ins{goto}\ j}
	  {\cstate{i, \rho, os} \rightsquigarrow \cstate{j, \rho, os}} \hfill \\
	  
\end{array}\\
\begin{array}{ccc} \hspace{0.3\textwidth} & \hspace{0.3\textwidth} & 
  \hspace{0.3\textwidth}\\
	
	\hfill 
	\dfrac{\begin{gathered}P_m[i] = \ins{store }\ x\ \ x \in \fn{dom}{\rho}\end{gathered}}
	{\begin{gathered}\cstate{i, \rho, v :: os} \rightsquigarrow 
	  \cstate{i+1, \rho \oplus \{x \mapsto v\}, os}\end{gathered}} \hfill &  
  \hfill \dfrac{P_m[i] = \ins{load }\ x}
	{\begin{gathered}\cstate{i, \rho, os} \rightsquigarrow 
		\cstate{i + 1, \rho, \rho(x) :: os}\end{gathered}} \hfill &
  \hfill \dfrac{P_m[i] = \ins{binop }\ op\ \ \ \ n_2\ \underline{op}\ n_1 = n}
  {\cstate{i, \rho, n_1 :: n_2 :: os} \rightsquigarrow \cstate{i + 1,\rho, n :: os}}
  \hfill \\
		
\end{array}\\
\begin{array}{ccc}	\hspace{0.35\textwidth} & \hspace{0.25\textwidth} & 
  \hspace{0.25\textwidth}\\
	 
  \hfill \dfrac{P_m[i] = \ins{swap}}
  {\cstate{i, \rho, v_1 :: v_2 :: os} \rightsquigarrow 
    \cstate{i+1, \rho, v_2 :: v_1 :: os}} \hfill & 
	\hfill \dfrac{P_m[i] = \ins{ifeq}\ j\ \ n \neq 0}
	{\cstate{i, \rho, n :: os} \rightsquigarrow \cstate{i + 1, \rho, os}} \hfill  &
	\hfill \dfrac{P_m[i] = \ins{ifeq }\ j\ \ \ \ n = 0}
	{\cstate{i, \rho, n :: os} \rightsquigarrow \cstate{j, \rho, os}} \hfill \\
	
\end{array}\\\\
\end{array}}\] 

\caption{JVM Operational Semantic (Selected)}
\label{figure:javaOperationalSemanticSelected}
\end{figure*}

The instruction that may throw an exception primarily are method invocation and the 
object/array manipulation instructions. \{\np\} is used as the class for null pointer 
exceptions, with the associated exception handler being 
$\fnn{RuntimeExceptionHandling}$. The transitions are also parameterized by a tag 
$\tau \in \{\norm\} + \mathcal{C}$ to
describe whether the transition occurs normally or some exception is thrown. 

Some last remarks: firstly, because of method invocation, the operational semantics 
will also be mixed with a big step semantics style $\rightsquigarrow^{+}_{m}$ 
from method invocations of method $m$ and its associated result, to be more precise
$\rightsquigarrow^{+}_{m}$ is a transitive closure of $\rightsquigarrow_m$. 
Then, for instructions that may not throw an exception, we remove the subscript 
$\{m, \norm\}$ from $\rightsquigarrow$ because it is clear that they have no exception 
throwing operational semantic counterpart. 
A list of operational 
semantics are contained in Figure~\ref{figure:javaOperationalSemanticSelected}.
We do not show the full list of operational semantics due to space limitations.
However, the interested reader can see Figure~\ref{figure:javaOperationalSemanticFull}
in Appendix~\ref{section:appendix_jvm} for the full list of JVM operational semantics.

\textbf{Successor Relation} The successor relation 
$\mapsto \subseteq \mathcal{PP} \times \mathcal{PP}$ of a program $P$ are tagged with 
whether the execution is normal or throwing an exception. 
According to the types of instructions at program point $i$, 
there are several possibilities:
\begin{itemize}
  \item $P_m[i] = \ins{goto}\ t$. The successor relation is $i \mapsto^{\norm} t$
  
  \item $P_m[i] = \ins{ifeq}\ t$. In this case, there are 2 successor relations 
  denoted by $i \mapsto^{\norm} i+1$ and $i \mapsto^{\norm} t$.
  
  \item $P_m[i] = \ins{return}$. In this case it is a return point denoted by 
  $i \mapsto^{\norm}$
  
  \item $P_m[i]$ is an instruction
  throwing a null pointer exception, and there is a handler for it 
  $(\fn{Handler}{i, \np} = t)$. In this case, the successor is $t$ denoted by 
  $i \mapsto^{\np} t$.
  
  \item $P_m[i]$ is an instruction
  throwing a null pointer exception, and there is no handler for it 
  $(\fn{Handler}{i, \np} \uparrow)$. In this case it is a return point denoted by 
  $i \mapsto^{\np}$.  
  
  \item $P_m[i] = \ins{throw}$, throwing an exception $C \in \fn{classAnalysis}{m, i}$,
  and $\fn{Handler}{i, C} = t$. The successor relation is $i \mapsto^{C} t$.
  
  \item $P_m[i] = \ins{throw}$, throwing an exception $C \in \fn{classAnalysis}{m, i}$,
  and $\fn{Handler}{i, C} = t$. It is a return point and the successor relation is 
  $i \mapsto^{C}$.
  
  \item $P_m[i] = \ins{invoke}\ m_\mathrm{ID}$, throwing an exception 
  $C \in \fn{excAnalysis}{m_\mathrm{ID}}$, and $\fn{Handler}{i, C} = t$. 
  The successor relation is $i \mapsto^{C} t$.  
  
  \item $P_m[i] = \ins{invoke}\ m_\mathrm{ID}$, throwing an exception 
  $C \in \fn{excAnalysis}{m_\mathrm{ID}}$, and $\fn{Handler}{i, C} \uparrow$. 
  It is a return point and the successor relation is $i \mapsto^{C}$.  
  
  \item $P_m[i]$ is any other cases. The successor is its immediate instruction denoted
  by $i \mapsto^{norm} i+1$
\end{itemize}

\begin{figure*}
\[\boxed{
\begin{array}{c}
\begin{array}{ccc}
  \hspace{0.3\textwidth} & \hspace{0.3\textwidth} & \hspace{0.3\textwidth}\\
  
  \hfill \dfrac{P_m[i] = \ins{load}\ x}
    {se, i \vdash st \Rightarrow \big(\vec{k_v}(x) \sqcup se(i)\big) :: st} \hfill 
  & \hfill \dfrac{P[i]_m = \ins{store}\ x\ \ \ \ se(i) \sqcup k \leq \vec{k_a}(x)}
    {se, i \vdash k :: st \Rightarrow st} \hfill	
  & \hfill \dfrac{P_m[i] = \ins{swap}}
    { i \vdash k_1 :: k_2 :: st \Rightarrow k_2 :: k_1 :: st} \hfill \\
    
\end{array}\\
\begin{array}{ccc}
  \hspace{0.5\textwidth} & \hspace{0.2\textwidth} & \hspace{0.2\textwidth}\\	
  
  \hfill 
  \dfrac{P[i]_m=\ins{ifeq}\ j\ \ \ \ \forall j'\in\fn{region}{i,\norm},k\leq se(j')}
  {\fnn{region}, se, i \vdash k :: st \Rightarrow \fn{lift_k}{st}} \hfill &  
  
  \hfill \dfrac{P_m[i] = \ins{goto}\ j}{i \vdash st \Rightarrow st} \hfill & 
  
  \hfill 
    \dfrac{P_m[i] = \ins{push}\ n}{i \vdash st \Rightarrow se(i) :: st}
  \hfill \\
  
\end{array}\\
\begin{array}{ccc}
  \hspace{0.4\textwidth} & \hspace{0.25\textwidth} & \hspace{0.25\textwidth}\\
  
  \hfill \dfrac{P[i]_m = \ins{binop}\ op}
  {se, i \vdash k_1 :: k_2 :: st \Rightarrow (k_1 \sqcup k_2 \sqcup se(i)) :: st} \hfill
  
  & \hfill
    \dfrac{P_m[i] = \ins{return}\ \ \ \ se(i) \sqcup k \leq k_r[n]}
    {\vec{k_a} \overset{k_h}{\rightarrow} \vec{k_r}, se, i \vdash k :: st \Rightarrow}  
  \hfill
   
  & \hfill \dfrac{P_m[i] = \ins{pop}}{i \vdash k :: st \Rightarrow st} \hfill \\
  
\end{array}\\\\
\end{array}}\]
\caption{JVM Transfer Rule (Selected)}
\label{figure:javaTypeRuleSelected}
\end{figure*}

\textbf{Typing Rules} The security level is defined as a partially ordered set 
($\mathcal{S}, \leq$) of security levels $\mathcal{S}$ that form a lattice. 
$\sqcup$ denotes the lub of two security levels, and for 
every $k \in \mathcal{S}$, $\fnn{lift_k}$ is a point-wise extension to stack types of 
$\lambda l. k \sqcup l$. The policy of a method is 
also defined relative to a security level $\kobs$ which denotes the capability of an 
observer to observe values from local variables, fields, and return values whose 
security level are below $\kobs$.The typing rules are defined in terms of stack types, 
that is a stack that associates a value in the operand stack to the set of security 
levels $\mathcal{S}$. The stack type itself takes the form of a stack with 
corresponding indices from the operand stack, as shown below. 

\begin{center}
	\includegraphics[scale=0.5]{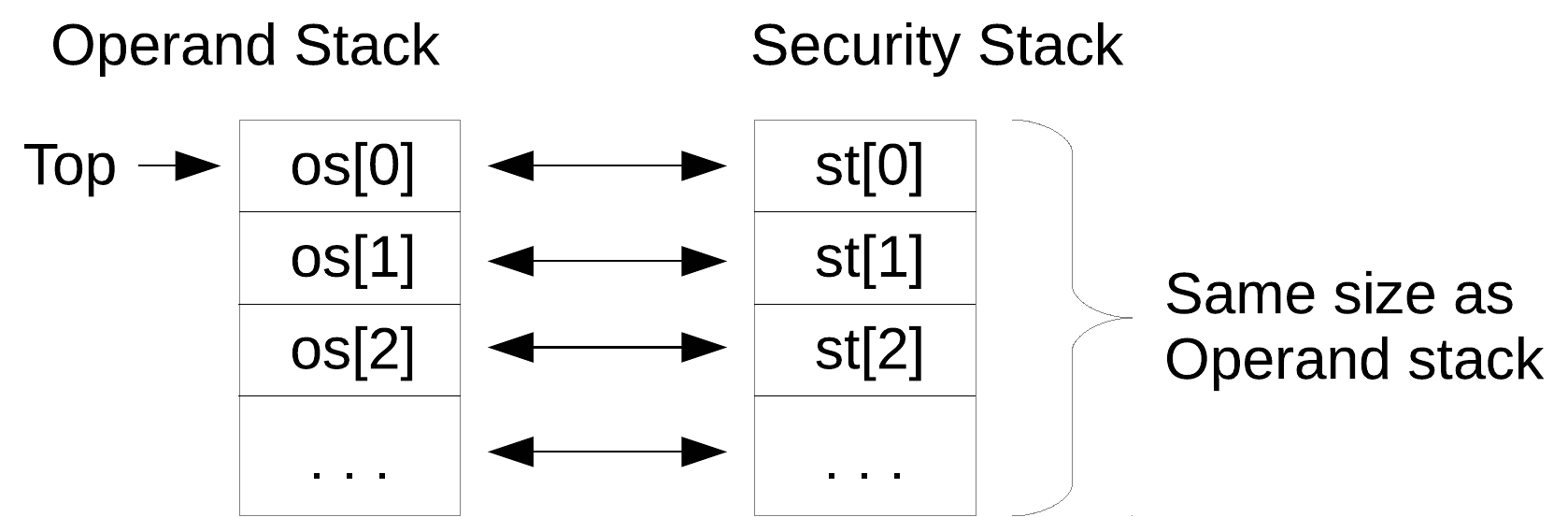}
\end{center}

We assume that a method comes with its security policy of the form 
$\vec{k_a} \overset{k_h}{\rightarrow} \vec{k_r}$ 
where $\vec{k_a}$ represents a list $\{v_1 : k_1, \dots, v_n : k_n\}$ with 
$k_i \in \mathcal{S}$ being the security level of local variables 
$v_i \in \mathcal{R}$, $k_h$ is the effect of the method on the heap and 
$\vec{k_r}$ is the return signature, i.e. the security levels of the 
return value. The return signature is of the form of a list to cater for the 
possibility of an uncaught exception on top of the normal return value. The $\vec{k_r}$ 
is a list of the form $\{\norm : k_n, e_1 : k_{e_1}, \dots, e_n : k_{e_n}\}$ 
where $k_n$ is the security level for the normal return value, and $e_i$ is the class 
of the uncaught exception thrown by the method and  $k_{e_i}$ is the associated 
security level.
In the sequel, we write $\vec{k_r}[n]$ to stand for $k_n$ and 
$\vec{k_r}[e_i]$ to stand for $k_{e_i}$. 
An example of this policy can be 
$\{1 : L, 2 : H\} \overset{H}{\rightarrow} \{ \norm : L \}$ where 
$L, H \in \mathcal{S}, L \leq \kobs, H \nleq \kobs$ which indicates that the method 
will return a low value, and that throughout the execution of the method, the security 
level of local variable $1$ will be low while the security level of local variable $2$ 
will be high.

Arrays have an extended security level than that of the usual object or value to 
cater for the security level of the contents. 
The security level of an array will be of the form 
$k[k_c]$ where $k$ represents the security level of an array and $k_c$ will represent 
the security level of its content 
(this implies that all array elements have the same security level $k_c$). 
Denote $\sext$ as the extension of security levels 
$\mathcal{S}$ to define the array's security level. The partial order on $\mathcal{S}$ 
will also be extended with $\leqext$ :
\[\begin{array}{cc}
	\hfill \dfrac{k\leq k'\ \ \ \ k,k'\in\mathcal{S}}
	  {k\leqext k'} \hspace{0.5cm} \hfill &
	\hfill \dfrac{k \leq k'\ \ \ \ k, k' \in \mathcal{S}\ \ \ \ k_c \in \sext}
	  {k[k_c] \leqext k'[k_c]} \hfill
\end{array}\]
Generally, in the case of a comparison between extended level $k[k_c] \in \sext$ and a 
standard level $k' \in \mathcal{S}$, we only compare $k$ and $k'$ w.r.t. the partial 
order on $\mathcal{S}$. In the case of comparison with \kobs, since 
$\kobs \in \mathcal{S}$ an extended security $k[k_c]$ is considered low 
(written $k[k_c] \leq \kobs$) if $k \leq \kobs$. 
Only $\kobs$ and $se$ (defined later) will stay in the form of $\mathcal{S}$, 
everything else will be extended to also include the extended level \sext.

The transfer rules come equipped with a security policy for fields 
$\fnn{ft} : \mathcal{F} \rightarrow \sext$ and $\textmd{at} : \mathcal{PP} \rightarrow 
\sext$ that maps the creation point of an array with the security level of its content. 
$\fn{at}{a}$ will also be used to denote the security level of the content of array $a$ 
at its creation point.

The notation $\Gamma$ is used to define the table of method signatures which will 
associate a method identifier $m_\mathrm{ID}$ and a security level 
$k \in \mathcal{S}$ (of the object invoked) to 
a security signature $\Gamma_{m_\mathrm{ID}}[k]$. 
The collection of security signatures of a method $m$ is defined as 
$\fn{Policies_{\Gamma}}{m_\mathrm{ID}} = 
\{\Gamma_{m_\mathrm{ID}}[k]\ |\ k \in \mathcal{S}\}$.

A method is also parameterized by a control dependence region (CDR) which is defined in 
terms of two functions: $\fnn{region}$ and $\fnn{jun}$. The function
$\fnn{region} : \mathcal{PP} \rightarrow \wp(\mathcal{PP})$ can be seen as all the 
program points executing under the guard of the instruction at the specified program 
point, i.e. in the case of $\fn{region}{i}$ the guard will be program point $i$. 
The function $\fn{jun}{i}$ itself can be seen as the nearest program point which all 
instructions in $\fn{region}{i}$ have to execute (junction point). A CDR is safe if it 
satisfies the following SOAP (Safe Over APproximation) properties.
\begin{definition}
\label{SOAP_G}
	A CDR structure (region, jun) satisfies the SOAP properties if the following 
	properties hold :
	\begin{description}
		\item[SOAP1.] $\forall i, j, k \in \mathcal{PP}$ and tag $\tau$ such that 
		$i \mapsto j$ and $i \mapsto^{\tau} k$ and $j \neq k$ ($i$ is hence a branching 
		point), $k \in \fn{region}{i, \tau}$ or $k = \fn{jun}{i, \tau}$.
		
		\item[SOAP2.]$\forall i, j, k \in \mathcal{PP}$ and tag $\tau$, if 
		$j \in \fn{region}{i, \tau}$ and $j \mapsto k$, then either 
		$k \in \fn{region}{i, \tau}$ or $k = \fn{jun}{i, \tau}$.
		
		\item[SOAP3.] $\forall i, j \in \mathcal{PP}$ and tag $\tau$, if 
		$j \in \fn{region}{i, \tau}$ and $j$ is a return point then $\fn{jun}{i, \tau}$ is 
		undefined.
		
		\item[SOAP4.] $\forall i \in \mathcal{PP}$ and tags $\tau_1, \tau_2$ if 
		$\fn{jun}{i, \tau_1}$ and $\fn{jun}{i, \tau_2}$ are defined and 
		$\fn{jun}{i, \tau_1} \neq \fn{jun}{i, \tau_2}$ then $\fn{jun}{i, \tau_1} \in 
		\fn{region}{i, \tau_2}$ or $\fn{jun}{i, \tau_2} \in \fn{region}{i, \tau_1}$.
		
		\item[SOAP5.] $\forall i, j \in \mathcal{PP}$ and tag $\tau$, if 
		$j \in \fn{region}{i, \tau}$ and $j$ is a return point then for all tags $\tau'$ 
		such that $\fn{jun}{i, \tau'}$ is defined, $\fn{jun}{i, \tau'} \in 
		\fn{region}{i, \tau}$.
		
		\item[SOAP6.] $\forall i \in \mathcal{PP}$ and tag $\tau_1$, if $i\mapsto^{\tau_1}$ 
		then for all tags $\tau_2, \fn{region}{i, \tau_2}$ $\subseteq$ 
		$\fn{region}{i, \tau_1}$ and if $\fn{jun}{i, \tau_2}$ is defined, 
		$\fn{jun}{i, \tau_2} \in \fn{region}{i, \tau_1}$.
		
	\end{description}
\end{definition}

The security environment function $se : \mathcal{PP} \rightarrow \mathcal{S}$ 
is a map from a program point to a security level. 
The notation $\Rightarrow$ represents a relation between the 
stack type before execution and the stack type after execution of an instruction. 

The typing system is formally parameterized by :
\begin{description}
	\item[$\Gamma$:] a table of method signatures, needed to define the 
	transfer rules for method invocation;
	\item[$\fnn{ft}$:] a map from fields to their global policy level;
	\item[CDR:] a structure consists of ($\fnn{region}$, $\fnn{jun}$).
	\item[$se$:] security environment
	\item[$sgn$:] method signature of the current method
\end{description}
thus the complete form of a judgement parameterized by a tag 
$\tau \in \{\norm + \mathcal{C}\}$ is 
  \[ \Gamma, \fnn{ft}, \fnn{region}, se, sgn, i \vdash^{\tau} S_i \Rightarrow st \] 
although in the case where some elements are unnecessary, we may omit 
some of the parameters e.g. $i \vdash S_i \Rightarrow st$

As in the operational semantics, wherever it is clear that the instructions may not 
throw an exception, we remove the tag $\norm$ to reduce clutter. The transfer rules are 
contained in Figure~\ref{figure:javaTypeRuleSelected} 
(for the full list of transfer rules, see Figure~\ref{figure:javaTypeRuleFull} 
in Appendix~\ref{section:appendix_jvm}). 
Using these transfer rules, we can then define the notion of typability:


\begin{definition}[Typable method]
	A method $m$ is typable w.r.t. a method signature table $\Gamma$, a global field 
	policy $\fnn{ft}$, a policy $sgn$, and a CDR $\fnn{region_m} : \mathcal{PP} 
	\rightarrow \wp(\mathcal{PP})$ if there exists a security environment 
	$se : \mathcal{PP} \rightarrow \mathcal{S}$ and a function 
	$S : \mathcal{PP} \rightarrow \mathcal{S}^*$ s.t. $S_1 = \epsilon$ and for all 
	$i, j \in \mathcal{PP}$, and exception tags $e \in \{\norm + \mathcal{C}\}$:
	\begin{enumerate}
		\item[(a)] $i \mapsto^{e} j$ implies there exists $st \in \mathcal{S}^*$ such that 
		$\Gamma, \fnn{ft}, \fnn{region}$, $se$, $sgn$, $i$ $\vdash^e S_i \Rightarrow st$ 
		and $st \sqsubseteq S_j$;
		
		\item[(b)] $i \mapsto^e$ implies $\Gamma, \fnn{ft}, \fnn{region}, se, sgn, i 
		\vdash^e S_i \Rightarrow$
	\end{enumerate}
	where $\sqsubseteq$ denotes the point-wise partial order on type stack w.r.t. the
	partial order taken on security levels.
\end{definition}

The Non-interference definition relies on the notion of indistinguishability. Loosely 
speaking, a method is non-interferent whenever given indistinguishable inputs, it 
yields indistinguishable outputs. To cater for this definition, first there are 
definitions of indistinguishability. 

To define the notions of location, object, and array indistinguishability itself 
Barthe et. al. define the notion of a $\beta$ mapping. 
$\beta$ is a bijection on (a partial set of ) 
locations in the heap. The bijection maps low objects (objects whose 
references might be stored in low fields or variables) allocated in the heap of the 
first state to low objects allocated in the heap of the second state. The object might 
be  indistiguishable, even if their locations are different during execution.
\begin{definition}[Value indistinguishability]
	Letting $v, v_1, v_2 \in \mathcal{V}$, and given a partial function 
	$\beta \in \mathcal{L} \rightharpoonup \mathcal{L}$, the relation
	$\sim_{\beta} \subseteq \mathcal{V} \times \mathcal{V}$ is defined by the clauses :
\[
\begin{array} {ccc}
	null \sim_{\beta} null  & \dfrac{v \in \mathcal{N}}{v \sim_{\beta} v} & 
	\dfrac{v_1, v_2 \in \mathcal{L}\ \ \ \ \beta(v_1) = v_2}{v_1 \sim_{\beta} v_2}
\end{array}
\]
\end{definition}

\begin{definition}[Local variables indistinguishability] 
	For $\rho, \rho' : \mathcal{X} \rightharpoonup \mathcal{V}$, we have 
	$\rho \sim_{\kobs, \vec{k_a}, \beta} \rho'$ 
	if $\rho$ and $\rho'$ have the same domain and 
	$\rho(x) \sim_{\beta} \rho'(x)$ for all $x \in \fn{dom}{\rho}$ such that 
	$\vec{k_a}(x) \leq \kobs$.
\end{definition}

\begin{definition}[Object indistinguishability] 
	Two objects $o_1, o_2 \in \mathcal{O}$ are indistinguishable with respect to a 
	partial function $\beta \in \mathcal{L} \rightharpoonup \mathcal{L}$ 
	(noted by $o_1 \sim_{\kobs, \beta} o_2$) if and only if $o_1$ and 
	$o_2$ are objects of the same class and $o_1.f \sim_{\beta} o_2.f$
	for all fields $f \in \fn{dom}{o_1}$ s.t. $\fn{ft}{f} \leq \kobs$.
\end{definition}

\begin{definition}[Array indistinguishability]
	Two arrays $a_1, a_2 \in \mathcal{A}$ are indistinguishable w.r.t. an attacker level 
	$\kobs$ and a partial function $\beta \in \mathcal{L} \rightharpoonup \mathcal{L}$ 
	(noted by $a_1 \sim_{\kobs, \beta} o_2$)
	if and only if $a_1.\field{length} = a_2.\field{length}$ and, moreover, if 
	$\fn{at}{a_1} \leq \kobs$, then $a_1[i] \sim_{\beta} a_2[i]$
	for all $i$ such that $0 \leq i < a_1.\field{length}$.
\end{definition}

\begin{definition}[Heap indistinguishability] 
	Two heaps $h_1$ and $h_2$ are indistinguishable, written 
	$h_1 \sim_{\kobs, \beta} h_2$, 
	with respect to an attacker level \kobs and a partial function 
	$\beta \in \mathcal{L} \rightharpoonup \mathcal{L}$ iff:
	\begin{itemize}
		\item $\beta$ is a bijection between $\fn{dom}{\beta}$ and $\fn{rng}{\beta}$;
		
		\item $\fn{dom}{\beta} \subseteq \fn{dom}{h_1}$ and $\fn{rng}{\beta} \subseteq 
		\fn{dom}{h_2}$;
		
		\item $\forall l \in \fn{dom}{\beta}, h_1(l) \sim_{\kobs, \beta} h_2(\beta(l))$ 
		where $h_1(l)$ and $h_2(\beta(l))$ are either two objects or two arrays.
	\end{itemize}
\end{definition}

\begin{definition}[Output indistinguishability]
	Given an attacker level \kobs, a partial function $\beta \in \mathcal{L} 
	\rightharpoonup \mathcal{L}$, an output level $\vec{k_r}$, the
	indistinguishability of two final states in method $m$ is defined by the clauses :
	\[ \begin{array}{c}
		\dfrac{h_1 \sim_{\kobs, \beta} h_2\ \ \ \ 
		  \vec{k_r}[n] \leq \kobs \rightarrow v_1 \sim_{\beta} v_2}
		{(v_1, h_1) \sim_{\kobs, \beta, \vec{k_r}} (v_2, h_2)} \\\\
		
		\dfrac{h_1 \sim_{\kobs, \beta} h_2\ \ (\fn{class}{h_1(l_1)} : k) \in \vec{k_r}\ \ \ 
		k \leq \kobs\ \ l_1 \sim_{\beta} l_2}
		{(\langle l_1 \rangle, h_1) \sim_{\kobs, \beta, \vec{k_r}} 
		  (\langle l_2 \rangle, h_2)} \\\\
			
		\dfrac{h_1 \sim_{\kobs, \beta} h_2\ \ \ \ (\fn{class}{h_1(l_1)} : k) \in \vec{k_r}
		  \ \ \ \ k \nleq \kobs}
		{(\langle l_1 \rangle, h_1) \sim_{\kobs, \beta,\vec{k_r}}(v_2, h_2)} \\					

	\end{array}\]
	\[\begin{array}{c}

		\dfrac{h_1 \sim_{\kobs, \beta} h_2\ \ \ \ (\fn{class}{h_2(l_2)} : k) \in \vec{k_r}
		  \ \ \ \ k \nleq \kobs}
		{(v_1, h_1) \sim_{\kobs, \beta, \vec{k_r}} (\langle l_2 \rangle, h_2)} \\\\			

		\dfrac{\begin{gathered} h_1 \sim_{\kobs, \beta} h_2\ \ \ \ 
		  (\fn{class}{h_1(l_1)} : k_1) \in \vec{k_r}\ \ \ \ k_1 \nleq \kobs\\
			(\fn{class}{h_2(l_2)} : k_2) \in \vec{k_r}\ \ \ \ k_2 \nleq \kobs \end{gathered}}
		{(\langle l_1 \rangle, h_1) \sim_{\kobs, \beta, \vec{k_r}} 
		  (\langle l_2 \rangle, h_2)} \\	
	\end{array}
	\]
	where $\rightarrow$ indicates logical implication. 
\end{definition}

At this point it is worth mentioning that whenever it is clear from the usage, we may
drop some subscript from the indistinguishability relation, e.g. 
for two indistinguishable objects $o_1$ and $o_2$ w.r.t. a partial function
$\beta \in \mathcal{L} \rightharpoonup \mathcal{L}$ and observer level \kobs, 
instead of writing $o_1 \sim_{\kobs, \beta} o_2$ 
we may drop \kobs \ and write $o_1 \sim_{\beta} o_2$ if \kobs is obvious.
We may also drop $k_h$ from a policy $\vec{k_a}\overset{k_h}{\rightarrow}\vec{k_r}$
and write $\vec{k_a}\rightarrow\vec{k_r}$ if $k_h$ is irrelevant to the discussion. 

\begin{definition}[Non-interferent JVM method]
	A method $m$ is \emph{non-interferent} w.r.t. a policy $\vec{k_a} \rightarrow 
	\vec{k_r}$, if for every attacker level \kobs, every partial function 
	$\beta \in \mathcal{L} \rightharpoonup \mathcal{L}$ and every $\rho_1, \rho_2 \in 
	\mathcal{X} \rightharpoonup \mathcal{V}, h_1, h_2, h'_1, h'_2 \in \textmd{Heap}, 
	r_1, r_2 \in \mathcal{V} + \mathcal{L}$ s.t. 
	\[\begin{array}{cc}
		\langle 1, \rho_1, \epsilon, h_1 \rangle \rightsquigarrow^{+}_m r_1, h'_1 & h_1 \sim_{\kobs, \beta} h_2 \\
		\langle 1, \rho_2, \epsilon, h_2 \rangle \rightsquigarrow^{+}_m r_2, h'_2 & \rho_1 \sim_{\kobs, \vec{k_a}, \beta} \rho_2\\
	\end{array}\]
	there exists a partial function $\beta' \in \mathcal{L} \rightharpoonup \mathcal{L}$ 
	s.t. $\beta \subseteq \beta'$ and 
	\[(r_1, h'_1) \sim_{\kobs, \beta', \vec{k_a}} (r_2, h'_2)\]
\end{definition}

Because of method invocation, there will be a notion of a side effect preorder for the 
notion of safety.
\begin{definition}[Side effect preorder]
	Two heaps $h_1, h_2 \in \object{Heap}$ are \emph{side effect preordered} 
	(written as $h_1 \preceq_k h_2$) with respect 
	to a security level $k \in \mathcal{S}$  if and only 
	if $\fn{dom}{h_1} \subseteq \fn{dom}{h_2}$ and $ h_1(l).f = h_2(l).f$
	for all location $l \in \fn{dom}{h_1}$ 
	and all fields $f \in \mathcal{F}$ such that $k \nleq \fn{ft}{f}$.
\end{definition}

From which we can define a \textit{side-effect-safe} method.

\begin{definition}[Side effect safe]
	A method $m$ is \emph{side-effect-safe} with respect to a security level $k_h$ if 
	for all local variables $x \in \fn{dom}{\rho}$, 
	$\rho \in \mathcal{X} \rightharpoonup \mathcal{V}$, all 
	heaps $h, h' \in \object{Heap}$ and value $v \in \mathcal{V}, \langle 1, \rho, 
	\epsilon, h \rangle \rightsquigarrow^{+}_m v, h'$ implies $h \preceq_{k_h} h'$. 
\end{definition}

\begin{definition}[Safe JVM method]
	A method $m$ is \emph{safe} w.r.t. a policy $\vec{k_a} \overset{k_h}{\rightarrow} 
	\vec{k_r}$ if $m$ is side-effect safe w.r.t. $k_h$ and $m$ is non-interferent w.r.t. 
	$\vec{k_a} \rightarrow \vec{k_r}$.
\end{definition}

\begin{definition}[Safe JVM program]
	A program is \emph{safe} w.r.t. a table $\Gamma$ of method signature if every 
	method $m$ is safe w.r.t. all policies in $\fn{Policies_{\Gamma}}{m}$.
\end{definition}

\begin{theorem}
	Let $P$ be a JVM typable program w.r.t. safe CDRs ($\fnn{region_m}, \fnn{jun_m}$) and 
	a table $\Gamma$ of method signatures. Then $P$ is safe w.r.t. $\Gamma$.
\end{theorem}

\section {DEX Type System}\label{dexTypeSystem}

\begin{figure}
\[\boxed{\begin{array}{lll}
  \ins{binop}\ op& r, r_a, r_b & \rho(r) = \rho(r_a) op\ \rho(r_b). \\
  \ins{const } & r, v & \rho(r) = v \\
  \ins{move} & r, r_s & \rho(r) = \rho(r_s) \\
  \ins{ifeq} & {r, t} & \text{conditional jump if $\rho(r) = 0$} \\
  \ins{ifneq} & {r, t} & \text{conditional jump if s$\rho(r) \neq 0$} \\
  \ins{goto} & {t} & \text{unconditional jump} \\
  \ins{return} & {r_s} & \text{return the value of $\rho(r_s)$} \\
  \ins{new} & r, c & \rho(r) = \text{new object of class $c$} \\
  \ins{iget} & {r, r_o, f} & \rho(r) = \rho(r_o).f\\
  \ins{iput} & {r_s, r_o, f} & \rho(r_o).f = \rho(r) \\
  \ins{newarray} & r, r_l, t & r = \text{new array of type $t$ with $r_l$} \\
    && \text{number of elements} \\	
  \ins{arraylength} & r,r_a & \rho(r) = \rho(r_a).\field{length} \\ 				
  \ins{aput } & {r_s, r_a, r_i} & \rho(r_a)[\rho(r_i)] = \rho(r_s)\\		
  \ins{invoke} & n, m, \vec{p} & \text{invoke } \rho(\vec{p}[0]).m \text{ with $n$} \\
    && \text{arguments stored in $\vec{p}$} \\	
  \ins{moveresult} & {r} & \text{store invoke's result to $r$. Must} \\
    && \text{be placed directly after invoke} \\
  \ins{throw} & {r} & \text{throw the exception in $r$}\\
  \ins{move-} & {r} & \text{store exception in $r$. Have to} \\
    \ins{exception}&& \text{be the first in the handler.} \\
  \hline
  \\
  \multicolumn{3}{l}{\text{where }op \in \{+, -, \times, /\}, v \in \mathbb{Z}, 
    \{r, r_a, r_b, r_s\} \in \mathcal{R}, t \in \mathcal{PP}, }\\

  \multicolumn{3}{l}{c \in \mathcal{C}, f \in \mathcal{F}\text{ and } \rho : 
    \mathcal{R} \rightarrow \mathbb{Z}.}\\

\end{array}
}\]
\caption{DEX Instruction List}
\label{figure:dexInstructionList}
\end{figure}

A program \textit{P} is given by its list of instructions in 
Figure~\ref{figure:dexInstructionList}. 
The set $\mathcal{R}$ is the set of DEX virtual registers, $\mathcal{V}$ is the set of values, and 
$\mathcal{PP}$ is the set of program points. 
Since the DEX translation involves 
simulation of the JVM which uses a stack, we will also distinguish the registers :
\begin{itemize}
	\item registers used to store the local variables,
	\item registers used to store parameters, 
	\item and registers used to simulate the stack. 
\end{itemize}
In practice, there is no difference between registers used to simulate the stack and
those that are used to store local variables.
The translation of a JVM method 
refers to code which assumes that the parameters are already copied to the local 
variables.

\begin{figure*}
\[\boxed{
\begin{array}{c}
\begin{array}{ccc}
	\hspace{0.3\textwidth} & \hspace{0.3\textwidth} & \hspace{0.3\textwidth}\\
	
	\hfill \dfrac{P_m[i] = \dexins{const}{r, v}\ \ r \in \fn{dom}{\rho}}
	{\langle i, \rho, h \rangle\ \rightsquigarrow\ \langle i+1, \rho \oplus 
	  \{r \mapsto v\}, h \rangle} \hfill & 
	\hfill \dfrac{P[i]_m = \dexins{ifeq}{r, j}\ \ \ \rho(r) = 0}
	{\langle i, \rho, h \rangle\ \rightsquigarrow\ \langle t, \rho, h \rangle} \hfill &
	\hfill \dfrac{P_m[i] = \dexins{ifeq}{r, t}\ \ \ \rho(r) \neq 0}
	{\langle i,\rho,h \rangle\ \rightsquigarrow\ \langle i+1, \rho, h \rangle} \hfill\\\\
		
	\hfill \dfrac{P[i]_m = \dexins{return}{r_s}\ \ r_s\in \fn{dom}{\rho}}
	{\langle i, \rho, h \rangle\ \rightsquigarrow\ \rho(r_s), h} \hfill &	
	\hfill \dfrac{P_m[i] = \dexins{move}{r, r_s}\ \ r \in \fn{dom}{\rho}}
	{\langle i, \rho, h \rangle\ \rightsquigarrow\ \langle i+1, \rho \oplus 
	  \{r \mapsto \rho(r_s)\}, h \rangle} \hfill & 
	\hfill \dfrac{P_m[i] = \dexins{goto}{t}}
	{\langle i, \rho, h \rangle\ \rightsquigarrow\ \langle t, \rho, h \rangle} \hfill \\\\

\end{array}\\
\begin{array}{c}
				
	\dfrac{P_m[i] = \dexins{binop}{op, r, r_a, r_b}\ \ r, r_a, r_b \in \fn{dom}{\rho}
	  \ \ n = \rho(r_a)\ op\ \rho(r_b)}
	{\langle i, \rho, h \rangle\ \rightsquigarrow\ \langle i+1, \rho \oplus 
	  \{r \mapsto n\}, h \rangle} \\\\
	  
\end{array}
\end{array}}\]

\caption{DEX Operational Semantic (Selected)}
\label{figure:dexOperationalSemanticSelected}
\end{figure*}
As in the case for JVM, we assume that the program comes equipped with 
the set of class names $\mathcal{C}$ and the set of fields $\mathcal{F}$.
The program will also be extended with array 
manipulation instructions, and the program will come parameterized by the set of 
available DEX types $\mathcal{T}_D$ analogous to Java type $\mathcal{T}_J$.
The DEX language also deals with method invocation. As for JVM, DEX 
programs will also come with a set $m$ of method names. The method name and signatures 
themselves are represented explicitly in the DEX file, as such the lookup function 
required will be different from the JVM counterpart in that we do not need the class 
argument, thus in the sequel we will remove this lookup function and overload 
that method ID to refer to the code as well. DEX uses two special registers. We 
will use $ret$ for the first one which can hold the return value of a method 
invocation. In the case of a $\ins{moveresult}$, the instruction behaves like a 
$\ins{move}$ instruction with the special register $ret$ as the source register. 
The second special register is $ex$ 
which stores an exception thrown for the next instruction. 
Figure~\ref{figure:dexInstructionList} contains the list of DEX instructions.

\textbf{Operational Semantics} A state in DEX is just $\cstate{i, \rho, h}$ where the 
$\rho$ here is a mapping from registers to values and $h$ is the heap. 
As for the JVM in handling the method invocation, 
operational semantics are also extended to have a big 
step semantics for the method invoked. 
Figure~\ref{figure:dexOperationalSemanticSelected}
shows some of the operational semantics for DEX instructions.
Refer to Figure~\ref{figure:dexOperationalSemanticFull} in 
Appendix~\ref{section:appendix_dex} for a full list of DEX operational semantics.

The successor relation closely resembles that of the JVM, 
instructions will have its next instruction as the successor, except jump instructions, 
return instructions, and instructions that throw an exception.

\textbf{Type Systems} The transfer rules of DEX are defined in terms of registers typing 
$rt:\RTSpec$ instead of stack typing. 
Note that this registers typing is total w.r.t. the registers used in the method. To be more concrete,
if a method only uses 16 registers then $rt$ is a map for these 16 registers to security levels,
as opposed to the whole number of 65535 registers.

The transfer rules also come equipped with a security policy for fields 
$\fnn{ft} : \mathcal{F} \rightarrow \sext$ and 
$\fnn{at} : \mathcal{PP} \rightarrow \sext$. Some of the transfer rules for DEX 
instructions are contained in Figure~\ref{figure:dexTypeRuleSelected}. Full transfer 
rules are contained in Appendix~\ref{section:appendix_dex}.

\begin{figure*}
\[
\boxed{
\begin{array}{c}
\begin{array}{cc}
	\hspace{0.45\textwidth} & \hspace{0.45\textwidth}\\

	\hfill \dfrac{P_m[i] = \dexins{const}{r, v}}
	{i \vdash rt \Rightarrow rt \oplus \{r \mapsto se(i)\} } \hfill & 
	
	\dfrac{P_m[i] = \dexins{return}{r_s}\ \ \ \ se(i) \sqcup rt(r_s) \leq \vec{k_r}[n]}
	{i \vdash rt \Rightarrow}\\
	  
\end{array}\\
\begin{array}{cc}\hspace{0.45\textwidth} & \hspace{0.45\textwidth}\\

  \dfrac{P[i] = \dexins{move}{r, r_s}}
	{i \vdash rt \Rightarrow rt \oplus \{r \mapsto \big(rt(r_s) \sqcup se(i)\big)\}} 
	 &

  \dfrac{P_m[i] = \dexins{binop}{op, r, r_a, r_b}}
  {i \vdash rt \Rightarrow rt \oplus 
      \{r \mapsto \big(rt(r_a) \sqcup rt(r_b) \sqcup se(i)\big)\}} \\
  
\end{array}\\
\begin{array}{cc}
  \hspace{0.3\textwidth} & \hspace{0.6\textwidth} \\
	\dfrac{P_m[i] = \dexins{goto}{j}}
	  {i \vdash rt \Rightarrow rt} &

	\dfrac{P_m[i] = \dexins{ifeq }{r, t}\ \ \ 
	  \forall_{j'} \in \fn{region}{i, \norm}, se(i) \sqcup rt(r) \leq se(j')}
	{i \vdash rt \Rightarrow rt} \\

\end{array}\\\\

\end{array}}\]

\caption{DEX Transfer Rule (Selected)}
\label{figure:dexTypeRuleSelected}
\end{figure*}

The typability of the DEX closely follows that of the JVM, except that the relation 
between program points is $i \vdash RT_i \Rightarrow rt, rt \sqsubseteq RT_j$.
The definition of $\sqsubseteq$ is also defined in terms of point-wise registers.
For now we assume the existence of safe CDR with the same definition as that of 
the JVM side. We shall see later how we can construct a safe CDR for DEX from a
safe CDR in JVM. Formal definition of typable DEX method:
\begin{definition}[Typable method]
	A method $m$ is typable w.r.t. a method signature table $\Gamma$, a global field 
	policy $\fnn{ft}$, a policy $sgn$, and a CDR $\fnn{region_m} : \mathcal{PP} 
	\rightarrow \wp(\mathcal{PP})$ if there exists a security environment 
	$se : \mathcal{PP} \rightarrow \mathcal{S}$ and a function $\fnn{RT} : \mathcal{PP} 
	\rightarrow \RTSpec$ s.t. 
	$RT_1 = \vec{k_a}$ and for all $i, j \in \mathcal{PP}, e \in \{\norm + 
	\mathcal{C}\}$:
	\begin{itemize}
		\item $i \mapsto^{e} j$ implies there exists $rt \in \RTSpec$ 
		such that $\Gamma$, $\fnn{ft}$, $\fnn{region}$, $se$, $sgn$, $i \vdash^e 
		RT_i \Rightarrow rt$ and $rt \sqsubseteq RT_j$;
		
		\item $i \mapsto^e$ implies $\Gamma, \fnn{ft}, \fnn{region}, se, sgn, i \vdash^e 
		RT_i \Rightarrow$
	\end{itemize}
\end{definition}

Following that of the JVM side, what we want to establish here is not just the 
typability, but also that typability means non-interference. As in the JVM, the notion 
of non-interference relies on the definition of indistinguishability, while the notion 
of value indistinguishability is the same as that of JVM.

\begin{definition}[Registers indistinguishability]
  For $\rho, \rho' : \RTVal$ and $rt, rt' : \RTSpec$, we have 
  $\rho \sim_{\kobs, rt, rt', \beta} \rho'$ iff $\forall x \notin locR,$
  $rt(x) = rt'(x) = k, k \nleq \kobs$ or $rt(x) = k,$ $rt'(x) = k',$ $k \leq \kobs,$ $k' \leq \kobs,$
  $ \rho(x) = \rho'(x) = v$.
%
	where 
	$v \in \mathcal{V}$, and $k, k' \in \mathcal{S}$.
\end{definition}


\begin{definition}[Object indistinguishability] 
	Two objects $o_1, o_2 \in \mathcal{O}$ are indistinguishable with respect to a 
	partial function $\beta \in \mathcal{L} \rightharpoonup \mathcal{L}$ 
	(noted by $o_1 \sim_{\kobs, \beta} o_2$) if and only if $o_1$ and 
	$o_2$ are objects of the same class and $o_1.f \sim_{\beta} o_2.f$
	for all fields $f \in \fn{dom}{o_1}$ s.t. $\fn{ft}{f} \leq \kobs$.
\end{definition}

\begin{definition}[Array indistinguishability]
	Two arrays $a_1, a_2 \in \mathcal{A}$ are indistinguishable w.r.t. an attacker level 
	$\kobs$ and a partial function $\beta \in \mathcal{L} \rightharpoonup \mathcal{L}$ 
	(noted by $a_1 \sim_{\kobs, \beta} o_2$)
	if and only if $a_1.\field{length} = a_2.\field{length}$ and, moreover, if 
	$\fn{at}{a_1} \leq \kobs$, then $a_1[i] \sim_{\beta} a_2[i]$
	for all $i$ such that $0 \leq i < a_1.\field{length}$.
\end{definition}

\begin{definition}[Heap indistinguishability] 
	Two heaps $h_1$ and $h_2$ are indistinguishable with respect to an attacker level 
	\kobs and a partial function $\beta \in \mathcal{L} \rightharpoonup \mathcal{L}$, 
	written $h_1 \sim_{\kobs, \beta} h_2$, if and only if :
	\begin{itemize}
		\item $\beta$ is a bijection between $\fn{dom}{\beta}$ and $\fn{rng}{\beta}$;
		
		\item $\fn{dom}{\beta} \subseteq \fn{dom}{h_1}$ and $\fn{rng}{\beta} \subseteq 
		\fn{dom}{h_2}$;
		
		\item $\forall l \in \fn{dom}{\beta}, h_1(l) \sim_{\kobs, \beta} h_2(\beta(l))$ 
		where $h_1(l)$ and $h_2(\beta(l))$ are either two objects or two arrays.
	\end{itemize}
\end{definition}

\begin{definition}[Output indistinguishability]
	Given an attacker level \kobs, a partial function 
	$\beta \in \mathcal{L} \rightharpoonup \mathcal{L}$, 
	an output level $\vec{k_r}$, the indistinguishability of 
	two final states in method $m$ is defined by the clauses :
	\[ \begin{array}{c}
		\dfrac{h_1 \sim_{\kobs, \beta} h_2\ \ \ \ \vec{k_r}[n] \leq \kobs \Rightarrow 
		  v_1 \sim_{\beta} v_2}
		{(v_1, h_1) \sim_{\kobs, \beta, \vec{k_r}} (v_2, h_2)}\\\\

		\dfrac{h_1 \sim_{\kobs, \beta} h_2\ \ \ (\fn{class}{h_1(l_1)} : k) \in \vec{k_r}
		  \ \ \ k \leq \kobs\ \ l_1 \sim_{\beta} l_2}
		{(\langle l_1 \rangle, h_1) \sim_{\kobs, \beta, \vec{k_r}} 
		  (\langle l_2 \rangle, h_2)} \\\\

\end{array} \]	
\[		\begin{array}{c}

		\dfrac{h_1 \sim_{\kobs, \beta} h_2\ \ \ \ (\fn{class}{h_1(l_1)} : k) \in \vec{k_r}
		  \ \ \ \ k \nleq \kobs}
		{(\langle l_1 \rangle, h_1) \sim_{\kobs, \beta, \vec{k_r}} (v_2, h_2)} \\\\

		\dfrac{h_1 \sim_{\kobs, \beta} h_2\ \ \ \ (\fn{class}{h_2(l_2)} : k) \in \vec{k_r}
		  \ \ \ \ k \nleq \kobs}
		{(v_1, h_1) \sim_{\kobs, \beta, \vec{k_r}} (\langle l_2 \rangle, h_2)} \\\\		

		\dfrac{\begin{gathered} h_1 \sim_{\kobs, \beta} h_2\ \ \ \ 
		  (\fn{class}{h_1(l_1)} : k_1) \in \vec{k_r}\ \ \ \ k_1 \nleq \kobs\\
		  (\fn{class}{h_2(l_2)} : k_2) \in \vec{k_r}\ \ \ \ k_2 \nleq \kobs \end{gathered}}
		{(\langle l_1 \rangle, h_1) \sim_{\kobs, \beta, \vec{k_r}} 
		  (\langle l_2 \rangle, h_2)} \\\\
	  \end{array}
	\]
	where $\rightarrow$ indicates logical implication. 	
\end{definition}

\begin{definition}[Non-interferent DEX method]
	A method $m$ is \emph{non-interferent} w.r.t. a policy $\vec{k_a} \rightarrow 
	\vec{k_r}$, if for every attacker level \kobs, every partial function $\beta \in 
	\mathcal{L} \rightharpoonup \mathcal{L}$ and every $\rho_1, \rho_2 \in \mathcal{X} 
	\rightharpoonup \mathcal{V}, h_1, h_2, h'_1, h'_2 \in \object{Heap}, v_1, v_2 \in 
	\mathcal{V} + \mathcal{L}$ s.t. 
	\[\begin{array}{cc}
		\langle 1, \rho_1, h_1 \rangle \rightsquigarrow^{+}_m v_1, h'_1  & 		h_1 \sim_{\kobs, \beta} h_2\\
		\langle 1, \rho_2, h_2 \rangle \rightsquigarrow^{+}_m v_2, h'_2 & \rho_1 \sim_{\kobs, \vec{k_a}, \beta} \rho_2
	\end{array}\]
	there exists a partial function $\beta' \in \mathcal{L} \rightharpoonup \mathcal{L}$ 
	s.t. $\beta \subseteq \beta'$ and 
	\[(v_1, h'_1) \sim_{\kobs, \beta', \vec{k_a}} (v_2, h'_2)\]
\end{definition}

\begin{definition}[Side effect preorder]
	Two heaps $h_1, h_2 \in \text{Heap}$ are \textit{side effect preordered} with respect 
	to a security level $k \in \mathcal{S}$ (written as $h_1 \preceq_k h_2$) if and only 
	if $\fn{dom}{h_1} \subseteq \fn{dom}{h_2}$ and $h_1(l).f = h_2(l).f$ 
	for all location $l \in \fn{dom}{h_1}$ 
	and all fields $f \in \mathcal{F}$ such that $k \nleq \fn{ft}{f}$.
\end{definition}

\begin{definition}[Side effect safe]
	A method $m$ is \textit{side-effect-safe} with respect to a security level $k_h$ if 
	for all registers in $\rho \in \mathcal{R} \rightharpoonup \mathcal{V}, 
	\fn{dom}{\rho} = locR$, for all heaps $h, h' \in \object{Heap}$ and value 
	$v \in \mathcal{V}, \langle 1, \rho, h \rangle \rightsquigarrow^{+}_m v, h'$ implies 
	$h \preceq_{k_h} h'$.
\end{definition}

\begin{definition}[Safe DEX method]
	A method $m$ is \emph{safe} w.r.t. a policy $\vec{k_a} \overset{k_h}{\rightarrow} 
	\vec{k_r}$ if $m$ is side-effect safe w.r.t. $k_h$ and $m$ is non-interferent w.r.t. 
	$\vec{k_a} \rightarrow \vec{k_r}$.
\end{definition}

\begin{definition}[Safe DEX program]
	A program is \emph{safe} w.r.t. a table $\Gamma$ of method signatures if every
	method $m$ is safe w.r.t. all policies in $\text{Policies}_{\Gamma}(m)$.
\end{definition}

\begin{theorem}
\label{thm:DEXSoundness}
	Let $P$ be a DEX typable program w.r.t. safe CDRs ($\fnn{region_m}, \fnn{jun_m}$) and 
	a table $\Gamma$ of method signatures. Then $P$ is safe w.r.t. $\Gamma$.
\end{theorem}

\section{Examples}

Throughout our examples we will use two security levels $L$ and $H$ to indicate low and 
high security level respectively. We start with a simple example where a high guard is 
used to determine the value of a low variable.

\begin{example}
Assume that local variable 1 is $H$ and local variable 2 is $L$. For now also assume 
that $r_3$ is the start of the registers used to simulate the stack in the DEX 
instructions. Consider the following JVM bytecode and its translation.
	\[\begin{array}{rlrl}
		& \dots & & \dots\\
		& \ins{push}\ 0& & \dexins{const}{r_3, 0}\\
		& \ins{store}\ 2 & & \dexins{move}{r_2, r_3}\\
		& \ins{load}\ 1 & & \dexins{move}{r_3, r_1}\\
		& \ins{ifeq}\ l_1 & & \dexins{ifeq}{r_3, l_1}\\
		& \ins{push}\ 1 & & \dexins{const}{r_3, 1}\\
		& \ins{store}\ 2 &  & \dexins{move}{r_2, r_3}\\
		l_1: & \dots  & l_1:&  \dots
	\end{array}\]
In this case, the type system for the JVM bytecode will reject this example because 
there is a violation in the last $\ins{store}\ 2$ constraint. 
The reasoning is that $se(i)$ for $\ins{push}\ 1$ will be $H$, 
thus the constraint will be $H \sqcup H \leq L$ which can not be 
satisfied. The same goes for the DEX instructions. Since $r_3$ gets its value from 
$r_1$ which is $H$, the typing rule for $\dexins{ifeq}{r_3, l_1}$
states that $se$ in the last instruction will 
be $H$. Since the last $\ins{move}$ instruction is targetting variables in the local 
variables side, the constraint applies, which states that $H \sqcup H \leq L$ which 
also can not be satisfied, thus this program will be rejected by the DEX type system.
\end{example}


The following example illustrates one of the types of the interference caused by 
modification of low fields of a high object aliased to a low object.

\begin{example}
Assume that $\vec{k_a} = \{r_1 \mapsto H, r_2 \mapsto H, r_3 \mapsto L\}$ (which means 
local variable 1 is high, local variable 2 is high and local variable 3 is low). Also 
the field $f$ is low ($\fn{ft}{f} = L$).
	\[\begin{array}{rlrl}
		& \dots & & \dots\\
		& \ins{new}\ C& & \dexins{new}{r_4, C}\\
		& \ins{store}\ 3 & & \dexins{move}{r_3, r_4}\\
		& \ins{load}\ 2& & \dexins{move}{r_4, r_2} \\
		l_1:& \ins{ifeq}\ l_2& l_1:& \dexins{ifeq}{r_4, l_2}\\
		& \ins{new}\ C& & \dexins{new}{r_4, C}\\
		& \ins{goto}\ l_3& & \dexins{goto}{l_3}\\
		l_2: & \ins{load}\ 3& l_2:& \dexins{move}{r_4, r_3}\\
		l_3: & \ins{store}\ 1& l_3:& \dexins{move}{r_1, r_4}\\
		& \ins{load}\ 1& & \dexins{move}{r_4, r_1}\\
		& \ins{push}\ 1& & \dexins{const}{r_5, 1}\\
		& \ins{putfield}\ f& & \dexins{iput}{r_5, r_4, f} \\
		& \dots & & \dots
	\end{array}\]
The above JVM bytecode and its translation 
will be rejected by the type system for the JVM bytecode because for 
$\ins{putfield}\ f$ at the last line 
there is a constraint with the security level of the object. 
In this case, the $\ins{load}\ 1$ instruction will push 
a reference of the object with high security level,
therefore, the constraint that $L \sqcup H \sqcup L \leq L$ can not be satisfied. 
The same goes for the DEX type system, it will also reject the translated program. 
The reasoning is that the $\dexins{move}{r_4, r_1}$ instruction will copy a 
reference to the object stored in $r_1$ which has a high security level, therefore 
$rt(r_4) = H$.
Then, at the $\dexins{iput}{r_5, r_4, f}$ we won't be able 
to satisfy $L \sqcup H \sqcup L \leq L$.
\end{example}

This last example shows that the type system also handles information flow through 
exceptions.

\begin{example}
Assume that $\vec{k_a} = \{r_1 \mapsto H, r_2 \mapsto L, r_3 \mapsto H\}$ and
$\vec{k_r} = \{n \mapsto L, e \mapsto H, \np \mapsto H\}$. 
$\fn{Handler}{l_2, \np} = l_h$, and for any $e, \fn{Handler}{l_2, e} \uparrow$. 
The following JVM bytecode and its translation 
will be rejected by the typing system for the JVM bytecode. 
	\[\begin{array}{rlrl}
		& \dots & & \dots\\
		& \ins{load}\ 1& & \dexins{move}{r_4, r_1}\\
		l_1:& \ins{ifeq}\ l_2& l_1:& \dexins{ifeq}{r_4, l_2}\\
		& \ins{new}\ C& & \dexins{new}{r_4, C}\\
		& \ins{store}\ 3& & \dexins{move}{r_3, r_4}\\
		& \ins{load}\ 3& & \dexins{move}{r_4, r_3}\\
		l_2:& \ins{invokevirtual}\ m & l_2:& \dexins{invoke}{1, m, r_4}\\
		& \ins{push}\ 0& & \dexins{const}{r_4, 0} \\
		& \ins{return}& ret & \dexins{move}{r_0, r_4}\\ 
		& & & \dexins{return}{r_0}\\ 
		l_h:& \ins{push}\ 1& l_h:& \dexins{const}{r_4, r_1} \\
		& \ins{return}& & \dexins{move}{r_0, r_4}\\ 
		& & & \dexins{goto}{ret}\\
		& \dots & & \dots
	\end{array}\]
The reason is that the typing constraint for the 
$\ins{invokevirtual}$ will be separated into several tags, 
and on each tag of execution we will have $se$ as high 
(because the local variable 3 is high). 
Therefore, when the program reaches $\ins{return}\ 2$ (line  8 and 10) 
the constraint is violated since we have $\vec{k_r}[n] = L$, 
thus the program is rejected. 
Similar reasoning holds for the DEX type system as well, 
in that the $\ins{invoke}$ will have $se$ high because the 
object on which the method is invoked upon is high, 
therefore the typing rule will reject the program because it can not 
satisfy the constraint when the program is about to store the value in local variable 2 
(constraint $H \leq L$ is violated, where $H$ comes from lub with $se$).
\end{example}

\section{Translation Phase}
\label{section:Translation}

Before we continue to describe the translation processes, 
we find it helpful to first define a construct called the basic block. 
The Basic block is a construct containing a group of 
code that has one entry point and one exit point 
(not necessarily one successor/one parent), has parents list, successors list, 
primary succesor, and its order in the output phase.
There are also some auxilliary functions :

\noindent\begin{tabular}{cp{6.8cm}}
  BMap & is a mapping function from program
  pointer in JVM bytecode to a DEX basic block.\\
  
  SBMap & Similar to BMap, this function takes a program pointer 
  in JVM bytecode and returns whether that instruction is the start of a DEX basic 
  block. \\
  TSMap & A function that maps a program pointer in JVM bytecode 
  to an integer denoting the index to the top of the stack. Initialized with the number 
  of local variables as that index is the index which will be used by DEX to simulate 
  the stack.\\

  NewBlock & A function to create a new Block and associate it with the 
  given instruction.
\end{tabular}

The DEX bytecode is simulating the JVM bytecode although 
they have different infrastructure. DEX is register-based
whereas JVM is stack-based. To bridge this gap, DEX uses 
registers to simulate the stack. The way it works :
\begin{itemize}
  \item $l$ number of registers are used to hold local variables.
    ($1, \dots, l$). We denote these registers with $locR.$
  \item Immediately after $l$, there are $s$ number of registers which are used to 
    simulate the stack ($l+1, \dots, l+s$).
\end{itemize}
Note that although in principal stack can grow indefinitely, it is impossible to write
a program that does so in Java, due to the strict stack discipline in Java.
Given a program in JVM bytecode, it is possible to statically determine the height of
the operand stack at each program point. This makes it possible to statically map
each operand stack location to a register in DEX (cf. TSMap above and
Appendix \ref{apx:translate}); see \cite{Davis03IVME} for a discussion on how this can be done.

There are several phases involved to translate JVM bytecode into DEX bytecode:\\

  \noindent\textbf{StartBlock}: Indicates the program point at which the
  instruction starts a block, then creates a new block for each 
  of these program points and associates it with a new empty block.

  \noindent\textbf{TraceParentChild}: Resolves the parents successors (and primary 
  successor) relationship between blocks. 
  Implicit in this phase is a step creating a temporary return 
  block used to hold successors of the block containing return instruction. At this 
  point in time, assume there is a special label called $ret$ to address this temporary 
  return block. 
  
  The creation of a temporary return block depends on whether the function 
  returns a value.
    If it is return void, then this block contains only the instruction 
    return-void.   
    Otherwise depending on the type returned (integer, wide, object, etc), 
    the instruction is translated into the corresponding $\ins{move}$ and 
    $\ins{return}$. The \ins{move} instruction moves the value from the register 
    simulating the top of the stack to register 0. Then \ins{return} will just return 
    $r_0$.
  
  \noindent\textbf{Translate}: Translate JVM instructions into DEX 
  instructions.
   
  \noindent\textbf{PickOrder}: Order blocks according to ``trace analysis''.
  
  \noindent\textbf{Output}: Output the instructions in order. 
  During this phase, \ins{goto} will be added for each block whose next block
  to output is not its successor. After the compiler has output all blocks, it will
  then read the list of DEX instructions and fix up the targets of jump instructions.
  Finally, all the information about exception handlers is collected and put in 
  the section that deals with exception handlers in the DEX file structure.

\begin{definition}[Translated JVM Program] The translation of a JVM program $P$ 
into blocks and have their JVM instructions translated into DEX instructions is 
denoted by $\translate{ P }$, where
  \[ \translate{ P } = \fn{Translate}{\fn{TraceParentChild}{\fn{StartBlock}{P}}}.\]
\end{definition}

\begin{definition}[Output Translated Program] The output of the translated JVM program 
$\translate{P}$ in which the blocks are ordered and then output into DEX program is 
denoted by $\ordout{ \translate{ P }}$, where
  \[ \ordout{\translate{P}} = \fn{Output}{\fn{PickOrder}{\translate{P}}}.\]
\end{definition}

\begin{definition}[Compiled JVM Program] The compilation of a JVM program $P$ is 
denoted by $\compile{ P }$, where
  \[ \compile{ P } = \ordout{ \translate{P} }. \]
\end{definition}

Details for each of the phase can be seen in appendix~\ref{section:TranslationAppendix}.

\section {Proof that Translation Preserves Typability}

\subsection{Compilation of CDR and Security Environments}

Since now we will be working on blocks, we need to know how the CDRs 
of the JVM and that of the translated DEX are related. 
First we need to define the definition of the successor relation between blocks.

\begin{definition}[Block Successor]
\label{blockSuccessor}
	Suppose $a \mapsto b$ and $a$ and $b$ are on different blocks. Let $B_a$ be the block 
	containing $a$ and $B_b$ be the block containing $b$. Then $B_b$ will be the 
	successor of $B_a$ denoted by abusing the notation $B_a \mapsto B_b$.
\end{definition}

Before we continue on with the properties of CDR and SOAP, we first need to define the 
translation of $\fnn{region}$ and $\fnn{jun}$ since we assume that the JVM bytecode 
comes equipped with $\fnn{region}$ and $\fnn{jun}$. 

\begin{definition}[Region Translation and Compilation]
\label{regionTranslation}
  Given a JVM $\fn{region}{i, \tau}$ and $P[i]$ is a branching instruction, 
  let $i_b$ be the program point in $\translate{i}$ such that 
  $P_\mathrm{DEX}[i_b]$ is a branching instruction, then 
  \[\begin{array}{c}
    \translate{\fn{region}{i, \tau}} = \fn{region}{i_b, \translate{\tau}} = 
      \bigcup_{j \in \fn{region}{i, \tau}} \translate{j}\\
    \text{and}\\
    \compile{\fn{region}{i, \tau}} = \fn{region}{i_b, \compile{\tau}} = 
      \bigcup_{j \in \fn{region}{i, \tau}} \compile{j}\\
  \end{array}\]
\end{definition}

\begin{definition}[Region Translation and Compilation for $\ins{invoke}$]
\label{regionTranslationInvoke}
  $\forall i. P_{DEX}[i] = \ins{invoke}$, $i + 1 \in \fn{region}{i, \norm}$ ($i + 1$ 
  will be the program point for $\ins{moveresult}$).
\end{definition}

\begin{definition}[Region Translation and Compilation for handler]
\label{regionTranslationHandler}
  $\forall i, j.  j \in \fn{region}{i, \tau}$, let $i_e$ be the instruction in 
  $\translate{ P[i] }$ that possibly throws, then 
  \[ \begin{array}{c}
  \fn{handler}{i_e, \tau} \in \fn{region}{i_e, \tau} \text{ in } \translate{P} \\
  \text{and}\\
   \fn{handler}{\ordout{i_e},\tau}\in\fn{region}{\ordout{i_e},\tau}
  \text{ in }\compile{P} 
  \end{array}\]
  (note that the handler will point to $\ins{moveexception}$).
\end{definition}

\begin{definition}[Region for appended $\ins{goto}$ instruction]
\label{regionTranslationGoto}
  \[\begin{array}{ll}
    \forall b \in \translate{P}. & 
    P_\text{DEX}[\ordout{b.\field{lastAddress}} + 1] = \ins{goto} \\
    & \rightarrow \big( \forall. i \in \mathcal{PP}_\text{DEX}. 
    b.\field{lastAddress} \in \fn{region}{i, \tau} \\ 
    & \rightarrow (\ordout{b.\field{lastAddress}} + 1) \in \fn{region}{i, \tau} \big)
  \end{array}\]
  where $\rightarrow$ indicates logical implication.
\end{definition}

\begin{definition}[Junction Translation and Compilation]
\label{junTranslation}
  $\forall i, j. j = \fn{jun}{i, \tau}$, let $i_b$ be in $\translate{P[i]}$ that branch 
  then 
  \[\begin{array}{c} 
  \translate{j}[0] = \fn{jun}{\translate{i}[i_b], \tau} \text{ in }\translate{P} \\
  \text{and}\\
  \compile{\translate{j}[0]}=\fn{jun}{\compile{\translate{i}[i_b]}, \tau} \text{ in }
  \compile{P}.
  \end{array}\]
\end{definition}

\begin{definition}[Security Environment Translation and Compilation]
\label{def:se_translation}
  $\forall i \in \mathcal{PP}, j \in \translate{i}. se(j) = se(i)$ in \translate{P} and
  $\forall i \in \mathcal{PP}, j \in \translate{i}. se(\ordout{j}) = se(i)$ in 
  \compile{P}.
\end{definition}

\begin{lemma}[SOAP Preservation]
\label{lemma:preservedSOAP}
  The SOAP properties are preserved in the translation from JVM to DEX, i.e. if the JVM 
  program satisfies the SOAP properties, so does the translated DEX program.
\end{lemma}

\subsection{Compilation Preserves Typability}

There are several assumption we make for this compilation. 
  Firstly, the JVM program will not modify its self reference for an object. 
  Secondly, since now we are going to work in blocks, the notion of $se, S,$ and $RT$ 
  will also be defined in term of this addressing. 
  A new scheme for addressing $blockAddress$ 
  is defined from sets of pairs $(bi, j)$, $bi \in blockIndex$, a set of all block 
  indices (label of the first instruction in the block), where 
  $\forall i \in \mathcal{PP}. \exists bi, j.$ s.t.$bi + j = i$. 
  We also add additional relation $\Rightarrow^*$ to denote the reflexive and 
  transitive closure of $\Rightarrow$ to simplify the typing relation between blocks.

	We overload $\translate{.}$ and $\compile{ . }$ to also apply to stack type to 
	denote translation from 
	stack type into typing for registers. This translation basically just 
	maps each element of the stack to registers at the end of registers containing the 
	local variables (with the top of the stack with larger index, i.e. stack expanding to 
	the right), and fill the rest with high security level. 
	More formally, if there are $n$ local variables denoted by $v_1, \dots, 
	v_n$ and stack type with the height of $m$ ($0$ denotes the top of the stack),
	and the method has $o$ registers (which corresponds to the maximum depth of the stack), 
	then $\compile{st} = \{r_0 \mapsto \vec{k_a}(v_1), \dots,  
	r_{n - 1} \mapsto \vec{k_a}(v_n), r_{n} \mapsto st[m-1], \dots,  
	r_{n + m - 1} \mapsto st[0], r_{n + m} \mapsto H, \dots, r_{o} \mapsto H \}$.
	Lastly, the function $\ordout{ . }$ is also overloaded for addressing $(bi, i)$ to 
	denote abstract address in the DEX side which will actually be instantiated when 
	producing the output DEX program from the blocks.

  Due to the way stack type is translated to registers typing, we find it beneficial
  to introduce a simple lemma that can be proved trivially (by structural induction
  and definition) in regards to the $rt_1 \sqsubseteq rt_2$ relation. 
  In particular this lemma will relates the 
  registers metioned in $rt_1$ but are not mentioned in $rt_2$.

\begin{lemma}[Registers Not in Stack Less Equal]
\label{lemma:registersNotInStackLessEqual}
  Let the number of local variables be $locN$. For any two stack types $st_1, st_2, 
  \fn{length}{st_1} = n, \fn{length}{st_2} = m, m < n$, any register  
  $x \in \{r_{locN + m + 1}, \dots, r_{locN + m + n}\}$, and register types 
  $rt_1 = \translate{st_1}, rt_2 = \translate{st_2}$ we have that 
  $rt_1(x) \leq rt_2(x)$.
\end{lemma}

\begin{definition}[$\translate{\fnn{excAnalysis}}$ and $\compile{\fnn{excAnalysis}}$]
\label{def:excAnalysis_translation}
  \[\begin{array}{c}
  \forall m \in \mathcal{M}. \translate{\fn{excAnalysis}{m}} = \fn{excAnalysis}{m}
  \text{ in } \translate{P} \\
    
  \text{and}\\
  
  \forall m \in \mathcal{M}. \compile{\fn{excAnalysis}{m}} = \fn{excAnalysis}{m}
  \text{ in } \compile{P}.
  \end{array} \]
\end{definition}

\begin{definition}[$\translate{\fnn{classAnalysis}}$ and 
$\compile{\fnn{classAnalysis}}$]
\label{def:classAnalysis_translation}
  Let $e$ be the index of the throwing instruction from $\translate{i}$.
  \[\begin{array}{c} 
  \left( \begin{array}{c}\forall m \in \mathcal{M}, i \in \mathcal{PP}.
    \translate{\fn{classAnalysis}{m, \translate{i}[e]}} = \\ \fn{classAnalysis}{m, i}
    \text{ in } \translate{P} \end{array} \right) \\
  \text{and}\\
  \left(\begin{array}{c}\forall m \in \mathcal{M}, i \in \mathcal{PP}. 
    \translate{\fn{classAnalysis}{m, \ordout{\translate{i}[e]}}} = \\
    \fn{classAnalysis}{m, i} \text{ in } \compile{P}\end{array} \right).
  \end{array}\]
\end{definition}

\begin{definition}[$\translate{\Gamma}$ and $\compile{\Gamma}$]
\label{def:gamma_translation}
  $\forall m \in \mathcal{M}. \translate{\Gamma[\translate{m}]} = \Gamma[m]$ in 
  \translate{P}
  and
  $\forall m \in \mathcal{M}. \compile{\Gamma[\compile{m}]} = \Gamma[m]$ in
  \compile{P}.
\end{definition}

\begin{definition}
\label{def:st_translation}
  $\forall i \in \mathcal{PP}, RT_{\translate{i}[0]} = \translate{S_i}$.
\end{definition}

The idea of the proof that compilation from JVM bytecode to DEX bytecode preserves
typability is that 
any instruction that does not modify the block structure can be proved 
using Lemma~\ref{lemma:typable_rt}, Lemma~\ref{lemma:typable_rt_exception} and 
Lemma~\ref{lemma:typable_compile} to prove 
the typability of register typing. 

Initially we state lemmas saying that typable JVM instructions will yield typable DEX 
instructions. Paired with each normal execution is the lemma for the exception throwing
one. These lemmas are needed to handle the additional block of $\ins{moveexception}$
attached for each exception handler.

\begin{lemma}
\label{lemma:typable_rt}
  For any JVM program $P$ with instruction $Ins$ at address $i$ and tag $\norm$, 
  let the length of $\translate{ Ins }$ be $n$.
  Let $RT_{\translate{i}[0]}=\translate{ S_i }$.
  If according to the transfer rule for $P[i] = Ins$ there exists $st$ s.t. 
  $i \vdash^{\norm} S_{i} \Rightarrow st$ then
    \[ \begin{array}{c}
      \left( \begin{array}{l} 
        \forall 0 \leq j < (n-1). \exists rt'. \translate{i}[j] \vdash^{\norm}  \\  
        RT_{\translate{i}[j]} \Rightarrow rt', rt' \sqsubseteq RT_{\translate{i}[j+1]}
      \end{array} \right) \\ 
      \text{and} \\
      \exists rt. \translate{i}[n-1] \vdash^{\norm} RT_{\translate{i}[n-1]} \Rightarrow 
        rt, rt \sqsubseteq \translate{st} 
    \end{array}\] 
  according to the typing rule(s) of $\translate{ Ins }$.
\end{lemma}

\begin{lemma}
\label{lemma:typable_rt_exception}
  For any JVM program $P$ with instruction $Ins$ at address $i$ and tag 
  $\tau \neq \norm$ with exception handler at address $i_e$. 
  Let the length of $\translate{ Ins }$ until the instruction that 
  throws exception $\tau$ be denoted by $n$. 
  Let $(be, 0)=\translate{i_e}$ be the address of the handler for that particular 
  exception. If $i \vdash^{\tau} S_{i} \Rightarrow st$ according to the 
  transfer rule for $Ins$, then
    \[ \begin{array}{c}
      \left( \begin{array}{l}
        \forall 0 \leq j < (n-1). \exists rt'. \translate{i}[j] \vdash^{\norm} \\
        RT_{\translate{i}[j]} \Rightarrow rt', rt' \sqsubseteq RT_{\translate{i}[j+1]} 
      \end{array} \right)  \\
      \text{and} \\
      \exists rt. \translate{i}[n-1] \vdash^{\tau} RT_{\translate{i}[n-1]} \Rightarrow 
        rt, rt \sqsubseteq RT_{(be, 0)}
      \\ \text{and} \\
      \exists rt. (be, 0) \vdash^{\norm} RT_{(be, 0)} \Rightarrow rt, 
        rt \sqsubseteq \translate{st} 
    \end{array}\] 
  according to the typing rule(s) of the first $n$ instructions in 
  $\translate{ Ins }$ and $\ins{moveexception}$.
\end{lemma}

\begin{lemma}
\label{lemma:typable_compile}
  Let $Ins$ be an instruction at address $i$, $i \mapsto j$, $st$, $S_{i}$ and $S_{j}$ 
  are stack types such that $i \vdash S_{i} \Rightarrow st, st \sqsubseteq S_{j}$. 
  Let $n$ be the length of $\translate{ Ins }$. 
  Let $RT_{\translate{i}[0]} = \translate{ S_{i} }$, 
  let $RT_{\translate{j}[0]} = \translate{ S_{j}}$ 
  and $rt$ be registers typing obtained from the transfer rules involved in 
  $\translate{Ins}$. Then $rt \sqsubseteq RT_{\translate{j}[0]} $. 
\end{lemma}

We need an additional lemma to establish that the constraints 
in the JVM transfer rules are satisfied after the translation. 
This is because the definition of typability also relies on 
the constraint which can affect the existence of register typing.

\begin{lemma}
\label{lemma:constraintsatisfaction}
  Let $Ins$ be an instruction at program point $i$, $S_i$ its corresponding stack 
  types, and let $RT_{\translate{i}[0]} = \translate{ S_i }$. If $P[i]$ 
  satisfy the typing constraint for $Ins$ with the stack type $S_i$, then 
  $\forall (bj, j) \in \translate{i}. P_{DEX}[bj, j]$ 
  will also satisfy the typing constraints for all instructions in $\translate{ Ins }$ 
  with the initial registers typing $RT_{\translate{i}[0]}$.
\end{lemma}



Using the above lemmas, we can prove the lemma that all the resulting 
blocks will also be typable in DEX.

\begin{lemma}
\label{lemma:typable_blocks}
  Let $P$ be a Java program such that 
  \[ \forall i, j. i \mapsto j. \exists st. i \vdash S_i \Rightarrow st 
  \hspace{0.5cm} \text{ and } \hspace{0.5cm} st \sqsubseteq S_j \]
  Then $\translate{ P }$ will satisfy 
  \begin{enumerate}
    \item for all blocks $bi, bj$ s.t. $bi \mapsto bj$, $\exists rt_b.$ s.t.
      $RTs_{bi} \Rightarrow^* rt_b, rt_b \sqsubseteq RTs_{bj}$; and
    \item $\forall bi, i, j \in bi.$ s.t. $(bi, i) \mapsto (bi, j). \exists rt.$ s.t.
      $(bi, i) \vdash RT_{(bi, i)} \Rightarrow rt, rt \sqsubseteq RT_{(bi, j)}$
  \end{enumerate}
  where 
    \[ \begin{array}{lcl}
      RTs_{bi} = \translate{S_{i}} &\text{ with }& \translate{i} = (bi, 0) \\
      RTs_{bj} = \translate{S_{j}} &\text { with }& \translate{j} = (bj, 0), \\
      RT_{(bi, i)} = \translate {S_{i'}}&\text{ with }&\translate{i'} = (bi, i)\\
      RT_{(bi, j)} = \translate {S_{j'}} &\text{ with }& \translate{j'} = (bj, j).
      \end{array} \]  
\end{lemma}

After we established that the translation into DEX instructions in the form of blocks
preserves typability, 
we also need ensure that the next phases in the translation process also preserves
typability. The next phases are ordering the blocks, output the DEX code, 
then fix the branching targets. 

\begin{lemma}
\label{lemma:orderTypability}
  Let $\llfloor P \rrfloor$ be typable basic blocks resulting from translation of JVM
  instructions still in the block form, i.e.
  \[\llfloor P \rrfloor = \fn{Translate}{\fn{TraceParentChild}{\fn{StartBlock}{P}}}.\] 
  Given the ordering scheme to output the block contained in
  $\fnn{PickOrder}$, if the starting block starts with flag $0$ ($F_{(0,0)} = 0$) 
  then the output $\compile{ P }$ is also typable.
\end{lemma}

Finally, the main result of this paper in that the compilation of typable JVM bytecode
will yield typable DEX bytecode which can be proved from 
Lemma~\ref{lemma:typable_blocks} and Lemma~\ref{lemma:orderTypability}. Typable DEX 
bytecode will also have the non-interferent property because it is based on a safe CDR 
(Lemma~\ref{lemma:preservedSOAP}) according to DEX.

\begin{theorem}\label{theorem:main_theorem}
  Let $P$ be a typable JVM bytecode according to its safe CDR ($\fnn{region}, 
  \fnn{jun}$), PA-Analysis ($\fnn{classAnalysis}$ and $\fnn{excAnalysis}$), 
  and method policies $\Gamma$, then $\compile{ P }$ 
  according to the translation scheme has the property that 
    \[ \forall i, j \in PP_\mathrm{DEX}. \text{ s.t. } i \mapsto j. \exists rt. \text
    { s.t. } RT_i 
    \Rightarrow rt, rt \sqsubseteq RT_j \]
  according to a safe CDR ($\compile{ \fnn{region} }, \compile{ \fnn{jun} }$), 
  $\compile{ \mathit{PA-Analysis} }$, and $\compile{ \Gamma }$.  
\end{theorem}

\section{Conclusion and Future Work}

We presented the design of a type system for DEX programs and showed that
the non-optimizing compilation done by the dx tool preserves the typability
of JVM bytecode. Furthermore, the typability of the DEX program also
implies its non-interference.
We provide a proof-of-concept implementation illustrating the feasibility of the idea. 
This opens up
the possibility of reusing analysis techniques applicable to Java bytecode
for Android. As an immediate next step for this research, we plan to also take
into account the optimization done in the dx tool to see whether typability
is still preserved by the translation.

Our result is quite orthogonal to the Bitblaze project \cite{song2008bitblaze},
where they aim to unify different bytecodes into a common intermediate language, and then
analyze this intermediate language instead. At this moment, we
still do not see yet how DEX bytecode can be unified with this intermediate language
as there is a quite different approach in programming Android's applications, namely the
use of the message passing paradigm which has to be built into the Bitblaze 
infrastructure. This problem with message passing paradigm is essentially a limitation 
to our currentwork as well in that we still have not identified special object and 
method invocation for this message passing mechanism in the bytecode.

In this study, we have not worked directly with the dx tool; rather, we have written
our own DEX compiler in Ocaml based on our understanding of how the actual dx tool works.
This allows us to look at several sublanguages of DEX bytecode in isolation.
The output of our custom compiler
resembles the output from the dx compiler up to some details such as the size of register addressing.
Following the Compcert project 
\cite{leroy2006formal,blazy2006formal}, we would ultimately like to have 
a fully certified end to end compiler. We leave this as future work.

\bibliographystyle{IEEEtran}

\bibliography{biblio}

\appendices

\clearpage
\section{Translation Appendix}\label{section:TranslationAppendix}

This section details compilation phases described in Section~\ref{section:Translation} 
in more details. We first start this section by detailing the structure of basic block. 
$BasicBlock$ is a structure of \[\{parents;\ succs;\ pSucc;\ order;\ insn\}\] which 
denotes a structure of a basic block where $parents \subseteq \mathcal{Z}$ is a set of 
the block's parents, $succs \subseteq \mathcal{Z}$ is a set of the block's successors, 
$pSucc \in \mathcal{Z}$ is the primary successor of the block (if the block does not
have a primary successor it will have $-1$ as the value), $order \in \mathcal{Z}$ is 
the order of the block in the output phase, and $insn \cup DEXins$ is the DEX 
instructions contained in the block. The set of $BasicBlock$ is denoted as 
$BasicBlocks$.When instatiating the basic block, we denote the default object 
$NewBlock$, which will be a basic block with 
  \[\{parents=\emptyset;\ succs=\emptyset;\ pSucc=-1;
    \ order=-1;\ insn=\emptyset\}\]
    
Throughout the compilation phases, we also make use of two mappings 
$\fnn{PMap}:\mathcal{PP} \rightarrow \mathcal{PP}$ and 
$\fnn{BMap}:\mathcal{PP} \rightharpoonup BasicBlocks$. The mapping $\fnn{PMap}$ is a
mapping from a program point in JVM to a program point in JVM which starts a 
particular block, e.g. if we have a set of program points $\{2, 3, 4\}$ forming a basic
block, then we have that $\fn{PMap}{2}=2$, $\fn{PMap}{3}=2$, and $\fn{PMap}{4}=2$.
$\fnn{BMap}$ itself will be used to map a program point in JVM to a basic block.

\subsection{Indicate Instructions starting a Block ($\fnn{StartBlock}$)}

This phase is done by sweeping through the JVM instructions (still in the form of a 
list). In the implementation, this phase will update the mapping startBlock. Apart from 
the first instruction, which will be the starting block regardless of the instruction, 
the instructions that become the start of a block have the characteristics that either 
they are a target of a branching instruction, or the previous instruction ends a block. 

Case by case translation behaviour :
\begin{itemize}
	\item $P[i]$ is Unconditional jump (goto $t$) : the target instruction will be a 
	block starting point. There is implicit in this instruction that the next instruction 
	should also be a start of the block, but this will be handled by another jump. We do 
	not take care of the case where no jump instruction addresses this next instruction 
	(the next instruction is a dead code), i.e. 
	\begin{itemize} 
	  \item $\fnn{BMap} \oplus \{t \mapsto NewBlock\}$; and
	  \item $\fnn{PMap} \oplus \{t \mapsto t\}$ 
	\end{itemize}
	
	\item $P[i]$ is Conditional jump (ifeq $t$) : both the target instruction and the 
	next instruction will be the start of a block, i.e. 
	\begin{itemize}
    \item $\fnn{BMap} \oplus \{t \mapsto NewBlock, (i+1) \mapsto NewBlock\}$; and
    \item $\fnn{PMap} \oplus \{t \mapsto t, (i+1) \mapsto (i+1)\}$.
  \end{itemize}
	
	\item $P[i]$ is Return: the next instruction will be the start of a block. This 
	instruction will update the mapping of the next instruction for $\fnn{BMap}$ and 
	$\fnn{SBMap}$ if this instruction is not at the end of the instruction list. The 
	reason is that we already assumed that there is no dead code, so the next 
	instruction must be part of some execution path. To be more explicit, if there is 
	a next instruction $i + 1$ then 
	\begin{itemize}
	  \item $\fnn{BMap} \oplus \{(i + 1) \mapsto NewBlock\}$; and
	  \item $\fnn{PMap} \oplus \{(i + 1) \mapsto (i+1)\}$
	\end{itemize}
	
	\item $P[i]$ is an instruction which may throw an exception : just like return 
	instruction, the next instruction will be the start of a new block. During this 
	phase, there is also the setup for the additional block containing the sole 
	instruction of $\ins{moveexception}$ which serves as an intermediary between 
	the block with throwing instruction and its exception handler. 
	Then for each associated exception handler, its:
	\begin{itemize}
	  \item[startPC] program counter (pc) which serves as the starting point (inclusive) 
	  of which the exception handler is active;
	  \item[endPC] program counter which serves as the ending point (exclusive) of which 
	  the exception handler is active; and 
	  \item[handlerPC] program counter which points to the start of the exception handler
	\end{itemize}
	are indicated as starting of a block. 
	For handler $h$, the intermediary block will have label 
	$intPC = maxLabel + h.\field{handlerPC}$. To reduce clutter, we write $sPC$ to
	stand for $h.\field{startPC}$, $ePC$ to stand for $h.\field{endPC}$, and $hPC$ to
	stand for $h.\field{handlerPC}$.
	\begin{itemize}
    \item $\fnn{BMap} \oplus \{(i+1) \mapsto NewBlock\}$;
    \item $\fnn{BMap} \oplus \{sPC \mapsto NewBlock\}$;
    \item $\fnn{BMap} \oplus \{ePC \mapsto NewBlock\}$;
    \item $\fnn{BMap} \oplus \{hPC \mapsto NewBlock\}$;
    \item $\fnn{BMap} \oplus \{int \mapsto NewBlock\}$;   
    \item $\fnn{PMap} \oplus \{(i+1) \mapsto (i+1)\}$;
    \item $\fnn{PMap} \oplus \{sPC \mapsto sPC\}$;
    \item $\fnn{PMap} \oplus \{ePC \mapsto ePC\}$;
    \item $\fnn{PMap} \oplus \{hPC \mapsto hPC\}$;
    \item $\fnn{PMap} \oplus \{int \mapsto int\}$;
  \end{itemize}
	
  \item $P[i]$ is any other instruction : no changes to $\fnn{BMap}$
  and $\fnn{PMap}$.
	
\end{itemize}

\subsection{Resolve Parents Successors Relationship ($\fnn{TraceParentChild}$)}

Before we mention the procedure to establish the parents successors relationship, we 
need to introduce an additional function $\fnn{getAvailableLabel}$. Although defined 
clearly in the dx compiler itself, we'll abstract away from the detail and define the 
function as getting a fresh label which will not conflict with any existing label and 
labels for additional blocks before handler. These additional blocks before handlers 
are basically a block with a single instruction $\ins{moveexception}$ with the primary 
successor of the handler. Suppose the handler is at program point $i$, then this block 
will have a label of $maxLabel + i$ with the primary successor $i$. Furthermore, when
a block has this particular handler as one of its successors, the successor index is 
pointed to $maxLabel + i$ (the block containing $\ins{moveexception}$ instead of $i$). 
In the sequel, whenever we say to add a handler to a block, it means that adding 
this additional block as successor a of the mentioned block, e.g. in the JVM bytecode, 
block $i$ has exception handlers at $j$ and $k$, so during translation block $i$ will 
have successors of $\{maxLabel + j, maxLabel + k\}$, block $j$ and $k$ will have 
additional parent of block $maxLabel + j$ and $maxLabel + k$, and they each will have 
block $i$ as their sole parent. 

This phase is also done by sweeping through the JVM instructions but with the 
additional help of $\fnn{BMap}$ and $\fnn{PMap}$ mapping. Case by case 
translation behaviour :
\begin{itemize}
  \item $P[i]$ is Unconditional jump ($\ins{goto}\ t$) : update the successors of the 
  current block with the target branching, and the target block to have its parent list 
  include the current block, i.e. 
  \begin{itemize}
    \item $\fn{BMap}{\fn{PMap}{i}}.\field{succs} \cup \{t\}$; 
    \item $\fn{BMap}{\fn{PMap}{i}}.\field{pSucc} = t$; and 
    \item $\fn{BMap}{t}.\field{parents}$ $\cup$ $\{\fn{PMap}{i}\}$
  \end{itemize}

  \item $P[i]$ is Conditional jump ($\ins{ifeq}\ t$) : since there will be 2 successors 
  from this instruction, the current block will have additional 2 successors block and 
  both of the blocks will also update their parents list to include the current block, 
  i.e. 
  \begin{itemize}
    \item $\fn{BMap}{\fn{PMap}{i}}.\field{succs} \cup \{i+1, t\}$;
    \item $\fn{BMap}{\fn{PMap}{i}}.\field{pSucc} = i+1$; 
    \item $\fn{BMap}{i+1}.\field{parents} \cup \{\fn{PMap}{i}\}$; and
    \item $\fn{BMap}{t}.\field{parents} \cup \{\fn{PMap}{i}\}$
  \end{itemize}

  \item $P[i]$ is $\ins{Return}$ : just add the return block as the current block 
  successors, and also update the parent of return block to include the current block, 
  i.e. 
  \begin{itemize}
    \item $\fn{BMap}{\fn{PMap}{i}}.\field{succs} \cup \{ret\}$; 
    \item $\fn{BMap}{\fn{PMap}{i}}.\field{pSucc} = ret$; and 
    \item $\fn{BMap}{ret}.\field{parents} \cup \{\fn{PMap}{i}\}$
  \end{itemize}

  \item $P[i]$ is one of the object manipulation instruction. The idea is that the next 
  instruction will be the primary successor of this block, and should there be 
  exception handler(s) associated with this block, they will be added as successors as 
  well. We are making a little bit of simplification here where we add the next 
  instruction as the block's successor directly, i.e.
  \begin{itemize}
    \item $\fn{BMap}{\fn{PMap}{i}}.\field{succs} \cup \{i + 1\}$;
    \item $\fn{BMap}{\fn{PMap}{i}}.\field{pSucc} = i + 1$;	  
    \item $\fn{BMap}{i+1}.\field{parents} \cup \{\fn{PMap}{i}\}$; and
    \item for each exception handler $j$ associated with $i$, let 
    $intPC = maxLabel + j.\field{handlerPC}$ and
    $hPC = j.\field{handlerPC}$: 
    \begin{itemize} 
      \item $\fn{BMap}{\fn{PMap}{i}}.\field{succs} \cup \{intPC\}$;
      \item $\fn{BMap}{\fn{PMap}{i}}.\field{handlers} \cup \{j\}$;
      \item $\fn{BMap}{intPC}.\field{parents} \cup \{\fn{PMap}{i}\}$
      \item $\fn{BMap}{intPC}.\field{succs} \cup \{hPC\}$
      \item $\fn{BMap}{intPC}.\field{insn} =$ \\*
        $\{\ins{moveexception}\}$
      \item $\fn{BMap}{hPC}.\field{parents} \cup \{intPC\}$
    \end{itemize}
  \end{itemize} 
	In the original dx tool, they add a new block to contain a pseudo instruction in 
	between the current instruction and the next instruction, which will be removed 
	anyway during translation
	
  \item $P[i]$ is method invocation instruction. The treatment here is similar to that 
  of object manipulation, where the next instruction is primary successor, and the 
  exception handler for this instruction are added as successors as well. The 
  difference lies in that where the additional block is bypassed in object manipulation 
  instruction, this time we really add a block with an instruction $\ins{moveresult}$ 
  (if the method is returning a value) with a fresh label $l = \fnn{getAvailableLabel}$ 
  and the sole successor of $i + 1$. The current block will then have $l$ as it's 
  primary successor, and the next instruction ($i + 1$) will have $l$ added to its list 
  of parents, i.e.
  \begin{itemize}
    \item $l = \fnn{getAvailableLabel}$;
    \item $\fn{BMap}{\fn{PMap}{i}}.\field{succs} \cup \{l\}$;
    \item $\fn{BMap}{\fn{PMap}{i}}.\field{pSucc} = l$;
    \item $\fn{BMap}{\fn{PMap}{i}}.\field{parents} = \{i\}$;	  
    \item $\fnn{BMap} \oplus \{l \mapsto NewBlock\}$;
    \item $\fn{BMap}{l}.\field{succs} = \{i+1\}$;
    \item $\fn{BMap}{l}.\field{pSucc} = (i+1)$;
    \item $\fn{BMap}{l}.\field{insn} =$ \\*
      $\{\ins{moveresult}\}$
    \item $\fn{BMap}{i+1}.\field{parents} \cup \{l\}$; and
    \item for each exception handler $j$ associated with $i$, let 
    $intPC = maxLabel + j.\field{handlerPC}$ and
    $hPC = j.\field{handlerPC}$: 
    \begin{itemize} 
      \item $\fn{BMap}{\fn{PMap}{i}}.\field{succs} \cup \{intPC\}$;
      \item $\fn{BMap}{\fn{PMap}{i}}.\field{handlers} \cup \{j\}$;
      \item $\fn{BMap}{intPC}.\field{parents} \cup \{\fn{PMap}{i}\}$
      \item $\fn{BMap}{intPC}.\field{succs} \cup \{hPC\}$
      \item $\fn{BMap}{intPC}.\field{insn} =$ \\*
        $\{\ins{moveexception}\}$
      \item $\fn{BMap}{hPC}.\field{parents} \cup \{intPC\}$
    \end{itemize}
  \end{itemize} 

  \item $P[i]$ is throw instruction. This instruction only add the exception handlers 
  to the block without updating other block's relationship, i.e. if the current block
  is $i$, then for each exception handler $j$ associated with $i$, let 
  $intPC = maxLabel + j.\field{handlerPC}$ and
  $hPC = j.\field{handlerPC}$: 
  \begin{itemize} 
    \item $\fn{BMap}{\fn{PMap}{i}}.\field{succs} \cup \{intPC\}$;
    \item $\fn{BMap}{\fn{PMap}{i}}.\field{handlers} \cup \{j\}$;
    \item $\fn{BMap}{intPC}.\field{parents} \cup \{\fn{PMap}{i}\}$
    \item $\fn{BMap}{intPC}.\field{succs} \cup \{hPC\}$
    \item $\fn{BMap}{intPC}.\field{insn} =$ \\*
      $\{\ins{moveexception}\}$
    \item $\fn{BMap}{hPC}.\field{parents} \cup \{intPC\}$
  \end{itemize}
	
  \item $P[i]$ is any other instruction : depending whether the next 
  instruction is a start of a block or not. 
  \begin{itemize}

  \item If the next instruction is a start of a block, then update the successor of 
  the current block to include the block of the next instruction and the parent of 
  the block of the next instruction to include the current block i.e.
  \begin{itemize} 
    \item $\fn{BMap}{\fn{PMap}{i}}.\field{succs} \cup \{i+1\}$; and 
    \item $\fn{BMap}{i+1}.\field{parents} \cup \{\fn{PMap}{i}\}$  
  \end{itemize}

  \item If the next instruction is not start of a block, then just point the next 
  instruction to have the same pointer as the current block, i.e. $\fn{PMap}{i+1} = 
  \fn{PMap}{i}$

  \end{itemize}
\end{itemize}

\subsection{Reading Java Bytecodes ($\fnn{Translate}$)}
\label{apx:translate}

\begin{table*}
\[\begin{array}{| lcl | c |}
  \hline
  \multicolumn{3}{|c|}{\textbf{Translation}} & \textbf{Side effect} \\
  \hline
  \compile{\ins{push}} & = & \dexins{const}{r(\fnn{TS}_{i}), n} & 
    \fn{TS}{i+1} = \fn{TS}{i}+1\\

  \compile{\ins{pop}}& = & \emptyset & \fn{TS}{i+1} = \fn{TS}{i} - 1\\

  \compile{\ins{load}\ x} & = & \dexins{move}{r(\fnn{TS}_{i}), r_x} & 
  \fn{TS}{i+1} = \fn{TS}{i} + 1\\

  \compile{\ins{store}\ x } & = & \dexins{move}{r_x, r(\fnn{TS}_{i} - 1)} & 
  \fn{TS}{i+1} = \fn{TS}{i} - 1\\

  \compile{\ins{binop}\ op}&=&\ins{binop}(op, r(\fnn{TS}_{i}-2), r(\fnn{TS}_{i}-2)
  , & \fn{TS}{i+1} = \fn{TS}{i} - 1 \\
  & & r(\fnn{TS}_{i} - 1) ) & \\

  \compile{ \ins{swap} } & = & \dexins{move}{r(\fnn{TS}_{i}), r(\fnn{TS}_{i} - 2}) & 
  \fn{TS}{i+1} = \fn{TS}{i} \\
    && \dexins{move}{r(\fnn{TS}_{i} + 1), r(\fnn{TS}_{i} - 2) } & \\
    && \dexins{move}{r(\fnn{TS}_{i} - 1), r(\fnn{TS}_{i} + 1) } & \\
    && \dexins{move}{r(\fnn{TS}_{i} - 2), r(\fnn{TS}_{i}) } &  \\

  \compile{ \ins{goto}\ t } & = & \emptyset & \fn{TS}{t} = \fn{TS}{i} \\

  \compile{ \ins{ifeq}\ t} & = & \dexins{ifeq}{r(\fnn{TS}_{i} - 1), t} & 
  \fn{TS}{i+1} = \fn{TS}{i} - 1 \\
    &&& \fn{TS}{t} = \fn{TS}{i} - 1 \\
		
  \compile{ \ins{return} } & = & \dexins{move}{r_0, r(\fnn{TS}_{i} - 1) } & \\
    & & \dexins{return}{r_0} & \\
    & & \text{or} & \\
    & & \dexins{goto}{ret} & \\
		
  \compile{ \ins{new}\ C }& = & \dexins{new}{r(\fnn{TS}_{i} - 1), C} & 
    \fn{TS}{i+1} = \fn{TS}{i} + 1\\

  \compile{\ins{getfield}\ f}&=&\dexins{iget}{r(\fnn{TS}_{i}-1), 
    r(\fnn{TS}_{i}-1}, f) & 
  \fn{TS}{i+1} = \fn{TS}{i} + 1\\

  \compile{\ins{putfield}\ f}&=&\dexins{iput}{r(\fnn{TS}_{i}-1), 
    r(\fnn{TS}_{i}-2), f} & 
  \fn{TS}{i+1} = \fn{TS}{i} - 2 \\

  \compile{\ins{newarray}\ t} & = & 
  \dexins{newarray}{r(\fnn{TS}_{i}-1), r(\fnn{TS}_{i}-1), t} & 
  \fn{TS}{i+1} = \fn{TS}{i}\\
  
  \compile{ \ins{arraylength} } & = & 
  \dexins{arraylength}{r(\fnn{TS}_{i} - 1), r(\fnn{TS}_{i} - 1) } & 
  \fn{TS}{i+1} = \fn{TS}{i}\\
  
  \compile{ \ins{arrayload} } & = & 
  \dexins{aget}{r(\fnn{TS}_{i} - 2), r(\fnn{TS}_{i} - 2), r(\fnn{TS}_{i} - 1) } & 
  \fn{TS}{i+1} = \fn{TS}{i} - 1 \\
  
  \compile{ \ins{arraystore} } & = & 
  \dexins{aput}{r(\fnn{TS}_{i} - 1), r(\fnn{TS}_{i} - 3), r(\fnn{TS}_{i} - 2) } & 
  \fn{TS}{i+1} = \fn{TS}{i} - 3 \\
  
  \compile{ \ins{invoke}\ m } & = & 
  \dexins{invoke}{n, m, \vec{p}} & l = \fnn{getAvailableLabel} \\
    & & \dexins{moveresult}{r(\fnn{TS}_{i} - n)} \text{ at block $l$} & 
    \fn{TS}{i+1} = \fn{TS}{i} - n \\
	  
  \compile{ \ins{throw} } & = & \dexins{throw}{r(\fnn{TS}_{i} - 1)} & \\

  \hline
\end{array}\]
\caption{Instruction Translation Table}
\label{instructionTranslationTable}
\end{table*}

Table~\ref{instructionTranslationTable} list the resulting DEX translation for each of 
the JVM bytecode instruction listed in section~\ref{javaTypeSystem}. The full 
translation scheme with their typing rules can be seen in table~\ref{TranslationTable} 
in the appendix. A note about these instructions is that during this parsing of JVM 
bytecodes, the dx translation will also modify the top of the stack for the next 
instruction. Since the dx translation only happens in verified JVM bytecodes, we can 
safely assume that these top of the stacks will be consistent (even though an 
instruction may have a lot of parents, the resulting top of the stack from the parent 
instruction will be consistent with each other). To improve readability, we abuse the
notation $r(x)$ to also mean $r_x$.

\subsection{Ordering Blocks ($\fnn{PickOrder}$)}

The ``trace analysis'' itself is quite simple in essence, that is for each block we 
assign an integer denoting the order of appearance of that particular block. Starting 
from the initial block, we pick the first unordered successor and then keep on tracing 
until there is no more successor.

After we reached one end, we pick an unordered block and do the ``trace analysis'' 
again. But this time we trace its source ancestor first, by tracing an unordered parent 
block and stop when there is no more unordered parent block or already forming a loop.
Algorithm~\ref{PickOrder_Algorithm} describes how we implement this
``trace analysis''.
\begin{algorithm}
\caption{PickOrder$(blocks)$}
\label{PickOrder_Algorithm}
\begin{algorithmic}

\STATE $order := 0;$ 
\WHILE {there is still block $x \in blocks$ without order;}

\STATE $var := PickStartingPoint(x, \{x\});$
\STATE $order = TraceSuccessors(source, order);$

\ENDWHILE

\RETURN {order;}

\end{algorithmic}
\end{algorithm}


\begin{itemize}
\item \textbf{Pick Starting Point}\\
This function is a recursive function with an auxiliary data structure to prevent 
ancestor loop from viewpoint of block $x$. On each recursion, we pick a parent $p$ from 
$x$ which primary successor is $x$, not yet ordered, and not yet in the loop. The 
function then return $\fn{PickStartingPoint}{p}$.
\begin{algorithm}
\caption{PickStartingPoint$(x, loop)$}
\label{PickStartingPoint_Algorithm}
\begin{algorithmic}

\FORALL {$p \in BMap(x).parents$}
\IFLINE{$p \in loop$}{\textbf{return} $x;$}
\STATE {bp = BMap(p);}
\IF {$bp.pSucc = x \textbf{ and } bp.order = -1$}
\STATE {$loop = loop \cup \{p\};$}
\RETURN {PickStartingPoint(p, loop)}
\ENDIF
\ENDFOR

\RETURN {order;}

\end{algorithmic}
\end{algorithm}

\item \textbf{Trace Successors}\\
This function is also a recursive function with an argument of block $x$. It starts by 
assigning the current order $o$ to $x$ then increment $o$ by 1. Then it does recursive 
call to $\fnn{TraceSuccessors}$ giving one successor of $x$ which is not yet ordered as 
the argument (giving priority to the primary successor of $x$ if there is one).
\begin{algorithm}
\caption{TraceSuccessors$(x, order)$}
\label{TraceSuccessors_Algorithm}
\begin{algorithmic}

\STATE {$BMap(x).order = order;$}
\IF{$BMap(x).psucc \neq -1$}
  \STATE {pSucc = BMap(x).pSucc;}
  \IFLINE{$BMap(pSucc).order = -1$}
    {\textbf{return} $TraceSuccessors(pSucc, order + 1);$}
\FORALL {$s \in BMap(x).succs$}
  \IFLINE{$BMap(pSucc).order = -1$}
    {\textbf{return} $TraceSuccessors(s, order + 1);$}
\ENDFOR
\ENDIF

\RETURN {order;}

\end{algorithmic}
\end{algorithm}
\end{itemize}

\subsection{Output DEX Instructions ($\fnn{Output}$)}
Since the translation phase already translated the JVM instruction and ordered the 
block, this phase basically just output the instructions in order of the block. 
Nevertheless, there are some housekeeping to do alongside producing output of 
instructions. 
\begin{itemize}
  \item Remember the program counter for the first instruction in the block within DEX 
  program. This is mainly useful for fixing up the branching target later on. 

  \item Add gotos to the successor when needed for each of the block that is not 
  ending in branch instruction like $\ins{goto}$ or if. The main reason to do this is
  to maintain the successor relation in the case where the next block in order is not
  the expected block. More specifically, this is step here is in order to satisfy the
  property~\ref{fixTarget}.
  
  \item Instantiate the return block.
  
  \item Reading the list of DEX instructions and fix up the target of jump instructions.
  
  \item Collecting information about exception handlers. It is done by sweeping through
  the block in ordered fashion, inspecting the exception handlers associated with each
  block. We assume that the variable $DEXHandler$ is a global variable that store the
  information about exception handler in the DEX bytecode. The function 
  $\fn{newHandler}{cS, cE, hPC, t}$ will create a new handler (for DEX) with $cS$ as
  the start PC, $cE$ as the end PC, $hPC$ as the handler PC, and $t$ as the type of
  exception caught by this new handler.
  
\begin{algorithm}
\caption{makeHandlerEntry$(cH, cS, cE)$}
\label{makeHandlerEntry_Algorithm}
\begin{algorithmic}

\FORALL {handler $h \in cH$}
\STATE {$hPC = h.handlerPC;$}
\STATE {$t = h.catchType;$}
\STATE {$DEXHandler = DEXHandler + newHandler(cS, cE, hPC, t);$}
\ENDFOR

\end{algorithmic}
\end{algorithm}


  The only information that are needed to produce the information about exception 
  handlers in DEX is the basic blocks contained in $\fnn{BMap}$. The procedure 
  $\fnn{translateExceptionHandlers}$ 
  (Algorithm~\ref{translateExceptionHandlers_Algorithm})
  take these basic blocks $blocks$ and make use
  the procedure $\fnn{makeHandlerEntry}$ to create the exception handlers in DEX.

\begin{algorithm}
\caption{translateExceptionHandlers$(blocks)$}
\label{translateExceptionHandlers_Algorithm}
\begin{algorithmic}

\STATE {$cH = \emptyset;$} // current handler
\STATE {$cS$} // current start PC
\STATE {$cE$} // current start PC
\FORALL {block $x$ in order}
\IF {$x.handlers$ is not empty}
  \IF{$cH = x.handlers;$}
    \STATE {$cE = x.endPC;$}
  \ELSIF{$cH \neq x.handlers$}
    \STATE {$makeHandlerEntry(cH, cS, cE);$}
    \STATE {$cS = x.startPC;$}
    \STATE {$cE = x.endPC;$}
    \STATE {$cH = x.handlers;$}
  \ENDIF
\ENDIF
\ENDFOR
\STATE {$makeHandlerEntry(cH, cS, cE);$}

\end{algorithmic}
\end{algorithm}

 
  A note about the last make entry is that the algorithm will leave one set of handlers
  hanging at the end of loop, therefore we need to make that set of handlers into entry
  in the DEX exception handlers.
\end{itemize}

For simplicity, we overload the length of instructions list to also mean the total
length of instructions contained in the list. The operator $+$ here is also taken
to mean list append operation. The function $\fnn{oppositeCondition}$ takes an 
$\dexins{ifeq}{r, t}$ and returns its opposite $\dexins{ifneq}{r, t}$.
Finally, we assume that the target of jump instruction can be accessed using 
the field $\field{target}$, e.g. $\dexins{ifeq}{r, t}.\field{target} = t$. 
The details of the steps in this phase
is contained in Algorithm~\ref{output_Algorithm}.
\begin{algorithm}[H]
\caption{output}
\label{output_Algorithm}
\begin{algorithmic}

\STATE {$blocks = $ ordered blocks $\in BMap$;}
\STATE {$lbl = \emptyset;$} // label mapping
\STATE {$out = \emptyset;$} // list of DEX output
\STATE {$pc = 0;$} // DEX program counter

\FORALL {block $x$ in order}
  \STATE{$next = $ next block in order;}
  \STATE{$lbl[x] = pc;$}
  \STATE{$pc = pc + x.insn.length;$}
  \STATE{$out = out + x.insn;$}
  \IF {$p.pSucc \neq next$}
    \IF {$x.insn.last$ is $ifeq$}
      \STATE {$t = x.insn.last.target;$}
      \IF {$t = next$}
        \STATE {$out.last = oppositeCondition(x.insn.last);$}
      \ELSE \STATE {$out = out + goto(next);$}
      \ENDIF
    \ELSE
      \STATE{ $out = out + goto(next);$}
    \ENDIF
  \ENDIF
\ENDFOR
\FORALL {index $i$ in $out$}
  \IF {$out[i]$ is a jump instruction}
    \STATE {$out[i].target = lbl[out[i].target];$}
  \ENDIF
\ENDFOR
\STATE{$translateExceptionHandlers(blocks);$}

\end{algorithmic}
\end{algorithm}

The full translation scheme from JVM to DEX can be seen in table~\ref{TranslationTable}.

\begin{table*}
\begin{center}
\begin{tabular}{|c|c|c|c|}
  \hline
  &&& \\
  \textbf{JVM} & \textbf{DEX} &
  \textbf{Original Transfer Rule} &
    \textbf{Related DEX Transfer Rule}\\
  &&& \\
  \hline
  &&&\\
  Push & Const & 
  $\dfrac{P[i]=\ins{Push }v}{se, i \vdash^{\norm} st \Rightarrow se(i) :: st}$ & 
  $\dfrac{P[i]= \ins{Const} (r, n)}
    {se, i \vdash^{\norm} rt \Rightarrow rt \oplus \{r \mapsto se(i) \}}$\\
  &&&\\
  \hline
  &&&\\
  Pop & None & 
  $\dfrac{P[i]= \ins{Pop}}{i \vdash st \Rightarrow st}$ &
  None \\
  &&&\\
  \hline
  &&&\\
  Load & Move & 
  $\dfrac{P[i]= \ins{Load }x}{se, i \vdash^{\norm} st \Rightarrow 
    \big(se(i) \sqcup \vec{k_a}(x)\big) :: st}$ &
  $\dfrac{P[i]= \ins{Move }(r, r_s)}{se, i \vdash^{\norm}
    rt \Rightarrow rt\oplus\{r\mapsto\big(se(i)\sqcup rt(r_s)\big)\}}$\\
  &&&\\
  \hline
  &&&\\
  Store & Move & 
  $\dfrac{P[i]=\ins{Store }x\ \ k \sqcup se(i) \leq \vec{k_a}(x)}{
    \vec{k_a} \overset{k_h}{\rightarrow} \vec{k_r}, se, i \vdash^{\norm} 
    k :: st \Rightarrow st}$ &
  $\dfrac{P[i]=\ins{Move }(r, r_s)}
    {se, i \vdash^{\norm} rt \Rightarrow 
      rt \oplus \{r \mapsto se(i) \sqcup rt(r_s)\}}$ \\
  &&&\\
  \hline
  &&&\\
  Binop & Binop &
  $\dfrac{P[i]=\ins{Binop}}{se, i \vdash^{\norm} a :: b :: st \Rightarrow 
    \big(se(i) \sqcup a \sqcup b\big) :: st}$ &
  $\dfrac{P[i]=\ins{Binop } (r, r_a, r_b)}{se, i \vdash^{\norm} rt \Rightarrow 
    rt \oplus \{r \mapsto \big(se(i) \sqcup rt(r_a) \sqcup rt(r_b)\big)\}}$ \\
  &&&\\
  \hline
  &&&\\
  Swap & Move &
  $\dfrac{P[i]= \ins{Swap}}
    {i \vdash^{\norm} k_1 :: k_2 :: st \Rightarrow k_2 :: k_1 :: st}$ &
  $\dfrac{P[i]= \ins{Move }(r, r_s)}{se, i \vdash^{\norm}  rt \Rightarrow 
    rt \oplus \{r \mapsto \big(se(i) \sqcup rt(r_s)\big)\}}$ \\
  &&&\\
  \hline
  &&&\\
  Goto & Goto & 
  $\dfrac{P[i]=\text{Goto } t}{i\vdash st\Rightarrow st}$& 
  $\dfrac{P[i]=\text{Goto } t}{i\vdash rt \Rightarrow rt}$\\
  &&&\\
  &&&*) Not directly translated \\
  &&&\\
  \hline
  &&&\\
  & Ifeq &
  $\dfrac{P[i]=\ins{ifeq }t\ \ \forall j' \in \fn{region}{i}, k \leq se(j')}
    {\fnn{reigon}, se, i \vdash^{\norm} k :: st \Rightarrow \text{lift}_k(st)}$ &
  $\dfrac{P[i]=\ins{ifeq }(r, t)\ \ 
    \forall j' \in \mathrm{region}(i), se(i) \sqcup rt(r) \leq se(j')}
    {\fnn{region}, se, i \vdash^{\norm} rt \Rightarrow rt}$ \\
  Ifeq &&&\\
  & Ifneq & Ifeq may be translated into Ifneq on certain condition &
  $\dfrac{P[i]=\ins{ifneq }(r, t)\ \ 
    \forall j' \in \mathrm{region}(i), se(i) \sqcup rt(r) \leq se(j')}
    {\fnn{region}, se, i \vdash^{\norm} rt \Rightarrow rt}$ \\  
  &&&\\
  \hline
  &&&\\
  New & New &
  $\dfrac{P[i]=\ins{new } C}{se, i \vdash^{\norm} st \Rightarrow se(i) :: st}$ &
  $\dfrac{P[i]=\ins{new }(r, c)}
    {se, i \vdash^{\norm} rt \Rightarrow rt \oplus \{r \mapsto se(i) \}}$ \\
  &&&\\
  \hline
  &&&\\
  &&
  $\dfrac{P[i]=\ins{getfield }f\ \ \ \ k\in\mathcal{S}\ \ \ \ 
      \forall j \in \fn{region}{i, \norm}, k \leq se(j) }
    {\fnn{ft}, \fnn{region}, se, i \vdash^{\norm} k :: st \Rightarrow 
    \fn{lift_k}{(k \sqcup \fn{ft}{f} \sqcup se(i)) :: st}}$ &
  $\dfrac{\begin{gathered} P[i] = \ins{iget } (r, r_o, f)\ \ \ \ 
      rt(r_o) \in \mathcal{S}\\
      \forall j \in \fn{region}{i, \norm}, rt(r_o) \leq se(j) \end{gathered}}
    {\fnn{ft}, se, i \vdash^{\norm} rt \Rightarrow 
      rt \oplus\{r\mapsto rt(r_o)\sqcup\fn{ft}{f}\sqcup se(i)\}}$ \\
  &&&\\
  Getfield & Iget &  
  $\dfrac{\begin{gathered}
      P[i]=\ins{getfield }f\ \ \ \ k\in\mathcal{S}\ \ \ \ 
      \forall j \in \fn{region}{i, \np}, k \leq se(j) \\
      \fn{Handler}{i, \np} = t \end{gathered}}
    {\fnn{ft}, \fnn{region}, se, 
      i \vdash^{\np} k :: st \Rightarrow (k \sqcup se(i)) :: \epsilon}$ &
  $\dfrac{\begin{gathered} P[i] = \ins{iget } (r, r_o, f)\ \ \ \ 
      rt(r_o) \in \mathcal{S}\\
      \forall j \in \fn{region}{i, \np}, rt(r_o) \leq se(j) 
      \ \ \ \ \fn{Handler}{i, \np} = t\end{gathered}}
    {\fnn{ft}, se, i \vdash^{\np} rt \Rightarrow 
      \vec{k_a} \oplus\{ex\mapsto rt(r_o) \sqcup se(i)\}}$ \\
  &&&\\
  &&
    $\dfrac{\begin{gathered}
      P[i]=\ins{getfield }f\ \ \ \ k\in\mathcal{S}\ \ \ \ 
      \forall j \in \fn{region}{i, \np}, k \leq se(j) \\
      \fn{Handler}{i, \np} \uparrow \ \ \ \ k \leq \vec{k_r}[\np] \end{gathered}}
    {\fnn{ft}, \fnn{region}, se, i \vdash^{\np} k :: st \Rightarrow }$ &
  $\dfrac{\begin{gathered} P[i] = \ins{iget } (r, r_o, f)\ \ \ \ 
      rt(r_o) \in \mathcal{S}\ \ \ \ se(i) \sqcup rt(r_o) \leq \vec{k_r}[\np]\\
      \forall j \in \fn{region}{i, \np}, rt(r_o) \leq se(j) 
      \ \ \ \ \fn{Handler}{i, \np} = t\end{gathered}}
    {\fnn{ft}, \vec{k_a}\overset{k_h}{\rightarrow}\vec{k_r}, se, 
      i \vdash^{\np} rt \Rightarrow}$ \\
  &&&\\
  \hline
  &&&\\
  &&
  $\dfrac{\begin{gathered}P[i]=\ins{putfield }f\ \ \ \ k_h \leq \fn{ft}{f}\ \ \ \ 
    k_1 \sqcup se(i) \sqcup k_2 \leq \fn{ft}{f} \\
    k_1\in\sext\ \ \ \ k_2 \in \mathcal{S}\ \ \ \ 
    \forall j \in \fn{region}{i, \norm}, k_2 \leq se(j)\end{gathered}}
  {\fnn{ft}, \vec{k_a}\overset{k_h}{\rightarrow}\vec{k_r}, \fnn{region}, se, 
    i \vdash^{\norm} k_1 :: k_2 :: st \Rightarrow \fn{lift_{k_2}}{st}}$ &
  $\dfrac{\begin{gathered}P[i] = \ins{iput } (r_s, r_o, f)\ \ \ \ k_h \leq \fn{ft}{f}
      \ \ \ \ rt(r_o) \in \mathcal{S}\ \ \ \ rt(r_s) \in \sext\\
      rt(r_o) \sqcup se(i) \sqcup rt(r_s) \leq \fn{ft}{f}\\
      \forall j \in \fn{region}{i, \norm}, rt(r_o) \leq se(j)\end{gathered}}
    {\fnn{ft}, \vec{k_a}\overset{k_h}{\rightarrow}\vec{k_r}, se, i \vdash^{\norm}
      rt \Rightarrow rt}$ \\
  &&&\\
  Putfield & Iput &
  $\dfrac{\begin{gathered}P[i]=\ins{putfield }f\ \ \ \
    k_1 \sqcup se(i) \sqcup k_2 \leq \fn{ft}{f} \ \ \ \ \fn{Handler}{i, \np} = t\\
    k_1 \in \sext\ \ \ k_2 \in \mathcal{S}\ \ \ 
    \forall j \in \fn{region}{i, \np}, k_2 \leq se(j)\end{gathered}}
  {\fnn{ft}, \vec{k_a}\overset{k_h}{\rightarrow}\vec{k_r}, \fnn{region}, se, 
    i \vdash^{\np} k_1 :: k_2 :: st \Rightarrow (k_2 \sqcup se(i)) :: \epsilon}$ &
  $\dfrac{\begin{gathered}P[i] = \ins{iput } (r_s, r_o, f)\ \ \ \ k_h \leq \fn{ft}{f}
      \ \ \ \ rt(r_o) \in \mathcal{S}\ \ \ \ rt(r_s) \in \sext\\
      rt(r_o) \sqcup se(i) \sqcup rt(r_s) \leq \fn{ft}{f}\ \ \ \ 
      \fn{Handler}{i, \np} = t\\
      \forall j \in \fn{region}{i, \np}, rt(r_o) \leq se(j)\end{gathered}}
    {\fnn{ft}, \vec{k_a}\overset{k_h}{\rightarrow}\vec{k_r}, se, i \vdash^{\np}
      rt \Rightarrow \vec{k_a} \oplus \{ex \mapsto rt(r_o) \sqcup se(i)\} }$ \\
  &&&\\
  &&
  $\dfrac{\begin{gathered}P[i]=\ins{putfield }f\ \ \ \
    k_1 \sqcup se(i) \sqcup k_2 \leq \fn{ft}{f} \ \ \ \ \fn{Handler}{i, \np} \uparrow\\
    k_1 \in \sext\ \ \ k_2 \in \mathcal{S}\ \ \ 
    \forall j \in \fn{region}{i, \np}, k_2 \leq se(j)\ \ \ k_2 \leq \vec{k_r}[\np]
    \end{gathered}}
  {\fnn{ft}, \vec{k_a}\overset{k_h}{\rightarrow}\vec{k_r}, \fnn{region}, se, 
    i \vdash^{\np} k_1 :: k_2 :: st \Rightarrow }$ &
  $\dfrac{\begin{gathered}P[i] = \ins{iput } (r_s, r_o, f)\ \ \ \ k_h \leq \fn{ft}{f}
      \ \ \ \ rt(r_o) \in \mathcal{S}\ \ \ \ rt(r_s) \in \sext\\
      rt(r_o) \sqcup se(i) \sqcup rt(r_s) \leq \fn{ft}{f}\ \ \ \ 
      \fn{Handler}{i, \np} \uparrow\\
      \forall j \in \fn{region}{i, \np}, rt(r_o) \leq se(j)\ \ \ \ 
      se(i) \sqcup rt(r_o) \leq \vec{k_r}[\np]\end{gathered}}
    {\fnn{ft}, \vec{k_a}\overset{k_h}{\rightarrow}\vec{k_r}, se, i \vdash^{\np}
      rt \Rightarrow }$ \\
  &&&\\
  \hline
  
\end{tabular}
\end{center}
\end{table*}

\begin{table*}
\begin{center}
\begin{tabular}{|c|c|c|c|}
  \hline
  &&& \\
  \textbf{JVM} & \textbf{DEX} &
  \textbf{Original Typing Rule} &
    \textbf{Related DEX Typing Rule}\\
  &&& \\
  \hline
  &&&\\
  Newarray & Newarray &
  $\dfrac{P[i] = \ins{newarray }t\ \ \ \ k \in \mathcal{S}}
  {\begin{gathered}i \vdash^{\norm}
    k :: st \Rightarrow  k[\fn{at}{i}] :: st\end{gathered}}$ &
  $\dfrac{P[i] = \ins{newarray }(r, r_l, t)\ \ \ \ rt(r_l) \in \mathcal{S}}
  {i\vdash^{\norm} rt \Rightarrow rt \oplus \{r \mapsto rt(r_l)[\fn{at}{i}]\}}$ \\
  &&&\\
  \hline
  &&&\\
  &&
  $\dfrac{\begin{gathered}
    P[i] = \ins{arraylength}\ \ \ \ \forall j \in \fn{region}{i, \norm}, k \leq se(j)
    \\ k \in \mathcal{S}\ \ \ \ k_c \in \sext \end{gathered}}
  {\begin{gathered}\fnn{region}, se, i \vdash^{\norm}
    k[k_c] :: st \Rightarrow \fn{lift_k}{k :: st}\end{gathered}}$ &
  $\dfrac{\begin{gathered}P[i] = \ins{arraylength }(r, r_a)\ \ \ \ k[k_c] = rt(r_a) 
    \ \ \ \ k \in \mathcal{S}\\ k_c \in \sext \ \ \ \ 
    \forall j \in \fn{region}{i, \norm}, k \leq se(j)\end{gathered}}
  {\fnn{region},se, i\vdash^{norm} rt \Rightarrow rt \oplus \{r \mapsto k\}}$\\
  &&&\\
  Arraylength & Arraylength &
  $\dfrac{\begin{gathered}
    P[i] = \ins{arraylength}\ \ \ \ \forall j \in \fn{region}{i, \np}, k \leq se(j)
    \\ k \in \mathcal{S}\ \ \ \ k_c \in \sext \ \ \ \ \fn{Handler}{i, \np} = t
    \end{gathered}}
  {\fnn{region}, se, i \vdash^{\np} 
    k[k_c] :: st \Rightarrow (k \sqcup se(i)) :: \epsilon}$ &
  $\dfrac{\begin{gathered}P[i] = \ins{arraylength }(r, r_a)\ \ \ \ k[k_c] = rt(r_a) 
    \ \ \ \ k \in \mathcal{S}\\ k_c \in \sext \ \ \ \ 
    \forall j \in \fn{region}{i, \np}, k \leq se(j)\\
    \fn{Handler}{i, \np} = t\end{gathered}}
  {\fnn{region},se, i\vdash^{np} rt \Rightarrow 
    \vec{k_a} \oplus \{ex \mapsto k \sqcup se(i)\}}$\\
  &&&\\
  &&
  $\dfrac{\begin{gathered}
    P[i] = \ins{arraylength}\ \ \ \ \forall j \in \fn{region}{i, \np}, k \leq se(j)
    \\ k \in \mathcal{S}\ \ \ \ k_c \in \sext \ \ \ \ \fn{Handler}{i, \np}\uparrow
    \ \ \ \ k \leq \vec{k_r}[\np]\end{gathered}}
  {\begin{gathered}\vec{k_a}\rightarrow k_r, \fnn{region}, se, i \vdash^{\np} 
    k[k_c] :: st \Rightarrow \end{gathered}}$ &
  $\dfrac{\begin{gathered}P[i] = \ins{arraylength }(r, r_a)\ \ \ \ k[k_c] = rt(r_a) 
    \ \ \ \ k \in \mathcal{S}\\ k_c \in \sext \ \ \ \ 
    \forall j \in \fn{region}{i, \np}, k \leq se(j)\\
    \fn{Handler}{i, \np} \uparrow \ \ \ \ se(i) \sqcup k \leq \vec{k_a}[\np]\end{gathered}}
  {\vec{k_a} \rightarrow \vec{k_r}, \fnn{region},se, i \vdash^{np} rt \Rightarrow}$\\
  &&&\\
  \hline
  &&&\\
  &&
  $\dfrac{\begin{gathered} P[i] = \ins{arrayload}\ \ \ \ 
    k_1, k_2 \in \mathcal{S} \ \ \ \ k_c \in \sext \\
    \forall j \in \fn{region}{i, \norm} k_2 \leq se(j) \end{gathered}}
  {\begin{gathered}\vec{k_a} \rightarrow k_r, \fnn{region}, se, i \vdash^{\norm}
    k_1 :: k_2[k_c]  :: st \Rightarrow \\
    \fn{lift_{k_2}}{\big( (k_1 \sqcup k_2) \sqcupext k_c \big) :: st}\end{gathered}}$ &
  $ \dfrac{\begin{gathered}P[i] = \ins{aget }(r, r_a, r_i)\ \ \ \ k[k_c] = rt(r_a)
    \ \ \ \ k_c \in \sext\\ 
    k, rt(r_i) \in \mathcal{S}\ \ \ \ 
    \forall j \in \fn{region}{i, \norm}, k \leq se(j)\end{gathered}}
  {\begin{gathered}\vec{k_a}\rightarrow k_r,\fnn{region},se, i \vdash^{norm} rt 
    \Rightarrow \\
    rt \oplus \{r \mapsto ((k \sqcup rt(r_i)) \sqcupext k_c)\} \end{gathered}}$ \\
  &&&\\
  Arrayload & Aget & 
  $\dfrac{\begin{gathered} P[i] = \ins{arrayload}\ \ \ \ 
    k_1, k_2 \in \mathcal{S} \ \ \ \ k_c \in \sext \\
    \forall j \in \fn{region}{i, \np} k_2 \leq se(j) 
    \ \ \ \ \fn{Handler}{i, \np} = t\end{gathered}}
  {\vec{k_a} \rightarrow k_r, \fnn{region}, se, i \vdash^{\np}
    k_1 :: k_2[k_c]  :: st \Rightarrow (k_2 \sqcup se(i)) :: \epsilon}$ &
  $ \dfrac{\begin{gathered}P[i] = \ins{aget }(r, r_a, r_i)\ \ \ \ k[k_c] = rt(r_a)
    \ \ \ \ k_c \in \sext\\ 
    k, rt(r_i) \in \mathcal{S}\ \ \ \ 
    \forall j \in \fn{region}{i, \np}, k \leq se(j)\\
    \fn{Handler}{i, \np} = t\end{gathered}}
  {\fnn{region},se, i \vdash^{np} rt \Rightarrow 
    \vec{k_a} \oplus \{ex \mapsto k \sqcup se(i)\} }$ \\
  &&&\\
  &&
  $\dfrac{\begin{gathered} P[i] = \ins{arrayload}\ \ \ \ 
    k_1, k_2 \in \mathcal{S} \ \ \ \ k_c \in \sext \ \ \ \ k_2 \leq \vec{k_r}[\np]\\
    \forall j \in \fn{region}{i, \np} k_2 \leq se(j) 
    \ \ \ \ \fn{Handler}{i, \np} \uparrow \end{gathered}}
  {\vec{k_a} \rightarrow k_r, \fnn{region}, se, i \vdash^{\np}
    k_1 :: k_2[k_c]  :: st \Rightarrow}$ &
  $ \dfrac{\begin{gathered}P[i] = \ins{aget }(r, r_a, r_i)\ \ \ \ k[k_c] = rt(r_a)
    \ \ \ \ k_c \in \sext\\ 
    k, rt(r_i) \in \mathcal{S}\ \ \ \ 
    \forall j \in \fn{region}{i, \np}, k \leq se(j)\\
    \fn{Handler}{i, \np} \uparrow\ \ \ \ se(i) \sqcup k \leq \vec{k_r}[\np]\end{gathered}}
  {\vec{k_a}\rightarrow k_r,\fnn{region},se, i \vdash^{np} rt \Rightarrow}$ \\
  &&&\\
  \hline
  &&&\\
  &&
  $\dfrac{\begin{gathered}P[i] = \ins{arraystore}\ \ \ \ k_1, k_c \in \sext
    \ \ \ \ k_2, k_3 \in \mathcal{S}\\
    ((k_2 \sqcup k_3) \sqcupext k_1) \leqext k_c\ \ \ \ 
    \forall j \in \fn{region}{i, \norm}, k_2 \leq se(j)
    \end{gathered}}
  {\begin{gathered}\vec{k_a} \rightarrow k_r, \fnn{region}, se, i \vdash^{\norm}
    k_1 :: k_2 :: k_3[k_c] :: st \Rightarrow \fn{lift_{k_2}}{st} \end{gathered}}$ &
  $\dfrac{\begin{gathered}P[i] = \ins{aput }(r_s, r_a, r_i)\ \ \ k[k_c] = rt(r_a)
    \ \ \ k, rt(r_i) \in \mathcal{S}\\ 
    ((k \sqcup rt(r_i)) \sqcupext rt(r_s)) \leqext k_c\ \ \ 
    k_c, rt(r_s) \in \sext \\
    \forall j \in \fn{region}{i, \norm}, k \leq se(i)\end{gathered}}
  {\vec{k_a} \rightarrow k_r, \fnn{region}, se, i \vdash^{\norm} rt \Rightarrow rt}$ \\
  &&&\\
  Arraystore & Aput &
  $\dfrac{\begin{gathered}P[i] = \ins{arraystore}\ \ \ \ \ \ \ k_1, k_c \in \sext \\
    ((k_2 \sqcup k_3) \sqcupext k_1) \leqext k_c\ \ \ \ k_2, k_3 \in \mathcal{S}\\
    \fn{Handler}{i, \np} = t\ \ \ \ \forall j \in \fn{region}{i, \np}, k_2 \leq se(j)
    \end{gathered}}
  {\begin{gathered}\vec{k_a} \rightarrow k_r, \fnn{region}, se, i \vdash^{\np}
    k_1 :: k_2 :: k_3[k_c] :: st \Rightarrow (k_2 \sqcup se(i)) :: \epsilon 
    \end{gathered}}$ &
  $\dfrac{\begin{gathered}P[i] = \ins{aput }(r_s, r_a, r_i)\ \ \ k[k_c] = rt(r_a)
    \ \ \ k, rt(r_i) \in \mathcal{S}\\ 
    ((k \sqcup rt(r_i)) \sqcupext rt(r_s)) \leqext k_c\ \ \ 
    k_c, rt(r_s) \in \sext \\
    \forall j \in \fn{region}{i, \np}, k \leq se(i)\ \ \ \ \fn{Handler}{i, \np} = t
    \end{gathered}}
  {\fnn{region}, se, i \vdash^{\np} rt \Rightarrow 
    \vec{k_a} \oplus \{ex \mapsto k \sqcup se(i)\}}$ \\
  &&&\\
  & &
  $\dfrac{\begin{gathered}P[i] = \ins{arraystore}\ \ \ \ k_1, k_c \in \sext
    \ \ \ \ k_2, k_3 \in \mathcal{S}\\
    ((k_2 \sqcup k_3) \sqcupext k_1) \leqext k_c\ \ \ 
    \forall j \in \fn{region}{i, \np}, k_2 \leq se(j)\\
    \fn{Handler}{i, \np} \uparrow\ \ \ \ k_2 \leq \vec{k_r}[\np]
    \end{gathered}}
  {\vec{k_a} \rightarrow k_r, \fnn{region}, se, i \vdash^{\np}
    k_1 :: k_2 :: k_3[k_c] :: st \Rightarrow }$ &
  $\dfrac{\begin{gathered}P[i] = \ins{aput }(r_s, r_a, r_i)\ \ \ k[k_c] = rt(r_a)
    \ \ \ k, rt(r_i) \in \mathcal{S}\\ 
    ((k \sqcup rt(r_i)) \sqcupext rt(r_s)) \leqext k_c\ \ \ 
    k_c, rt(r_s) \in \sext \\
    \forall j \in \fn{region}{i, \np}, k \leq se(i)\ \ \ \ \fn{Handler}{i, \np} = t\\
    se(i) \sqcup k \leq \vec{k_r}[\np]\end{gathered}}
  {\vec{k_a}\rightarrow\vec{k_r}\fnn{region}, se, i \vdash^{\np} rt \Rightarrow}$ \\
  &&&\\
  \hline
  &&&\\
  & Move &
  $\dfrac{P[i] = \ins{return}\ \ \ \ se(i) \sqcup k \leq k_r}
    {\vec{k_a}\overset{k_h}{\rightarrow}\vec{k_r}, se, i \vdash k :: st \Rightarrow}$ &
  $\dfrac{P[i]= \ins{Move }(r_0, r_s)}
    {se, i \vdash rt \Rightarrow rt \oplus 
      \{r \mapsto \big(se(i) \sqcup rt(r_s)\big) \}}$ \\
  
  & \textbf{and} &&\\	

  Return &&&\\
  & Goto &
  & $\dfrac{P[i] = \ins{goto }(t)}{i \vdash rt \Rightarrow rt}$ \\

  & \textbf{or} && \\
  
  & Return 
  && $\dfrac{P[i] = \ins{return }(r_s) \ se(i) \sqcup rt(r_s) \leq k_r}
    {\vec{k_a}\overset{k_h}{\rightarrow}\vec{k_r}, se, i \vdash rt \Rightarrow}$ \\
  &&&\\
  \hline
\end{tabular}
\end{center}
\end{table*}

\begin{table*}
\scriptsize
\begin{center}
\begin{tabular}{|c|c|c|c|}
  \hline
  &&& \\
  \textbf{JVM} & \textbf{DEX} &
  \textbf{Original Typing Rule} &
    \textbf{Related DEX Typing Rule}\\
  &&& \\
  \hline
  &&&\\

  &&&\\
  &&
  $\dfrac{\begin{gathered}P_m[i] = \ins{invoke } m_\mathrm{ID}\ \ \ \ 
    \fn{length}{st_1} = \text{nbArguments}(m_ID)\\
    k \leq \vec{k'_a}[0] \ \ \ \ 
    \forall i \in [0, \fn{length}{st_1} - 1] st_1[i] \leq \vec{k'_a}[i+1] \\
    k \sqcup k_h \sqcup se(i) \leq k'_h \ \ \ \ 
    k_e = \bigsqcup \{\vec{k'_r}[e]\ |\ e \in \fn{excAnalysis}{m_\mathrm{ID}}\} \\
    \Gamma_{m_{ID}}[k] = \vec{k'_a} \overset{k'_h}{\rightarrow} k'_r \ \ \ \ 
    \forall j \in \fn{region}{i, \norm}, k \sqcup k_e \leq se(j)\end{gathered}}
  {\begin{gathered}\Gamma, \fnn{region}, se, 
    \vec{k_a}\overset{k_h}{\rightarrow}\vec{k_r}, i \vdash^{\norm} 
    st_1 :: k :: st_2 \Rightarrow \\
    \fn{lift_{k \sqcup k_e}}{(k'_r \sqcup k \sqcup se(i)) :: st_2}\end{gathered}}$ &
  $\dfrac{\begin{gathered} P_m[i] = \text{invoke }(n, m, \vec{p})\ \ \ \ 
    \Gamma_{m'}[rt(\vec{p}[0])] = \vec{k_a'} \overset{k_h'}{\rightarrow} k_r' \\
    rt(\vec{p}[0]) \sqcup k_h \sqcup se(i) \leq k_h'\ \ \ \ 
    \forall_{0 \leq j < n} rt(\vec{p}[j]) \leq \vec{k_a'}[j]\\
    k_e = \bigsqcup \{\vec{k'_r}[e]\ |\ e \in \fn{excAnalysis}{m'}\}\\
    \forall j \in \fn{region}{i, \norm}, rt(\vec{p}[0]) \sqcup k_e \leq se(j)
  \end{gathered}}
  {\begin{gathered}\Gamma, \fnn{region}, se, \vec{k_a} 
    \overset{k_h}{\rightarrow} k_r, i \vdash^{\norm} rt \Rightarrow 
    \\ rt \oplus \{ret 
    \mapsto k'_r[n] \sqcup rt(\vec{p}[0]) \sqcup se(i)\} \end{gathered}}$ \\
  &&&\\
  Invoke & Invoke &
  $\dfrac{\begin{gathered}P_m[i] = \ins{invoke } m_\mathrm{ID}\ \ \ \ 
    \fn{length}{st_1} = \text{nbArguments}(m_ID)\\
    k \leq \vec{k'_a}[0] \ \ \ \ 
    \forall i \in [0, \fn{length}{st_1} - 1] st_1[i] \leq \vec{k'_a}[i+1] \\
    k \sqcup k_h \sqcup se(i) \leq k'_h \ \ \ \ 
    e \in \fn{excAnalysis}{m_\mathrm{ID}} \cup \{\np\} \\
    \Gamma_{m_{ID}}[k] = \vec{k'_a} \overset{k'_h}{\rightarrow} k'_r \ \ \ \ 
    \forall j \in \fn{region}{i, e}, k \sqcup k_e \leq se(j) \\
    \fn{Handler}{i, e} = t\end{gathered}}
  {\begin{gathered}\Gamma, \fnn{region}, se, 
    \vec{k_a}\overset{k_h}{\rightarrow}\vec{k_r}, i \vdash^{e} 
    st_1 :: k :: st_2 \Rightarrow (k \sqcup \vec{k'_r}[e]) :: \epsilon\end{gathered}}$ &
  $\dfrac{\begin{gathered} P_m[i] = \text{invoke }(n, m, \vec{p})\ \ \ \ 
    \Gamma_{m'}[rt(\vec{p}[0])] = \vec{k_a'} \overset{k_h'}{\rightarrow} k_r' \\
    rt(\vec{p}[0]) \sqcup k_h \sqcup se(i) \leq k_h'\ \ \ \ 
    \forall_{0 \leq j < n} rt(\vec{p}[j]) \leq \vec{k_a'}[j]\\
    e \in \fn{excAnalysis}{m'} \cup \{\np\}\}
    \ \ \ \ \fn{Handler}{i, e} = t\\
    \forall j \in \fn{region}{i, e}, rt(\vec{p}[0]) \sqcup \vec{k'_r}[e] \leq se(j)
  \end{gathered}}
  {\begin{gathered}\Gamma, \fnn{region}, se, \vec{k_a} 
    \overset{k_h}{\rightarrow} k_r, i \vdash^{e} rt \Rightarrow \\ 
    \vec{k_a} \oplus \{ex \mapsto k'_r[e] \sqcup sec(\vec{p}[0])\} \end{gathered}}$ \\
  &&&\\
  &&
  $\dfrac{\begin{gathered}P_m[i] = \ins{invoke } m_\mathrm{ID}\ \ \ \ 
    \fn{length}{st_1} = \text{nbArguments}(m_ID)\\
    k \leq \vec{k'_a}[0] \ \ \ \ 
    \forall i \in [0, \fn{length}{st_1} - 1] st_1[i] \leq \vec{k'_a}[i+1] \\
    k \sqcup k_h \sqcup se(i) \leq k'_h \ \ \ \ 
    e \in \fn{excAnalysis}{m_\mathrm{ID}} \cup \{\np\} \\
    \Gamma_{m_{ID}}[k] = \vec{k'_a} \overset{k'_h}{\rightarrow} k'_r \ \ \ \ 
    \forall j \in \fn{region}{i, e}, k \sqcup k_e \leq se(j) \\
    \fn{Handler}{i, e} \uparrow\ \ \ \ 
    k \sqcup se(i) \sqcup \vec{k'_r}[e] \leq \vec{k_r}[e]\end{gathered}}
  {\begin{gathered}\Gamma, \fnn{region}, se, 
    \vec{k_a}\overset{k_h}{\rightarrow}\vec{k_r}, i \vdash^{e} 
    st_1 :: k :: st_2 \Rightarrow \end{gathered}}$ &
  $\dfrac{\begin{gathered} P_m[i] = \text{invoke }(n, m, \vec{p})\ \ \ \ 
    \Gamma_{m'}[sec(\vec{p}[0])] = \vec{k_a'} \overset{k_h'}{\rightarrow} k_r' \\
    rt(\vec{p}[0]) \sqcup k_h \sqcup se(i) \leq k_h'\ \ \ \ 
    \forall_{0 \leq j < n} rt(\vec{p}[j]) \leq \vec{k_a'}[j]\\
    e \in \fn{excAnalysis}{m'} \cup \{\np\}\}
    \ \ \ \ \fn{Handler}{i, e} \uparrow\\
    \forall j \in \fn{region}{i, e}, rt(\vec{p}[0]) \sqcup k'_r[e] \leq se(j)\\
    rt(\vec{p}[0]) \sqcup se(i) \sqcup \vec{k'_r}[e] \leq \vec{k_r}[e]
  \end{gathered}}
  {\Gamma, \fnn{region}, se, \vec{k_a} \overset{k_h}{\rightarrow} k_r, i \vdash^{e} 
    rt \Rightarrow }$ \\
  &&&\\  	 		
  
  & Moveresult & 
  & $\dfrac{P_m[i]=\ins{moveresult } (r)}
  {i \vdash^{\norm} rt \Rightarrow rt \oplus \{r \mapsto rt(ret)\}}$ \\

  &&&\\
  \hline
\end{tabular}
\end{center}
\caption{Translation Table}
\label{TranslationTable}
\end{table*}

\newpage

\section{Proof of Lemmas}

\subsection{Proofs that Translation Preserves SOAP Satisfiability}

We first start this section with proofs of lemmas that are omitted in the paper due to 
space requirement.

\begin{lemma}
\label{preservedSuccessor}
  Let $P$ be a JVM program and $P[a] = Ins_a$ and $P[b] = Ins_b$ are two of its 
  instructions at program points $a$ and $b$ (both are non-invoke instructions). 
  Let $n_a > 0$ be the number of instructions translated from $Ins_a$.
  If $a \mapsto^{\norm} b$, then either
  \[\begin{array}{c} 
  \ordout{\translate{a}[n-1]} \mapsto^{\norm} \ordout{\translate{b}[0]} \\
  \text{or}\\
  \left( \begin{array}{c}
  \ordout{\translate{a}[n-1]} \mapsto^{\norm} (\ordout{\translate{a}[n-1]} + 1) \\
  \text{ and } \\
  (\ordout{\translate{a}[n-1]} + 1) \mapsto^{\norm} \ordout{\translate{b}[0]}
  \end{array}\right)
  \end{array}\]
  in $\compile{ P }$.
\end{lemma}
\begin{IEEEproof}
  To prove this lemma, we first unfold the definition of compilation. Using the 
  information that $a \mapsto b$, there are several possible cases to output the block 
  depending whether what instruction is $Ins_a$ and where $a$ and $b$ are located. 
  Either:
  \begin{itemize}
    \item $\translate{b}$ are placed directly after $\translate{a}$ and 
    $\translate{a}[n-1]$ is sequential instruction;\\
    In this case, appealing to the definition of successor relation this trivially
    holds as the first case.

    \item $\translate{Ins_a}$ ends in a sequential instruction and will have a 
    $\ins{goto}$ instruction appended that points to $\ordout{\translate{b}[0]}$;\\
    Again appealing to the definition of successor relation this trivially holds as
    the second case, where \\* $P_\text{DEX}[(\ordout{\translate{a}[n-1]} + 1)] = 
    \dexins{goto}{\ordout{\translate{b}[0]}}$. 

    \item $\translate{Ins_a}[n-1]$ is a branching instruction and $b$ is 
    one of its child, and $\translate{b}$ is placed directly after $\translate{a}$ 
    or is pointed to by the branching instruction;\\
    Either case, using the definition of successor relation to establish that we are in
    the first case.

    \item $\translate{Ins_a}[n-1]$ is a branching instruction and $b$ is one of its 
    child, nevertheless $\translate{b}$ is not placed directly after $\translate{a}$
    nor is pointed to by the branching instruction;\\
    In this case, according to the \textbf{Output} phase, a $\ins{goto}$ instruction
    will be appended in $(\ordout{\translate{a}[n-1]} + 1)$ and thus we are in the
    second case. Use the definition of successor relation to conclude the proof.
  \end{itemize}
\end{IEEEproof}

\begin{lemma}
\label{preservedSuccessorException}
  Let $P$ be a JVM program and $P[a] = Ins_a$ and $P[b] = Ins_b$ are two of its 
  instructions at program points $a$ and $b$ where $b$ is the address of the first 
  instruction in the exception handler $h$ for $a$ throwing exception $\tau$. 
  Let $e$ be the index to the instructions within $\translate{a}$ that throws 
  exception.
  If $a \mapsto^{\tau} b$, then
  $\ordout{\translate{a}[e]} \mapsto^{\tau} \ordout{\translate{h}}$ and 
  $\ordout{\translate{h}} \mapsto^{\norm} \ordout{\translate{b}[0]}$
\end{lemma}
\begin{IEEEproof}
  Trivial based on the unfolding definition of the compiler, where there is a block 
  containing sole instruction $\ins{moveexception}$, which will be pointed by the 
  exception handler in DEX, between possibly throwing 
  instruction and its handler for particular exception class $\tau$. The proof is then
  concluded by successor relation in DEX.
\end{IEEEproof}

\begin{lemma}
\label{preservedSuccessorInvoke}
  Let $P$ be a JVM program and $P[a] = \ins{invoke}$ and $P[b] = Ins_b$ are two of 
  its instructions at program points $a$ and $b$. 
  If $a \mapsto^{\norm} b$, then 
  $\ordout{\translate{a}[0]} \mapsto^{\norm} \ordout{\translate{a}[1]}$ and 
  $\ordout{\translate{a}[1]} \mapsto^{\norm} \ordout{\translate{b}[0]}$
\end{lemma}
\begin{IEEEproof}
  This is trivial based on the unfolding definition of the compiler since the primary 
  successor of $\ordout{\translate{a}[0]}$ is $\ordout{\translate{a}[1]}$, where 
  $\ordout{P}[\ordout{\translate{a}[1]}] = \ins{moveresult}$, and the primary 
  successor of $\ordout{\translate{a}[1]}$ is $\ordout{\translate{b}[0]}$. The 
  proof is then concluded by the definition of successor relation in DEX.
\end{IEEEproof}

\begin{lemma}
\label{specialPreservedSuccessor}
  Let $P$ be a JVM program and $P[a] = Ins_a$ and $P[b] = Ins_b$ are two of its 
  instructions at program points $a$ and $b$. 
  Suppose $Ins_b$ is translated to an empty sequence 
  (e.g. $Ins_b$ is pop or goto). Let $s$ be the first in the successor chain of $b$ 
  such that $\translate{ P[s] }$ is non-empty (we can justify this successor chain as 
  instruction that cause branching will never be translated into empty sequence).
  If $a \mapsto b$, then either 
  \[ \begin{array}{c}
  \ordout{\translate{a}[n-1]} \mapsto^{\norm} \ordout{\translate{s}[0]} \\
  \text{or}\\
  \left( \begin{array}{c}
  \ordout{\translate{a}[n-1]} \mapsto^{\norm} (\ordout{\translate{a}[n-1]} + 1) \\
  \text{ and } \\
  (\ordout{\translate{a}[n-1]} + 1) \mapsto^{\norm} \ordout{\translate{s}[0]}
  \end{array} \right)
  \end{array}\]
\end{lemma}
\begin{IEEEproof}
  We use induction on the length of successor's chain. In the base case where the 
  length is 0, we can use Lemma~\ref{preservedSuccessor} to establish that this 
  lemma holds. For the case where the length is $n + 1$, there are two possibilities 
  for the last instruction in the chain :
  \begin{itemize}
    \item 
    the successor is the next instruction\\
    In this case using the definition of successor relation we know that it will be
    in the first case.

    \item 
    the successor is not the next instruction
    Since there will be a $\ins{goto}$ appended, it will fall to the second case.
    Using the successor relation we know that the latter property holds.

  \end{itemize}
  then use IH to conclude.
\end{IEEEproof}

\begin{lem}[\ref{lemma:preservedSOAP}]
  SOAP properties are preserved in the translation from JVM to DEX.
\end{lem}
\begin{IEEEproof}
  We do prove by exhaustion, that is if the original JVM bytecode satisfies SOAP, then 
  the resulting translation to DEX instructions will also satisfy each of the property.
  \begin{description}
    \item[SOAP1.] Since the JVM bytecode satisfies SOAP, that means $i$ is a branching 
    point which will also be translated into a sequence of instruction. Denote 
    $i_b$ as the program point in the sequence and is a branching point. Let	
    $\translate{ P[k] }$ be the translation of instruction $P[k]$ and $k_1$ is the 
    address of its first instruction ($\translate{P[k]}[0]$). 
    Using the first case in the Lemma~\ref{preservedSuccessor}, 
    Lemma~\ref{preservedSuccessorException}, 
    Lemma~\ref{preservedSuccessorInvoke} and 
    Lemma~\ref{specialPreservedSuccessor}, 
    we know that $i_b \mapsto k_1$. In the case 
    that $k \in \fn{region}{i, \tau}$, we know that $k_1 \in \fn{region}{i_b, \tau}$ 
    using Definition~\ref{regionTranslation}. 
    In the case that $k = \fn{jun}{i_b, \tau}$, 
    we then will have $k_1 = \fn{jun}{i_b, \tau}$ using 
    Definition~\ref{junTranslation}.

    Special cases for the second case of Lemma~\ref{preservedSuccessor}, 
    Lemma~\ref{preservedSuccessorException} and 
    Lemma~\ref{preservedSuccessorInvoke} 
    in that they contain additional instruction in the lemma. 
    We argue that the property still holds using 
    Definition~\ref{regionTranslationInvoke}
    Definition~\ref{regionTranslationHandler}, 
    and Definition~\ref{regionTranslationGoto}. 
    Suppose $k'$ is the program point 
    that points to the extra instruction, then we have $k' \in \fn{region}{i_b, \tau}$ 
    from the 3 definitions we have mentioned. Following the argument from before, we
    can conclude that $k_1 \in \fn{region}{i, \tau}$ or $k_1 = \fn{jun}{i_b, \tau}$.

    \item[SOAP2.] Let $j_n$ be the last instruction in $\compile{ P[j] }$. 
    Denote $i_b$ as the program point in the sequence $\compile{i}$ and is a branching 
    point. Let	
    $\translate{ P[k] }$ be the translation of instruction $P[k]$ and $k_1$ is the 
    address of its first instruction ($\translate{P[k]}[0]$).
    Using 
    Definition~\ref{regionTranslation}, we obtain $j_n \in \fn{region}{i_b, \tau}$. 
    Using the first case of Lemma~\ref{preservedSuccessor}, 
    Lemma~\ref{preservedSuccessorException}, 
    Lemma~\ref{preservedSuccessorInvoke} and 
    Lemma~\ref{specialPreservedSuccessor} we will get that $j_n \mapsto k_1$. Now 
    since the JVM bytecode satisfies SOAP, we know that there are two cases we 
    need to take care of and $k$ will fall to one case or 
    the other. Assume $k \in \fn{region}{i, \tau}$, that means using 
    Definition~\ref{regionTranslation} we will have $k_1 \in \fn{region}{i_n, \tau}$. 
    Assume $k = \fn{jun}{i, \tau}$, we use Definition~\ref{junTranslation} and 
    obtain that $k_1 = \fn{jun}{i_n, \tau}$. 
    Either way, SOAP property is preserved for SOAP2. 
    Similar argument as SOAP1 to establish the second case of 
    Lemma~\ref{preservedSuccessor}, and that the property is still preserved in the 
    presence of $\ins{moveresult}$ and $\ins{moveexception}$.

    \item[SOAP3.] Trivial

    \item[SOAP4.] Let $k_1 = \fn{jun}{i, \tau_1}$ and $k_2 = 
    \fn{jun}{i, \tau_2}$ (this may be a bit confusing, this program point here refers 
    to the program point in JVM bytecode). Let $i_b$ be the instruction in 
    $\compile{ i }$ that branch and $k_{11}$ and $k_{21}$ be the first instruction in 
    $\compile{ P[k_1] }$ and $\compile{ P[k_2] }$ respectively. We proceed by using 
    Definition~\ref{regionTranslation} and the knowledge that the JVM bytecode satisfy 
    SOAP4 to establish that when $k_{11} \neq k_{21}$, then 
    $k_{11} \in \fn{region}{i_b,\tau_2}$ or $k_{21} \in \fn{region}{i_b, \tau_1}$ thus 
    the DEX program will also satisfy SOAP4.

    \item[SOAP5.] For any $\fn{jun}{i, \tau'}$ such that it is defined, let program 
    point $k = \fn{jun}{i, \tau'}$. Using Definition~\ref{junTranslation} we have 
    $k_1 = \fn{jun}{i_n, \tau'}$. Using Definition~\ref{regionTranslation}, 
    we know that $k_1 \in \fn{region}{i_n, \tau}$. 
    If we then set $k_1$ to be such point, where 
    $\fn{jun}{i_n, \tau'}$ and $\fn{jun}{i_n, \tau'} \in \fn{region}{i_n, \tau}$ for 
    any $\tau'$ with junction point defined, the property then holds.

    \item[SOAP6.] Is similar to the way proving SOAP5, with the addition of simple 
    property where the size of a code and its translation is covariant in a sense that 
    if an program $a$ has more codes than $b$, then $\compile{ a }$ also has more 
    codes than $\compile{ b }$.
  \end{description}
\end{IEEEproof}

\subsection{Proof that Translation Preserves Typability}

To prove the typability preservation of the compilation processes, we define an
intermediate type system closely resembles that of DEX, except that the addressing is
using block addressing. The purpose of this intermediate addressing is to know the
existence of registers typing to satisfy typability and the constraint satisfaction for
each instructions. We omit the details to avoid more clutters.

Following monotony lemma is useful in proving the relation of $\sqsubseteq$ between 
registers typing obtained from compiling stack types.

\begin{lemma}[Monotonicity of Translation]
\label{lemma:monotonicityOfTranslation}
  Let $rt$ be a register types and $S_1$ and $S_2$ stack types. If we have 
  $rt \sqsubseteq \translate{ S_1 }$ 
  and $S_1 \sqsubseteq S_2$, then $rt \sqsubseteq \translate{ S_2 }$ as well. 
\end{lemma}
\begin{IEEEproof} 
  Trivial based on the definition of $\translate{ . }$ and the $\sqsubseteq$ for 
  register types.  
\end{IEEEproof}

\begin{lem}[\ref{lemma:typable_rt}]
  For any JVM program $P$ with instruction $Ins$ at address $i$ and tag $\norm$, 
  let the length of $\translate{ Ins }$ denoted by $n$.
  Let $RT_{\translate{i}[0]}=\translate{ S_i }$.
  If according to the transfer rule for $P[i] = ins$ exists $st$ s.t. 
  $i \vdash^{\norm} S_{i} \Rightarrow st$ then
    \[ \begin{array}{c}
      \left( \begin{array}{c}
      \forall 0 \leq j < (n-1). \exists rt'. \translate{i}[j] \vdash^{\norm} 
        RT_{\translate{i}[j]} \Rightarrow rt', \\ 
        rt' \sqsubseteq RT_{\translate{i}[j+1]} 
      \end{array} \right)
      \\ \text{and} \\
      \exists rt. \translate{i}[n-1] \vdash^{\norm} RT_{\translate{i}[n-1]} 
        \Rightarrow rt, rt \sqsubseteq \translate{st} 
    \end{array}\] 
  according to the transfer rule(s) of $\translate{ Ins }$
\end{lem}
\begin{IEEEproof}
  It is case by case instruction, although for most of the instructions they are 
  straigthforward as they only translate into one instruction. 
  For the rest of the proof, 
  using definition~\ref{def:se_translation} to say that 
  the translated $se(\translate{i})$ have the same security level as $se(i)$.
  \begin{itemize}
    \item $\ins{Push}$\\
    We appeal directly to both of the transfer rule of $\ins{Push}$ and $\ins{Const}$. 
    In $\ins{Push}$ case, it only appends top of the stack with $se(i)$. Let such $rt$ 
    be 
      \[ rt=RT_{\translate{i}[0]} \oplus \{r(TS_{i}) \mapsto se(\translate{i}[0])\} \] 
    referring to $\ins{Const}$ transfer rule. 
    Since $\ins{Push}$ is translated into $\dexins{Const}{r(TS_{i})}$, where 
    $TS_{i}$ corresponds to the top of the stack, we know that $\translate{ st } = rt$
    because $RT_{\translate{i}[0]} = \translate{S_i}$ and the $rt$ we have is the same
    as $\translate{ se(i) :: S_i }$ thus $rt \sqsubseteq \translate{st}$
    	
    \item $\ins{Pop}$\\
    In this case, since the instruction does not get translated, this instruction  
    does not affect the lemma.
    
    \item $\ins{Load}\ x$\\
    Similar to $\ins{Push}$ except that the security value pushed on top of the stack 
    is $se(\translate{i}[0]) \sqcup \vec{k_a}(x)$. And although there are several 
    transfer rules for $\ins{move}$, there is only one applicable because the source 
    register comes from local variable register, and the target register is one of the 
    stack space. Using this transfer rule, we can trivially show that 
    $rt = \translate{st}$ where 
    $st = \big( se(i) \sqcup \vec{k_a}(x) \big) :: S_i$ and 
    $rt = RT_{\translate{i}[0]} \oplus 
    \{r(TS_{i}) \mapsto se(\translate{i}[0]) \sqcup \vec{k_a}(x)\}$, thus 
    $rt \sqsubseteq \translate{st}$.
    	
    \item $\ins{Store}\ x$\\
    This instruction is also translated as $\ins{move}$ except that the source 
    register is the top of the stack and the target register is one of the local 
    variable register. 
    The $rt$ in this case will be 
      \[\setlength{\arraycolsep}{1pt}\begin{array}{ll}
        rt = RT_{\translate{i}[0]} \oplus 
        \{ & r_x \mapsto se(\translate{i}[0]) \sqcup \\ 
        & RT_{\translate{i}[0]}\big(r(TS_{i} - 1)\big)\} 
      \end{array}\]
    This $rt$ coincides with the transfer rule for $\ins{move}$ where the target
    register is a register used to contain local variable.
    Since we know that the $x$ is in the range of local variable, we will have that 
    $rt \sqsubseteq \translate { st }$ based on
    the definition of $\sqsubseteq$, $\translate{ . }$ of flattening a stack. 
    	
    \item $\ins{Goto}$\\
    This instruction does not get translated just like $\ins{Pop}$, so this 
    instruction also does not affect the lemma.
    	
    \item $\ins{Ifeq}\ t$\\
    This instruction is translated to conditional branching in the DEX instruction. 
    There are two things happened to the stack types, one is that the removal of the 
    top value of the stack which is justified by the definition of 
    $\sqsubseteq$, and then lifting the value of the rest of the stack. 
    Since there is no lift involved in DEX, we know that this assignment will
    preserve typability as the registers are assigned higher security levels.
    	
    \item $\ins{Binop}$\\
    Translated as a DEX instruction for specified binary operator with the source 
    taken from the top two values from the stack, and then put the resulting value in 
    the then would be top of the stack. Let $rt$ in this case comes from 
      \[\setlength{\arraycolsep}{0pt}\begin{array}{ll}
        rt =&  RT_{\translate{i}[0]} \oplus \{r(TS_{i} - 2) \mapsto 
        se(\translate{i}[0]) \sqcup \\ 
        & RT_{\translate{i}[0]}\big(r(TS_{i} - 1)\big) \sqcup 
        \big(RT_{\translate{i}[0]}\big(r(TS_{i} - 2)\big) \big) \} 
      \end{array}\]
    This $rt$ corresponds to the scheme of DEX transfer rule for binary operation.
    Then we will have that $rt \sqsubseteq \translate{ st }$ where
    $st = se(i) \sqcup k_a \sqcup k_b :: st'$ and $S_i = k_a :: k_b :: st'$
    by Lemma~\ref{lemma:registersNotInStackLessEqual}
    
    \item $\ins{Swap}$\\
    In dx tool, this instruction is translated into 4 $\ins{move}$ instructions. In 
    this case, such $rt$ is 
    \[\setlength{\arraycolsep}{1pt}\begin{array}{lcl}
      rt & = RT_{\translate{i}[0]} & \oplus\{ \\
      & r(TS_{i}) & \mapsto se(\translate{i}[0]) \sqcup 
        RT_{\translate{i}[0]}\big(r(TS_{i} - 2)\big), \\
      & r(TS_{i} + 1) & \mapsto se(\translate{i}[1]) \sqcup 
        RT_{\translate{i}[1]}\big(r(TS_{i} - 1)\big), \\
      & r(TS_{i} - 2) & \mapsto se(\translate{i}[2]) \sqcup 
        RT_{\translate{i}[2]}\big(r(TS_{i} - 1)\big), \\
      & r(TS_{i} - 1) & \mapsto se(\translate{i}[3]) \sqcup 
        RT_{\translate{i}[3]}\big(r(TS_{i} - 2)\big) \}\\
    \end{array}\]
    justified by applying transfer rule for $\ins{move}$ 4 times. As before, appealing 
    to the definition of $\sqsubseteq$ to establish that this $rt \sqsubseteq 
    \translate{ st }$ where $st = k_b :: k_a :: st'$ and $S_i = k_a :: k_b :: st'$. 
    
    There's a slight subtlety here in that the relation might not hold 
    due to the presence of $se$ in $rt$ whereas there is no such occurrence in $st$. 
    But on a closer look, we know that in the case of swap instruction, the effect of 
    $se$ will be nothing. There are two cases to consider:
    \begin{itemize}
		  
      \item If the value in the operand stacks are already there before $se$ is 
      modified. We know that this can be the case only when there was a conditional 
      branch before, which also means that the operand stacks will be lifted to the 
      level of the guard and the level of $se$ is determined by this level of the 
      guard as well. So practically, they are the same thing
		  
      \item If the value in the operand stacks are put after $se$ is modified. Based 
      on the transfer rules of the instructions that put a value on top of the stack, 
      they will lub $se$ with the values, therefore another lub with $se$ will have 
      no effect.
    \end{itemize}
    
    For the first property, we have these registers typing
   \[\begin{array}{ccl}
      RT_{\translate{i}[1]} & = & RT_{\translate{i}[0]}
        \oplus \{ r(TS_{i}) \mapsto \\ && se(\translate{i}[0]) \sqcup 
        RT_{\translate{i}[0]}\big(r(TS_{i} - 2)\big) \}\\
      
      RT_{\translate{i}[2]} & = & RT_{\translate{i}[1]}
        \oplus \{ r(TS_{i} + 1) \mapsto \\ && se(\translate{i}[1]) \sqcup 
        RT_{\translate{i}[1]}\big(r(TS_{i} - 1)\big)\}\\
      
      RT_{\translate{i}[3]} & = & RT_{\translate{i}[2]}
        \oplus \{ r(TS_{i} - 2) \mapsto \\ && se(\translate{i}[2]) \sqcup 
        RT_{\translate{i}[3]}\big(r(TS_{i} - 1)\big)\}\\
    \end{array}\]
    which satisfy the property.
	  		  
    \item $\ins{New}$\\
    The argument that goes for this instruction is exactly the same as that of 
    $\ins{Push}$, where the $rt$ in this case is 
    $\translate{ S_{i} } \oplus \{r(TS_{i}) \mapsto se(\translate{i}[0])\}$.
       
    \item $\ins{Getfield}$\\
    In this case, the transfer rule for the translated instruction
    coincides with the transfer rule for $\ins{Getfield}$. Let 
      \[ rt = RT_{\translate{i}[0]} \oplus 
      \{r(TS_{i} - 1) \mapsto se(\translate{i}[0]) \sqcup \fn{ft}{f}\}\] 
    Then we have $rt = \translate{ st }$ which can be trivially shown with
    $st = se(i) \sqcup \fn{ft}{f} :: S_i$ thus giving us $rt\sqsubseteq\translate{st}$.
      
    \item $\ins{Putfield}$\\
    Since the JVM transfer rule for the operation itself only removes the top 2 stack, 
    and the transfer rule for DEX keep the registers typing, when we have
    $rt = \translate{ S_{i} }$, then by the definition of $\sqsubseteq$ we'll have 
    $rt \sqsubseteq \translate{ st }$ since $S_i = k_o :: k_v :: st$.
    As before, the registers that is not contained in the stack will by definition
    satisfy the $\sqsubseteq$ by Lemma~\ref{lemma:registersNotInStackLessEqual}.
      
    \item $\ins{Newarray}$\\
    Similar to the argument of $\ins{load}$, we have 
    \[\begin{array}{ll}
      rt = RT_{\translate{i}[0]} & \oplus \{r(TS_{i} - 1) \mapsto 
      \\ & RT_{\translate{i}[0]}\big(r(TS_{i}-1)\big)[\fn{at}{\translate{i}[0]}]\}
    \end{array}\] 
    , $rt = \translate{ st }$,
    where $st = k[at(i)] :: st', S_i = k :: st'$, which will give us
    $rt \sqsubseteq \translate{st}$.
        
    \item $\ins{Arraylength}$\\
    Let $k[k_c] = RT_{\translate{i}[0]}\big(r(TS_{i} - 1)\big) = S_i[0]$. In this case 
      $rt = RT_{\translate{i}[0]} \oplus \{r(TS_{i} - 1) \mapsto k\} = \translate{st}$
   then we will have $rt = \translate{st}$
   where $st = k :: st'$ and $S_i = k[k_c] :: st'$ which will give us 
   $rt \sqsubseteq \translate{st}$.
        
    \item $\ins{Arrayload}$\\
    Let $k[k_c] = RT_{\translate{i}[0]}\big(r(TS_{i} - 2)\big) = S_i[1]$. In this case 
      \[\setlength{\arraycolsep}{0pt}\begin{array}{ll}
        rt = & RT_{\translate{i}[0]} \oplus \{r(TS_{i} - 2) \mapsto   
        \\ & \big(se(i) \sqcup k \sqcup RT_{\translate{i}[0]}(r(TS_{i} - 1))\big) 
         \sqcupext k_c\} 
      \end{array}\] 
    which coincides with $\translate{st}$ where 
    $st = (k \sqcup k_i) \sqcupext k_c :: st'$ and $S_i = k_i :: k[k_c] :: st'$ except
    for lub with $se(i)$. The similar reasoning with $\ins{Swap}$ where lub with 
    $se(i)$ in this case will have no effect.
      
    \item $\ins{Arraystore}$\\  
    Similar argument with $\ins{putfield}$ where the JVM instruction remove top of the 
    stack and DEX instruction preserves the registers typing for $rt$. Thus appealing 
    to the definition of $\sqsubseteq$ we have that $rt \sqsubseteq \translate{ st }$.
      
    \item $\ins{Invoke}$\\
    This instruction itself yield 1 or 2 instructions depending whether the function 
    returns a value or not. Since the assumption for JVM type system is that functions 
    always return a value, the translation will be that $\ins{invoke}$ and 
    $\ins{moveresult}$ except that $\ins{moveresult}$ will always be in the region 
    $\norm$. Let $\vec{k'_a} \overset{k'_h}{\rightarrow} \vec{k'_r}$ be the policy for 
    method invoked. Type system wise, there will be 3 different cases for this 
    instruction, normal execution, caught, and uncaught exception. For this lemma, the 
    only one applicable is normal execution since it is the one tagged with $\norm$. 
    There will be 2 resulting instructions since it will also contain the instruction 
    $\ins{moveresult}$. Let $st_1$ be the stack containing the function's arguments,
    $t$ be the top of the stack after popping the function arguments from the stack 
    and the object reference $t = locN + (\fn{length}{S_i} - \fn{length}{st_1} - 1)$, 
    where $locN$ is the number of local variables. 
    Let $k$ be the security level of object referenced and 
    $k_e=\bigsqcup \big\{ \vec{k'_r}[e]\ |\ e \in \fn{excAnalysis}{m_\mathrm{ID}}$. 
    Since the method can also throw an exception, we have to also 
    include the lub of security level for possible exceptions, denoted by $k_e$. 
    In this case, such $rt$ can be 
      \[ \begin{array}{cl}
        RT_{\translate{i}[0]} \oplus 
        \{& ret \mapsto \big( \vec{k'_r}[n] \sqcup se(\translate{i}[0]) \big), 
        \\ & r_t \mapsto \big( \vec{k'_r}[n] \sqcup se(\translate{i}[1]) \big) \}
      \end{array}\]
    and by definition of $\sqsubseteq$ we will have that $rt\sqsubseteq\translate{st}$,
    where $st = \fn{lift_{k \sqcup k_e}}
    {\big( \vec{k_r'}[n] \sqcup se(i) \big) :: st_2}$ and $S_i = st_1 :: k :: st2$.    
    With that form of $rt$ in mind, then the registers typing for $\translate{i}[1]$ 
    can be
      \[ 
        RT_{\translate{i}[0]} \oplus 
        \{ret \mapsto \big( \vec{k'_r}[n] \sqcup se(\translate{i}[0]) \big)
      \]
    coming from the the transfer rule of $\ins{invoke}$ in DEX.
      
    \item $\ins{Throw}$\\
    This lemma will never apply to $\ins{Throw}$ since if the exception is caught, 
    then the successor will be in the tag $\tau \neq \norm$, but if the exception is 
    uncaught then the instruction is a return point.
    
  \end{itemize}
\end{IEEEproof}

\begin{lem}[\ref{lemma:typable_rt_exception}]
  For any JVM program $P$ with instruction $Ins$ at address $i$ and tag 
  $\tau \neq \norm$ with exception handler at address $i_e$. 
  Let the length of $\translate{ Ins }$ until the instruction that 
  throw exception $\tau$ denoted by $n$. Let $(be, 0)=\translate{i_e}$ be the address
  of the handler for that particular exception. If according to the transfer rule for 
  $Ins$ $i \vdash^{\tau} S_{i} \Rightarrow st$, then
    \[ \begin{array}{c}
      \left( \begin{array}{c}
      \forall 0 \leq j < (n-1). \exists rt'. \translate{i}[j] \vdash^{\norm} 
        RT_{\translate{i}[j]}\Rightarrow rt',\\ rt'\sqsubseteq RT_{\translate{i}[j+1]} 
      \end{array} \right)
      \\ \text{and} \\
      \exists rt. \translate{i}[n-1] \vdash^{\tau} RT_{\translate{i}[n-1]} \Rightarrow 
        rt, rt \sqsubseteq RT_{(be, 0)}
      \\ \text{and} \\
      \exists rt. (be, 0) \vdash^{\norm} RT_{(be, 0)} \Rightarrow rt, 
        rt \sqsubseteq \translate{st} 
    \end{array}\] 
  according to the transfer rule(s) of first $n$ instruction in $\translate{ Ins }$ 
  and $\ins{moveexception}$.
\end{lem}
\begin{IEEEproof}
  Case by case possibly throwing instructions:
  \begin{itemize}
    
    \item \ins{Invoke}\\
    We only need to take care of the case where the exception is caught, as uncaught 
    exception is a return point therefore there is no successor. In this case, $n = 1$ 
    as the instruction that may throw is the $\ins{invoke}$ itself, therefore the first
    property trivially holds ($\ins{moveexception}$ can't possibly throw an 
    exception). Let $locN$ in this case be 
    the number of local variables, and $e$ be the exception thrown. Let $k$ be the 
    security level of object referenced. In this case, the last $rt$ will take the 
    form 
      \[rt = \{ \vec{k_a}, ex \mapsto (k \sqcup \vec{k'_r}[e]), 
        r(locN) \mapsto (k \sqcup \vec{k'_r}[e]) \} \]
    Again with 
    this $rt$ we will have $rt \sqsubseteq \translate{ st }$, where 
    $st = (k \sqcup \vec{k_r'}[e]) :: \epsilon$. 
    Such $rt$ is obtained from the transfer rule for $\ins{invoke}$ where an exception 
    of tag $\tau$ is thrown, and the transfer rule for $\ins{moveexception}$.
    Then we have the registers typing for $(be, 0)$ as
      \[ RT_{(be, 0)} = \{ \vec{k_a}, ex \mapsto (k \sqcup \vec{k'_r}[e])\} \] 
    which fulfills the second property
    (transfer rule from $\ins{invoke}$) and the last property, which when joined with
    the transfer rule for $\ins{moveexception}$ will give us the $rt$ that we want.
      
    \item \ins{Throw}\\
    The argument follows that of $\ins{Invoke}$ for the caught and uncaught exception. 
    For uncaught exception, there is nothing to prove here as there is no resulting 
    $st$. For caught exception, let $k$ be the security level of the exception and 
    $locN$ be the number of local variable. Such $rt$ can be
      \[\setlength{\arraycolsep}{0pt}\begin{array}{ll}
        rt = \{ & \vec{k_a}, ex \mapsto \big(k \sqcup se(\translate{i}[0]) \big), 
        \\ & r(locN) \mapsto \big( k \sqcup se(\translate{i}[0]) \big) \} 
      \end{array}\]
    and it will make the relation $rt \sqsubseteq \translate{ st }$ holds, where 
    $st = (k \sqcup se(i)) :: \epsilon$ . This $rt$ 
    comes from the transfer rules for $\ins{throw}$ and $\ins{moveexception}$ combined.
    Registers typing for $(be, 0)$ takes the form of
    \[ RT_{(be, 0)}\{ \vec{k_a}, ex \mapsto (k \sqcup se(\translate{i}[0])\} \] 
    which will give us the 
    final $rt$ that we want after the transfer rule for $\ins{moveexception} $
      
    \item Other possibly throwing instruction\\
    Essentially they are the same as that of $\ins{throw}$ where the security level 
    that we are concerned with is the security level of the object lub-ed with its 
    security environment. The will also come from the transfer rule of each respective 
    instruction throwing a null pointer exception combined with the rule for 
    $\ins{moveexception}$.
    
  \end{itemize}
\end{IEEEproof}

\begin{lem}[\ref{lemma:typable_compile}]
  Let $ins$ be instruction at address $i$, $i \mapsto j$, $st$, $S_{i}$ and $S_{j}$ be 
  stack types such that $i \vdash S_{i} \Rightarrow st, st \sqsubseteq S_{j}$. 
  Let $n$ be the length of $\translate{ ins }$. 
  Let $RT_{\translate{i}[0]} = \translate{ S_{i} }$, 
  $RT_{\translate{j}[0]} = \translate{ S_{j}}$ 
  and $rt$ is obtained from the transfer rules involved in $\translate{ins}$. 
  Then $rt \sqsubseteq RT_{\translate{j}[0]} $. 
\end{lem}
\begin{IEEEproof}
  Using Lemma~\ref{lemma:typable_rt} and Lemma~\ref{lemma:typable_rt_exception} to 
  establish that we have $rt \sqsubseteq \compile{ st }$. Then we conclude by using 
  Lemma~\ref{lemma:monotonicityOfTranslation} to establish that 
  $rt \sqsubseteq RT_{\translate{j}[0]}$ because $st \sqsubseteq S_{j}$.
\end{IEEEproof}

\begin{lem}[\ref{lemma:constraintsatisfaction}]
  Let $Ins$ be instruction at program point $i$, $S_i$ its corresponding stack types, 
  and let $RT_{\translate{i}[0]} = \translate{ S_i }$. If $P[i]$ satisfy the typing 
  constraint for $Ins$ with the stack type $S_i$, then 
  $\forall (bj, j) \in \translate{i}. P_{DEX}[bj, j]$ 
  will also satisfy the typing constraints for all instructions in $\translate{ Ins }$ 
  with the initial registers typing $RT_{\translate{i}[0]}$.
\end{lem}
\begin{IEEEproof}
  We do this by case by case instruction:
  \begin{itemize}
    \item $\ins{Push}$ \\
      This instruction is translated into $\ins{Const}$ which
      does not have any constraints.
    	
    \item $\ins{Pop}$: does not get translated.
    	
    \item $\ins{Load}\ x$\\
      This instruction is translated into $\ins{Move}$ which does not have any 
      constraints.
    	
    \item $\ins{Store}\ x$\\
      This instruction is translated into $\ins{Move}$ which does not have any
      constraints.
    	
    \item $\ins{Goto}$: does not get translated
    	
    \item $\ins{Ifeq}\ t$\\
    This instruction will get translated to $\ins{ifeq}$ instruction where the 
    condition is based on top of the stack ($TS_i - 1$). 
    There is only one constraint of the form $\forall j' \in \fn{region}{i, \norm}, 
    rt\big(r(TS_i - 1)\big) \leq se(j')$, and we know that in the JVM bytecode the
    constraint $\forall j' \in \fn{region}{i, \norm}, k \leq se(j')$ is fulfilled.
    Based on the definition of $\translate{ . }$, we will have 
    $k = rt\big(r(TS_i - 1)\big)$. Thus we only need to prove that the difference
    in region will still preserve the constraint satisfaction.
    We do this by proof by contradiction.
    Suppose there exists such instruction at address 
    $(bj, j) \in \fn{region}{\translate{i}[n]}$ such that $k \nleq se(bj, j)$. 
    But according to definition~\ref{regionTranslation}, such instruction will come 
    from an instruction at address $i'$ s.t. $i' \in \fn{region}{i}$ thus it will 
    satisfy $k \leq se(i')$. By definition~\ref{def:se_translation}, 
    $se(bj, j) = se(i')$, thus we will have $k \leq se(bj, j)$. A plain contradiction.
    	
    \item $\ins{Binop}$\\
      This instruction is translated into $\ins{Binop}$ or $\ins{BinopConst}$ both of 
      which does not have any constraints.

    \item $\ins{Swap}$\\
    Trivially holds as well because all the 4 $\ins{move}$ instructions translated 
    from $\ins{swap}$ do not have any constraints.
	    
    \item $\ins{New}$\\ 
      Trivially holds as the $\ins{New}$ does not have any constraints.
      
    \item $\ins{Getfield}$\\
    There are different sets
    of constraints depending on whether the instruction executes normally, throw a 
    caught exception, or throw an uncaught exception.
    
    In the case of $\ins{Getfield}$ executing normally, there are only two constraints
    that we need to take care, one is that $rt(r_o) \in \mathcal{S}$ and 
    $\forall j \in \fn{region}{i, \norm}, rt(r_o) \leq se(j)$. The first constraint
    is trivial, since we already have that in JVM the constraint $k \in \mathcal{S}$
    is satisfied, where $S_i = k :: st$ for some stack type $st$. 
    We know that based on the definition of $\translate{ S_i }$ we
    have $rt(r_o) = k$, therefore we can conclude that $rt(r_o) \in \mathcal{S}$.
    The second constraint follows similar argument to the satisfaction of region
    constraint in $\ins{Ifeq}$.
    
    In the case of $\ins{Getfield}$ is throwing an exception, we then know that 
    based on the compilation scheme, depending on whether the exception is caught or 
    not, the same thing will apply to the translated instruction $\ins{iget}$, i.e.
    if $\ins{Getfield}$ has a handler for $\np$, so does $\ins{iget}$ and if
    $\ins{Getfield}$ does not have a handler for $\np$, $\ins{iget}$ does not either.
    Thus we only need to take care of one more constraint in that if this instruction
    does throw an uncaught exception, then it will satisfy 
    $rt(r_o) \leq \vec{k_r}[\np]$. This constraint is also trivially holds as the
    policy is translated directly, i.e. $\vec{k_r}[\np]$ is the same both in JVM type
    system and DEX type system, and that $rt(r_o) = k$. Since JVM typing satisfy 
    $k \leq \vec{k_r}[\np]$, then so does DEX typing.
      
    \item $\ins{Putfield}$\\
    To prove the constraint satisfaction for this instruction we appeal to the 
    translation scheme and the definition of $\translate{ . }$. We know from the 
    translation scheme that the resulting instruction is $\dexins{iput}
    {r(TS_i - 1), r(TS_i - 2), f}$, so the top of the stack ($TS_i - 1$) 
    corresponds to $r_s$ and the second to top of the stack ($TS_i - 2$) corresponds 
    to $r_o$. From the JVM transfer rule, we know that the security level of $S_i[0]$ 
    (denoted by $k_1$) is in the set of $\sext$ and the security level of $S_i[1]$ is 
    in the set of $\mathcal{S}$. Thus we know then know that the constraints 
    $rt(r_o) \in \mathcal{S}$ and $rt(r_s) \in \sext$ are fulfilled since we have 
    $rt(TS_i - 1) = S_i[0]$ and $rt(TS_i - 2) = S_i[1]$.   
    
    Now for constraints $k_h \leq \fn{ft}{f}$ and, 
    $(rt(r_o) \sqcup se(i)) \sqcupext rt(r_s) \leq \fn{ft}{f}$ we know that the 
    policies are translated directly, thus the constraint $k_h \leq \fn{ft}{f}$ 
    trivially holds. For the other constraint, we know that 
    $k_1 = rt(r_s)$, $k_2 = rt(r_o)$, and $se$ stays the same, therefore the 
    constraint $(rt(r_o) \sqcup se(i)) \sqcupext rt(r_s) \leq \fn{ft}{f}$ is also 
    satisfied because $(k_2 \sqcup se(i))\sqcupext k_1 \leq \fn{ft}{f}$ 
    is assumed to be satisfied. Lastly, for the rest of the constraints refer to 
    the proof in $\ins{Getfield}$ as they are essentially the same (the constraint
    for region, handler's existence~/~non-existence, and constraint against 
    $\vec{k_r}$ on uncaught exception).
      
    \item $\ins{Newarray}$\\
      Trivially holds as the instruction $\ins{Newarray}$ does not have any 
      constraints.
      
    \item $\ins{Arraylength}$\\
    We first deal with the constraints $k \in \mathcal{S}$ and 
    $k_c \in \sext$. From the definition of $\translate{ . }$, we know that 
    $rt(r_a) = k[k_c]$. Since JVM typing satisfies these constraints, it follows that 
    DEX typing also satisfies this constraints. For the rest of the constraints refer 
    to the proof in $\ins{Getfield}$ as they are essentially the same (the constraint 
    for region, handler's existence~/~non-existence, and constraint against 
    $\vec{k_r}$ on uncaught exception).
      
    \item $\ins{Arrayload}$\\
    We first deal with the constraints 
    $k, rt(r_i) \in \mathcal{S}$ and $k_c \in \sext$. From the definition of 
    $\translate{ . }$, we know that $rt(r_a) = k_2[k_c]$ and $rt(r_i) = k_1$.
    Since we know that JVM typing satisfies all the constraint, we know that 
    $rt(r_i) \in \mathcal{S}$ since $k_1 \in \mathcal{S}$, $k \in \mathcal{S}$ since
    $k_2 \in \mathcal{S}$, and $k_c \in \sext$ since in JVM typing $k_c \in \sext$.
    For the rest of the constraints refer 
    to the proof in $\ins{Getfield}$ as they are essentially the same (the constraint 
    for region, handler's existence~/~non-existence, and constraint against 
    $\vec{k_r}$ on uncaught exception).
      
    \item $\ins{Arraystore}$\\  
    Similar to that of $\ins{Putfield}$, where $rt(r_s) = k_1$, $rt(r_i) = k_2$, and 
    $k_3[k_c] = rt(r_a) = k'[k'_c]$. $k_2, k_3 \in \mathcal{S}$ gives us $k', rt(r_i) 
    \in \mathcal{S}$ and $k_1, k_c \in \sext$ gives us $k'_c, rt(r_s) \in \sext$.
    In this setting as well, it is easy to show that DEX typing satisfies
    $((k' \sqcup rt(r_i)) \sqcupext rt(r_s)) \leqext k_c'$ because JVM typing 
    satisfies $((k_2 \sqcup k_3) \sqcupext k_1) \leqext k_c$.
    For the rest of the constraints refer 
    to the proof in $\ins{Getfield}$ as they are essentially the same (the constraint 
    for region, handler's existence~/~non-existence, and constraint against 
    $\vec{k_r}$ on uncaught exception).
      
    \item $\ins{Invokevirtual}$\\
      There will be 3 different cases for this instruction,
        the first case is when method invocation executes normally.
        According to the translation scheme, the object reference will be put in 
        $\vec{p}[0]$ and the rest of parameters are arranged to match the arguments to 
        the method call. This way, we will have the correspondence that 
        $rt(\vec{p}[0]) = k$, and 
        $\forall i \in [0, \fn{length}{st_1} - 1]. \vec{p}[i + 1] = st_1[i]$. 
        Since the policies and $se$ are translated directly, we will have 
        $rt(\vec{p}[0]) \sqcup k_h \sqcup se(i) \leq k'_h$ since we know that the 
        original JVM instruction satisfy $k \sqcup k_h \sqcup se(i) \leq k'_h$. We 
        also know that $rt(\vec{p}[0])\leq \vec{k'_a}[0]$ since $k\leq\vec{k'_a}[0]$. 
        Similar argument applies to the rest of parameters to the method call to 
        establish that $\forall i \in [1, \fn{length}{st_1} - 1]. \vec{p}[i] \leq 
        \vec{k'_a}[i]$ that in turn will give us $\forall 0 \leq i \leq n. 
        rt(\vec{p}[i]) \leq \vec{k'_a}[i]$. For the last constraint, we know that 
        $\fnn{excAnalysis}$ also gets translated directly, thus yielding the same 
        $k_e$ for both JVM and DEX. Following the argument of $\ins{Getfield}$ for 
        the region constraint, we only need to make sure that 
        $rt(\vec{p}[0]) \sqcup k_e = k \sqcup k_e$ which is the case in our setting.
        Therefore, we will have that constraint
        $\forall j \in \fn{region}{i, \norm}. rt(\vec{p}[0]) \sqcup k_e \leq se(j)$ 
        is satisfied. 
          
        The second case is when method invocation thrown a caught exception.
        Basically the same arguments as that of normal execution, except that the 
        region condition is based upon particular exception
        $\forall j \in \fn{region}{i, e}.$ $rt(\vec{p}[0]) \sqcup k'_r[e] \leq 
        se(j)$. Since the policy stays the same, JVM instruction satisfy this 
        constraint will imply that the DEX instruction will also satisfy the 
        constraint. Since now the method is throwing an exception, we also need to 
        make sure that it is within the possible thrown exception defined in 
        $\fnn{excAnalysis}$. Again as the class stays the same and that 
        $\fnn{excAnalysis}$ is the same, the satisfaction of 
        $e \in \fn{excAnalysis}{m_\mathrm{ID}} \sqcup \{\np\}$ in JVM side implies the 
        satisfaction of $e \in \fn{excAnalysis}{m'} \sqcup \{\np\}$ in DEX side.
          
        The last case is when method invocation thrown an uncaught exception.
        Same argument as the caught exception with the addition that escaping 
        exception are contained within the method's policy. Since we have $k \sqcup 
        se(i) \sqcup \vec{k'_r}[e] \leq \vec{k_r}[e]$ in the JVM side, it will also 
        imply that $rt(\vec{p}[0])\sqcup se(i) \sqcup \vec{k'_r}[e]\leq\vec{k_r}[e]$ 
        in the DEX side since $rt(\vec{p}[0]) = k$ and everything else is the same.
          
      Actually there is a possibility that there is addition of $\ins{moveresult}$ 
      and/or $\ins{moveexception}$, except that the target of this instruction will be 
      in the stack space, therefore there will be no constraint involved to satisfy.
      
    \item $\ins{Throw}$\\
      Similar arguments to that of $\ins{Invokevirtual}$ addressing the similar form 
      of the constraints. 
        In the case of caught exception case, the constraint 
        $e \in \fn{classAnalysis}{i} \cup \{ \np \}$ is 
        satisfied because, as before, $\fnn{classAnalysis}$ and classes ($e$) are the 
        same. So, if JVM program satisfy the constraint the translated DEX program 
        will also satisfy it. The same with $\forall j \in \fn{region}{i, e} 
        rt(r) \leq se(j)$ since $rt(r) = k$.
          
        The case where exception is uncaught is the same as the caught case with 
        addition that the security level of thrown exception must be contained within 
        method's policy. In this case, we already have $rt(r) \leq \vec{k_r}[e]$ 
        since $rt(r) = k$ and policies stay the same.
      
  \end{itemize}
\end{IEEEproof}

This lemma states that a typable JVM program (block wise and within blocks) will 
translate into typable DEX program.
\begin{lem}[\ref{lemma:typable_blocks}]
  Let $P$ be a JVM program such that 
  \[ \forall i, j. i \mapsto j. \exists st. i \vdash S_i \Rightarrow st 
  \hspace{0.5cm} \text{ and } \hspace{0.5cm} st \sqsubseteq S_j \]
  Then $\translate{ P }$ will be 
  \begin{enumerate}
    \item for all blocks $bi, bj$ s.t. $bi \mapsto bj$, $\exists rt_b.$ s.t.
      $RTs_{bi} \Rightarrow^* rt_b, rt_b \sqsubseteq RTs_{bj}$; and
    \item $\forall bi, i, j \in bi.$ s.t. $(bi, i) \mapsto (bi, j). \exists rt.$ s.t.
      $(bi, i) \vdash RT_{(bi, i)} \Rightarrow rt, rt \sqsubseteq RT_{(bi, j)}$
  \end{enumerate}
  where 
    \[ \begin{array}{lcl}
      RTs_{bi} = \translate{S_{i}} &\text{ with }& \translate{i} = (bi, 0) \\
      RTs_{bj} = \translate{S_{j}} &\text { with }& \translate{j} = (bj, 0), \\
      RT_{(bi, i)} = \translate {S_{i'}}&\text{ when }&\translate{i'} = (bi, i)\\
      RT_{(bi, j)} = \translate {S_{j'}} &\text{ when }& \translate{j'} = (bj, j)
      \end{array} \]  
\end{lem}
\begin{IEEEproof}
  For the first property, they are mainly proved using 
  Lemma~\ref{lemma:typable_compile} because we know that if a DEX instruction is at 
  the end of a block, it is the last instruction in its translated JVM instruction, 
  except for $\ins{invoke}$ and throwing instructions. Based on 
  Lemma~\ref{lemma:typable_compile}, we have that 
  $rt \sqsubseteq RT_{(bj, 0)}$, where $RT_{(bj, 0)} = \translate{S_j}$. 
  Since by definition $rt_b$ is such $rt$ and $RTs_{bj} = RT_{(bj, 0)}$, the 
  property holds. For $\ins{invoke}$ we use the first case of 
  Lemma~\ref{lemma:typable_rt}, and for throwing instructions we use the first case
  of Lemma~\ref{lemma:typable_rt_exception}.
  
  For the second property, it is only possible if the DEX instruction at address $i$ 
  is non-invoke and non-throwing instruction. There are two possible cases here, 
  whether $i$ and $j$ comes from the same JVM instruction or not. If $i$ and $j$ comes 
  from the same JVM instruction, then we use the first case of 
  Lemma~\ref{lemma:typable_rt}. Otherwise, we use Lemma~\ref{lemma:typable_compile}.
  
\end{IEEEproof}

Before we proceed to the proof of Lemma~\ref{lemma:orderTypability}, we define a 
property which is satisfied after the ordering and output phase.

\begin{property}
\label{fixTarget}
For any block whose next order is not its primary successor, there are two possible 
cases. If the ending instruction is not $\ins{ifeq}$, then there will be a 
$\ins{goto}$ instruction appended after the output of that particular block. If the
ending instruction is $\ins{ifeq}$, check whether the next order is in fact the second 
branch. If it is the second branch, then we need to ``swap'' the $\ins{ifeq}$ 
instruction into $\ins{ifneq}$ instruction. Otherwise appends $\ins{goto}$ to the 
primary successor block.
\end{property}

\begin{lem}[\ref{lemma:orderTypability}]
  Let $\llfloor P \rrfloor$ be a typable DEX blocks resulted from translation of JVM
  instruction still in the block form, i.e.
  \[\llfloor P \rrfloor = \fn{Translate}{\fn{TraceParentChild}{\fn{StartBlock}{P}}}\] 
  Given the ordering scheme to output the block contained in
  $\fnn{PickOrder}$, if the starting block starts with flag $0$ ($F_{(0,0)} = 0$)  
  then the output $\compile{ P }$ is also typable.
\end{lem}
\begin{IEEEproof}
  The proof of this lemma is straightforward based on the definition of the property 
  and typability. Assuming that initially we have the blocks already typable, then 
  what's left is in ensuring that this successor relation is preserved in the output 
  as well. Since the output is based on the ordering, and the property ensures that 
  for any ordering, all the block will have correct successor, then the typability of 
  the program is preserved.

  To flesh out the proof, we go for each possible ending of a block and its program
  output.
  \begin{itemize}
    \item Sequential instruction\\
    There are two possible cases, the first case is that 
      the successor block is the next block in order. Let $bi$ indicate the 
      current block and $bj$ the successor block in question. Let $i_n$ be the last 
      instruction in $bi$, then we know that $\exists rt. RT_{(bi, i_n)} 
      \Rightarrow rt, rt \sqsubseteq RTs_{bj}$ where $RTs_{bj}$ will be the registers 
      typing for the next instruction (in another word $RT_{(bj, 0)}$). Therefore, the 
      typability property trivially holds.

      The second case is that 
      the successor block is not the next block in order. According to step
      performed in the \textbf{Output} phase, the property~\ref{fixTarget} will be 
      satisfied. Thus there will be a $\ins{goto}$ appended after instructions in
      the block output targetting the successor block. Let such block be $bi$ and the 
      successor block $bj$. Let $i_n$ be the last instruction in $bi$. From the 
      definition of typability, we know that if $bj$ is the next block to output, then 
      $\exists rt. RT_{(bi, i_n)} \Rightarrow rt, rt \sqsubseteq RTs_{bj}$. Now with 
      additional $\ins{goto}$ in the horizon, we appeal to the transfer rule to 
      establish that this instruction does not need to modify the registers typing, 
      i.e. $\exists rt. (bi, i_n) \vdash RT_{(bi, i_n)} \Rightarrow 
      RT_{(bi, i_n + 1)}, (bi, i_n + 1) \vdash RT_{(bi, i_n + 1)} \Rightarrow rt,  
      rt \sqsubseteq RTs_{bj}$ where $RT_{(bi, i_n + 1)} = rt$. 

    \item $\ins{ifeq}$\\
    There are three possible cases here, the first case is that
      the next block to output is its primary successor.
      It is trivial as the relationship is preserved in that the next block to output 
      is the primary successor.

      The next case is that 
      the next block to output is its secondary successor. We switch the 
      instruction to its complementary, i.e. $\ins{ifneq}$. Let $bi$ be the current 
      block, $bj$ be the primary successor (which is directly placed after this 
      block), and $bk$ the other successor. Let $i_n$ be the index to the last 
      instruction in $bi$. If $bi$ ends with $\ins{ifneq}$, then we know that it is 
      originally from the instruction $\ins{ifeq}$ and the blocks are typable, 
      therefore we have that for the two successors of $bi$ the following relation 
      holds: 
      $\exists rt_1. i_n \Rightarrow rt_1, rt_1 \sqsubseteq RTs_{bj}$ and
      $\exists rt_2. i_n \Rightarrow rt_2, rt_2 \sqsubseteq RTs_{bk}$, which defines 
      the typability for the output instructions.

      The last case is when 
      the next block to output is not its successor. The argument is the same 
      as  the sequential instruction one, where we know that adding $\ins{goto}$ can 
      maintain the registers typing thus preserving the typability by fixing the 
      successor relationship.

    For the secondary successor (target of branching), we know that there is a step in 
    the output that handles the branch addressing to maintain the successor 
    relations.  

    \item $\ins{invoke}$, yet the next block to output is not $\ins{moveresult}$\\
    Although superficially this seems like a possibility, the fact that 
    $\ins{moveresult}$ is added corresponding to a unique $\ins{invoke}$ renders the 
    case impossible. If $\ins{moveresult}$ is not yet ordered, we know that it will be 
    the next to output based on the ordering scheme. This is the only way that a 
    $\ins{moveresult}$ can be given an order, so it is impossible to order a 
    $\ins{moveresult}$ before ordering its unique $\ins{invoke}$.

  \end{itemize}
\end{IEEEproof}


\section{Full JVM Operational Semantics and Transfer Rules}
\label{section:appendix_jvm}

The following figure~\ref{figure:javaOperationalSemanticFull} is the full operational 
semantics for JVM in section~\ref{javaTypeSystem}.
The function $\fnn{fresh} : \text{Heap} \rightarrow \mathcal{L}$ is an 
allocator function that given a heap returns the location for that object. The function 
$\fnn{default} : \mathcal{C} \rightarrow \mathcal{O}$ returns for each class a default 
object of that class. For every field of that default object, the value will be $0$ if 
the field is numeric type, and $null$ if the field is of object type. Similarly 
$\fnn{defaultArray} : \mathbb{N} \times \mathcal{T}_J \rightarrow (\mathbb{N} 
\rightharpoonup \mathcal{V})$. The $\rightsquigarrow$ relation which defines transition 
between state is $\rightsquigarrow\subseteq \object{State} \times (\object{State} + 
\mathcal{V \times \object{Heap}})$. 

The operator $\oplus$ denotes the function 
where $\rho \oplus \{r \mapsto v\}$ means a new function $\rho'$ such that 
$\forall i \in \fn{dom}{\rho} \backslash \{r\}. 
\rho'(i) = \rho(i)$ and $\rho'(r) = v$. The operator $\oplus$ is overloaded to also 
mean the update of a field on an object, or update on a heap. 

For method invocation, program comes equipped with a set $\mathcal{M}$ of method names, 
and for each method $m$ there are associated list of instructions $P_m$. Each method is 
identified by method identifier $m_\mathrm{ID}$ which can refer to several methods in 
the case of overriding. Therefore we also need to know which class this method is 
invoked from, which can be identified by auxilliary function $\fnn{lookup_{P}}$ which 
returns the precise method to be executed based on the method identifier and class.

To handle exception, program will also comes equipped with two parameters 
$\fnn{classAnalysis}$ and $\fnn{excAnalysis}$. $\fnn{classAnalysis}$ contains 
information on possible classes of exception of a program point, and 
$\fnn{excAnalysis}$ contains possible escaping exception of a method.

There is also additional partial function for method $m$ $\fnn{Handler_m} : 
\mathcal{PP} \times \mathcal{C} \rightharpoonup \mathcal{PP}$ which gives the handler 
address for a given program point and exception. Given a program point $i$ and an 
exception thrown $c$, if $\fn{Handler_m}{i, c} = t$ then the control will be 
transferred to program point $t$, if the handler is undefined (noted 
$\fn{Handler_m}{i, c} \uparrow$) then the exception is uncaught in method $m$. 


The next figure~\ref{figure:javaTypeRuleFull} is the full version of 
figure~\ref{figure:javaTypeRuleSelected} in section~\ref{javaTypeSystem}.
The full typing judgement takes the form of $\Gamma, \fnn{ft}, \fnn{region}, 
sgn, se, i \vdash^{\tau} st \Rightarrow st'$
where $\Gamma$ is the table of method policies, $\fnn{ft}$ is the global policy for
fields, $\fnn{region}$ is the CDR information for the current method, 
$sgn$ is the policy for the current method taking the form of 
$\vec{k_a}\overset{k_h}{\rightarrow}\vec{k_r}$, $se$ is the security environment,
$i$ is the current program point, 
$st$ is the stack typing for the current instruction, and $st'$ is the stack
typing after the instruction is executed.

As in the main paper, we may not write the full notation whenever it is clear from the 
context. In the table of operational semantics, we may drop the subscript
$m, \norm$ from $\rightsquigarrow$, e.g. we may write $\rightsquigarrow$ instead of
$\rightsquigarrow_{m, \norm}$ to mean the same thing. In the table of transfer rules,
we may drop the superscript of tag from $\vdash^{\tau}$ and write $\vdash$ instead.
The same case applies to the typing judgement, we may write $i \vdash^{\tau} st 
\Rightarrow st'$ 
instead of $\Gamma, \fnn{ft}, \fnn{region}, 
\vec{k_a}\overset{k_h}{\rightarrow}\vec{k_r}, se, i \vdash^{\tau} st \Rightarrow st'$.


\begin{figure*}[ht!]
\[\setlength{\arraycolsep}{0pt}
\boxed{\begin{array}{c}
\begin{array}{ccc} \hspace{0.25\textwidth} & \hspace{0.45\textwidth} & 
    \hspace{0.3\textwidth}\\
  	\hfill \dfrac{P_m[i] = \ins{goto}\ j}
	  {\cstate{i, \rho, os} \rightsquigarrow_{m, \norm} \cstate{j, \rho, os}} \hfill &
	\hfill \dfrac{P_m[i] = \ins{swap}}
  {\cstate{i, \rho, v_1 :: v_2 :: os} \rightsquigarrow_{m, \norm} 
    \cstate{i+1, \rho, v_2 :: v_1 :: os}} \hfill &
	\hfill \dfrac{P_m[i] = \ins{goto}\ j}
	  {\cstate{i, \rho, os} \rightsquigarrow_{m, \norm} \cstate{j, \rho, os}} \hfill \\
\end{array}\\
\begin{array}{ccc}\hspace{0.3\textwidth} & \hspace{0.3\textwidth} & 
    \hspace{0.4\textwidth}\\
	  
	\hfill \dfrac{P_m[i] = \ins{ifeq}\ j\ \ \ \ n \neq 0}
	{\cstate{i, \rho, n :: os} \rightsquigarrow_{m, \norm} 
	  \cstate{i + 1, \rho, os}} \hfill &
	
	\hfill \dfrac{P_m[i] = \ins{ifeq }\ j\ \ \ \ n = 0}
	{\cstate{i, \rho, n :: os} \rightsquigarrow_{m, \norm} 
	  \cstate{j, \rho, os}} \hfill & 
	  
	\hfill 
	\dfrac{\begin{gathered}P_m[i] = \ins{store }\ x\ \ x \in \fn{dom}{\rho}\end{gathered}}
	{\begin{gathered}\cstate{i, \rho, v :: os} \rightsquigarrow_{m, \norm} 
	  \cstate{i+1, \rho \oplus \{x \mapsto v\}, os}\end{gathered}} \hfill \\
\end{array}\\

\begin{array}{ccc} \hspace{0.35\textwidth} & \hspace{0.4\textwidth} & 
    \hspace{0.25\textwidth}\\
	   
  \hfill \dfrac{P_m[i] = \ins{load }\ x}
	{\begin{gathered}\cstate{i, \rho, os} \rightsquigarrow_{m, \norm}
		\cstate{i + 1, \rho, \rho(x) :: os}\end{gathered}} \hfill &
		
  \hfill \dfrac{P_m[i] = \ins{binop }\ op\ \ \ \ n_2\ \underline{op}\ n_1 = n}
  {\cstate{i, \rho, n_1 :: n_2 :: os} \rightsquigarrow_{m, \norm} 
    \cstate{i + 1,\rho, n :: os}} \hfill &
  
  \hfill \dfrac{P_m[i] = \ins{return}}
	  {\cstate{i, \rho, v :: os} \rightsquigarrow_{m, \norm} v, h} \hfill
   \\
\end{array}\\
\begin{array}{cc} \hspace{0.5\textwidth} & \hspace{0.5\textwidth} \\  
  
  \dfrac{P_m[i] = \ins{new }\ C\ \ \ \ l = \fn{fresh}{h}}
  {\cstate{i, \rho, os, h} \rightsquigarrow
    \cstate{i + 1, \rho, l :: os, h \oplus \{l \mapsto \fn{default}{C}\}}} &
 
  \dfrac{P_m[i]=\ins{getfield }\ f\ \ \ \ l \in\fn{dom}{h}\ \ \ \ f \in\fn{dom}{h(l)}}
	{\cstate{i, \rho, l :: os, h} \rightsquigarrow_{m, \norm} 
	  \cstate{i + 1, \rho, h(l).f :: os, h}} \\  
\end{array}\\
\begin{array}{cc}	\hspace{0.6\textwidth} & \hspace{0.4\textwidth}\\  
	  
	\dfrac{P_m[i]=\ins{getfield }\ f\ \ \ \ l' = \fn{fresh}{h}}
	{\cstate{i, \rho, null :: os, h} \rightsquigarrow_{m, \np}
	  \fn{RuntimeExcHandling}{h, l', \np, i, \rho}} &
	\hfill \dfrac{P_m[i] = \ins{push }\ n}
	  {\cstate{i, \rho, os} \rightsquigarrow_{m, \norm} 
	    \cstate{i + 1, \rho, n :: os}} \hfill \\  
\end{array}\\
\begin{array}{cc}	 \hspace{0.6\textwidth} & \hspace{0.4\textwidth}\\	  
	\dfrac{P_m[i]=\ins{putfield }\ f\ \ \ \ l \in\fn{dom}{h}\ \ \ \ f \in\fn{dom}{h(l)}}
	{\cstate{i, \rho, v :: l :: os, h} \rightsquigarrow_{m, \norm} 
	  \cstate{i + 1, \rho, os, h \oplus \{l \mapsto h(l) \oplus \{f \mapsto v\}\}} } &
	\hfill \dfrac{P_m[i] = \ins{pop}}
	  {\cstate{i, \rho, v :: os} \rightsquigarrow_{m, \norm} 
	    \cstate{i + 1, \rho, os}} \hfill \\
\end{array}\\
\begin{array}{cc}	 \hspace{0.65\textwidth} & \hspace{0.35\textwidth}\\  
	\dfrac{P_m[i]=\ins{putfield }\ f\ \ \ \ l' = \fn{fresh}{h}}
	{\cstate{i, \rho, v :: null :: os, h} \rightsquigarrow_{m, \np} 
	  \fn{RuntimeExcHandling}{h, l', \np, i, \rho} } & 
	\hfill \dfrac{P_m[i] = \ins{return}}
	  {\cstate{i, \rho, v :: os} \rightsquigarrow_{m, \norm} v, h} \hfill \\  
	  
\end{array}\\
\begin{array}{c}\hspace{0.9\textwidth}\\
	  
  \colfrac{1}{P_m[i] = \ins{newarray }\ t\ \ \ \ l = \fn{fresh}{h}\ \ \ \ n \geq 0}
	{\begin{gathered}
	  \cstate{i, \rho, n :: os, h} \rightsquigarrow_{m, \norm} 
	  \cstate{i + 1, \rho, 
	  l :: os, h \oplus \{l \mapsto (n, \fn{defaultArray}{n, t}, i)\}} 
	  \end{gathered}} \\\\  
	 
	\colfrac{1}{P_m[i] = \ins{arraylength}\ \ \ \ l \in \fn{dom}{h}}
	{\cstate{i, \rho, l :: os, h} \rightsquigarrow_{m, \norm} 
	  \cstate{i + 1, \rho, h(l).\field{length} :: os, h}} \\\\

  \colfrac{1}{P_m[i] = \ins{arraylength}\ \ \ \ l' = \fn{fresh}{h}}
	{\cstate{i, \rho, null :: os, h} \rightsquigarrow_{m, \np} 
	  \fn{RuntimeExcHandling}{h, l', \np, i, \rho}} \\\\

	\dfrac{P_m[i]=\ins{arrayload}\ \ \ \  l\in\fn{dom}{h}\ \ \ \ 
	  0 \leq j<h(l).\field{length}}
	{\cstate{i, \rho, j :: l :: os, h} \rightsquigarrow_{m, \norm} 
	  \cstate{i + 1, \rho, h(l)[j] :: os, h}} \\\\
	
  \dfrac{P_m[i]=\ins{arrayload}\ \ \ \  l' = \fn{fresh}{h}}
	{\cstate{i, \rho, j :: null :: os, h} \rightsquigarrow_{m, \np} 
	  \fn{RuntimeExcHandling}{h, l', \np, i, \rho}}\\\\

	\dfrac{P_m[i] = \ins{arraystore}\ \ \ \ l \in \fn{dom}{h}\ \ \ \ 
	  0 \leq j < h(l).\field{length}}
	{\cstate{i, \rho, v :: j :: l :: os, h} \rightsquigarrow_{m, \norm}
		\cstate{i+1, \rho, os, h \oplus \{l \mapsto h(l) \oplus \{j \mapsto v\}\}} }\\\\
	  
	\dfrac{P_m[i] = \ins{arraystore}\ \ \ \ l' = \fn{fresh}{h}}
	{\cstate{i, \rho, v :: j :: null :: os, h} \rightsquigarrow_{m, \np} 
	  \fn{RuntimeExcHandling}{h, l', \np, i, \rho}}\\
\end{array}\\
\begin{array}{c}	\hspace{\textwidth}\\
  \colfrac{1}{\begin{gathered} Pm[i] = \ins{invoke }\ m_\mathrm{ID}\ \ \ \ 
	  m' = \fn{lookup_P}{m_\mathrm{ID}, \fn{class}{h(l)}} \ \ \ \ l \in \fn{dom}{h} \\ 
		\fn{length}{os_1} = \fn{nbArguments}{m_\mathrm{ID}}\ \ \ \ 
	  \langle 1, \{this \mapsto l, \vec{x} \mapsto os_1\}, \epsilon, h \rangle 
	  \rightsquigarrow^{+}_{m'} v, h' \end{gathered}}
	{\langle i, \rho, os_1 :: l :: os_2, h \rangle \rightsquigarrow_{m, \norm} 
	  \langle i+1, \rho, v :: os_2, h' \rangle}\\\\	 

  \colfrac{1}{\begin{gathered} P_m[i] = \ins{invoke }\ m_\mathrm{ID} \ \ \ \ 
	  m' = \fn{lookup_P}{m_\mathrm{ID}, \fn{class}{h(l)}}\ \ \ \ 
		\langle 1, \{this \mapsto l, \vec{x} \mapsto os_1\}, \epsilon, h \rangle 
		\rightsquigarrow^{+}_{m'} \langle l' \rangle, h'\\ 
		l \in \fn{dom}{h}\ \ \ \ 
		\fn{Handler_m}{i, e} = t\ \ \ \ e = \fn{class}{h'(l')}\ \ \ \ 
		e \in \fn{excAnalysis}{m_\mathrm{ID}} \end{gathered}}
	{\cstate{i, \rho, os_1 :: l :: os_2, h} \rightsquigarrow_{m, e} 
	  \langle t, \rho, l' :: \epsilon, h' \rangle} \\\\	    

  \dfrac{P_m[i] = \ins{invoke }\ m_\mathrm{ID}\ \ \ \ l' = \fn{fresh}{h}}
	{\cstate{i, \rho, os_1 :: null :: os_2, h} \rightsquigarrow_{m, \np} 
	  \fn{RuntimeExcHandling}{h, l', \np, i, \rho}} \\\\

\end{array}\\
\end{array}}\]
\end{figure*}

\begin{figure*}[ht!]
\[\setlength{\arraycolsep}{0pt}
\boxed{
\begin{array}{c}

\begin{array}{c}	\hspace{\textwidth}\\

  \dfrac{\begin{gathered} P_m[i] = \ins{invoke }\ m_\mathrm{ID}\ \ \ \ 
	  m' = \fn{lookup_P}{m_\mathrm{ID}, \fn{class}{h(l)}} \ \ \ \ 
		\langle 1, \{this \mapsto l, \vec{x} \mapsto os_1\}, \epsilon, h \rangle 
		\rightsquigarrow^{+}_{m'} \langle l' \rangle, h'\\
		l \in \fn{dom}{h}\ \ \ \ e = \fn{class}{h'(l')}\ \ \ \ 
		\fn{Handler_{m}}{i, e} \uparrow \ \ \ \ e \in \fn{excAnalysis}{m_\mathrm{ID}} 
		\end{gathered}}
	{\cstate{i, \rho, os_1 :: l :: os_2, h} \rightsquigarrow_{m, e} 
	  \langle l' \rangle, h'} \\\\

	\dfrac{P_m[i] = \ins{throw}\ \ \ \ l' = \fn{fresh}{h} }
	{\cstate{i, \rho, null :: os, h} \rightsquigarrow_{m, \np} 
	  \fn{RuntimeExcHandling}{h, l', \np, i, \rho}}\\\\
	
	\dfrac{\begin{gathered} P_m[i] = \ins{throw}\ \ \ \ l \in \fn{dom}{h}\ \ \ \ 
	  e = \fn{class}{h(l)}\ \ \ \ 
		\fn{Handler_m}{i, e} = t\ \ \ \ e \in \fn{classAnalysis}{m, i}\end{gathered}}			
	{\cstate{i, \rho, l :: os, h} \rightsquigarrow_{m, e} 
	  \cstate{t, \rho, l :: \epsilon, h}} \\\\										
	
	\dfrac{\begin{gathered} P_m[i] = \ins{throw}\ \ \ \ l \in \fn{dom}{h}\ \ \ \ 
	  e = \fn{class}{h(l)}\ \ \ \ \fn{Handler_m}{i, e} \uparrow\ \ \ \ 
	  e \in \fn{classAnalysis}{m, i}\end{gathered}}			
	{\cstate{i, \rho, l :: os, h} \rightsquigarrow_{m, e} \langle l \rangle, h} \\\\	
				
	\multicolumn{1}{l}{\text{with }\fnn{RuntimeExcHandling} : 
	\object{Heap} \times \mathcal{L} \times \mathcal{C} \times \mathcal{PP} \times 
	(\mathcal{X} \rightharpoonup \mathcal{V}) \rightarrow 
	\object{State} + (\mathcal{L} \times \object{Heap}) \text{ defined as}}\\\\
	
	\multicolumn{1}{l}{\fn{RuntimeExcHandling}{h, l', C, i, \rho} =
	\left\{ \begin{array}{ll} \cstate{t, \rho, l' :: \epsilon, h \oplus \{l' \mapsto 
	\fn{default}{C}\}} & \text{if }\fn{Handler_m}{i, C} = t \\
	\langle l' \rangle, h \oplus \{l' \mapsto \fn{default}{C}\} & \text{if }
	\fn{Handler_m}{i, C} \uparrow\end{array}\right.}	\\\\
	
\end{array}				 
\end{array}}\] 

\caption{Full JVM Operational Semantic}
\label{figure:javaOperationalSemanticFull}
\end{figure*}



\begin{figure*}[!htbp]
\[\boxed{
\begin{array}{c}
\begin{array}{ccc}
  \hspace{0.31\textwidth} & \hspace{0.31\textwidth} & \hspace{0.31\textwidth}\\
  
  \hfill \dfrac{P_m[i] = \ins{load}\ x}
    {se, i \vdash st \Rightarrow \big(\vec{k_a}(x) \sqcup se(i)\big) :: st} \hfill 
  & \hfill \dfrac{P_m[i] = \ins{store}\ x\ \ \ \ se(i) \sqcup k \leq \vec{k_a}(x)}
    {se, i \vdash k :: st \Rightarrow st} \hfill	
  & \hfill \dfrac{P_m[i] = \ins{swap}}
    { i \vdash k_1 :: k_2 :: st \Rightarrow k_2 :: k_1 :: st} \hfill \\
    
\end{array}\\
\begin{array}{ccc}
  \hspace{0.4\textwidth} & \hspace{0.13\textwidth} & \hspace{0.25\textwidth}\\	
  
  \hfill 
  \dfrac{P_m[i]=\ins{ifeq}\ j\ \ \ \ \forall j'\in\fn{region}{i,\norm},k\leq se(j')}
  {\fnn{region}, se, i \vdash k :: st \Rightarrow \fn{lift_k}{st}} \hfill &  
  
  \hfill \dfrac{P_m[i] = \ins{goto}\ j}{i \vdash st \Rightarrow st} \hfill & 
  
  \hfill \dfrac{P_m[i] = \ins{return}\ \ \ \ se(i) \sqcup k \leq k_r[n]}
  {\vec{k_a} \overset{k_h}{\rightarrow} \vec{k_r}, se, i \vdash k :: st \Rightarrow} 
  \hfill \\
  
\end{array}\\
\begin{array}{ccc}
  \hspace{0.4\textwidth} & \hspace{0.2\textwidth} & \hspace{0.2\textwidth}\\
  
  \hfill \dfrac{P_m[i] = \ins{binop}\ op}
  {se, i \vdash k_1 :: k_2 :: st \Rightarrow (k_1 \sqcup k_2 \sqcup se(i)) :: st} \hfill
  & \hfill \dfrac{P_m[i] = \ins{push}\ n}{se, i \vdash st \Rightarrow se(i)::st}\hfill 
  & \hfill \dfrac{P_m[i] = \ins{pop}}{i \vdash k :: st \Rightarrow st} \hfill \\
  
\end{array}\\
\begin{array}{cc}
  \hspace{0.4\textwidth} & \hspace{0.5\textwidth}\\
  
  \hfill \dfrac{P_m[i] = \ins{new}\ C}
  {\Gamma, \fnn{ft}, \vec{k_a} \overset{k_h}{\rightarrow} \vec{k_r}, \fnn{region}, se, i 
    \vdash st \Rightarrow se(i) :: st} \hfill &
    
  \hfill \dfrac{P_m[i] = \ins{newarray}\ t\ \ \ \ k \in \mathcal{S}}
  {\Gamma, \fnn{ft}, \vec{k_a} \overset{k_h}{\rightarrow} \vec{k_r}, \fnn{region}, se, i 
    \vdash^{\norm} k :: st \Rightarrow k[\fn{at}{i}] :: st} \hfill \\
    
\end{array}\\
\begin{array}{c}\hspace{0.9\textwidth}\\		
		
  \dfrac{P_m[i] = \ins{getfield}\ f\ \ \ \ k \in \mathcal{S} \ \ \ \ 
    \forall j \in \fn{region}{i, \norm}, k \leq se(j)}
  {\Gamma, \fnn{ft}, \vec{k_a} \overset{k_h}{\rightarrow} \vec{k_r}, \fnn{region}, se, i 
    \vdash^{\norm} k :: st \Rightarrow 
    \fn{lift_k}{\big((k \sqcup se(i)) \sqcupext \fn{ft}{f}\big) :: st}} \\\\
	
	\dfrac{P_m[i] = \ins{getfield}\ f\ \ \ \ k \in \mathcal{S} \ \ \ \ 
	  \forall j \in \fn{region}{i, \np}, k \leq se(j)\ \ \ \ \fn{Handler}{i, \np} = t}
  {\Gamma, \fnn{ft}, \vec{k_a} \overset{k_h}{\rightarrow} \vec{k_r}, \fnn{region}, se, i 
    \vdash^{\np} k :: st \Rightarrow (k \sqcup se(i)) :: \epsilon} \\\\
    
  \dfrac{P_m[i] = \ins{getfield}\ f\ \ \ \ k \in \mathcal{S} \ \ \ \ 
    \forall j \in \fn{region}{i, \np}, k \leq se(j)\ \ \ \ 
    \fn{Handler}{i, \np} \uparrow \ \ \ \ k \leq \vec{k_r}[\np]}
  {\Gamma, \fnn{ft}, \vec{k_a} \overset{k_h}{\rightarrow} \vec{k_r}, \fnn{region}, se, i 
    \vdash^{\np} k :: st \Rightarrow } \\\\  
			
  \dfrac{\begin{gathered}P_m[i] = \ins{putfield}\ f\ \ \ \ 
    (se(i) \sqcup k_2) \sqcupext k_1  \leq \fn{ft}{f}\ \ \ \ 
    k_1 \in \sext\ \ \ \ k_2 \in \mathcal{S}\ \ \ \ k_h \leq \fn{ft}{f} \\
    \forall j \in \fn{region}{i, \norm}, k_2 \leq se(j)\end{gathered}}
  {\Gamma, \fnn{ft}, \vec{k_a} \overset{k_h}{\rightarrow} \vec{k_r}, \fnn{region}, se, i 
    \vdash^{\norm} k_1 :: k_2 :: st \Rightarrow \fn{lift_{k_2}}{st}} \\\\

\end{array}\\
\end{array}}\]
\end{figure*}

\begin{figure*}[!htbp]
\[\boxed{
\begin{array}{c}

\begin{array}{c}\hspace{0.96\textwidth}\\		
	
	\dfrac{\begin{gathered}P_m[i] = \ins{putfield}\ f\ \ \ \ 
    (se(i) \sqcup k_2) \sqcupext k_1  \leq \fn{ft}{f}\ \ \ \ 
    k_1 \in \sext\ \ \ \ k_2 \in \mathcal{S} \\
    \forall j \in \fn{region}{i, \np}, k_2 \leq se(j)\ \ \ \ \fn{Handler}{i, \np} = t
    \end{gathered}}
  {\Gamma, \fnn{ft}, \vec{k_a} \overset{k_h}{\rightarrow} \vec{k_r}, \fnn{region}, se, i 
    \vdash^{\np} k_1 :: k_2 :: st \Rightarrow (k_2 \sqcup se(i)) :: \epsilon} \\\\
	
	\dfrac{\begin{gathered}P_m[i] = \ins{putfield}\ f\ \ \ \ 
    (se(i) \sqcup k_2) \sqcupext k_1  \leq \fn{ft}{f}\ \ \ \ 
    k_1 \in \sext\ \ \ \ k_2 \in \mathcal{S} \\
    \forall j \in \fn{region}{i, \np}, k \leq se(j)\ \ \ \ 
    \fn{Handler}{i, \np} \uparrow \ \ \ \ k_2 \leq \vec{k_r}[\np]\end{gathered}}
  {\Gamma, \fnn{ft}, \vec{k_a} \overset{k_h}{\rightarrow} \vec{k_r}, \fnn{region}, se, i 
    \vdash^{\np} k_1 :: k_2 :: st \Rightarrow } \\\\
	
  \dfrac{P_m[i] = \ins{arraylength}\ \ \ \ k \in \mathcal{S}\ \ \ \ k_c \in \sext
    \ \ \ \ \forall j \in \fn{region}{i, \norm}, k \leq se(j)}
  {\Gamma, \fnn{ft}, \vec{k_a} \overset{k_h}{\rightarrow} \vec{k_r}, \fnn{region}, se, i 
    \vdash^{\norm} k[k_c] :: st \Rightarrow \fn{lift_k}{k :: st}} \\\\

  \dfrac{P_m[i] = \ins{arraylength}\ \ \ \ k \in \mathcal{S}\ \ \ \ k_c \in \sext
    \ \ \ \ \forall j \in \fn{region}{i, \np}, k \leq se(j) \ \ \ \ 
    \fn{Handler}{i, \np} = t}
  {\Gamma, \fnn{ft}, \vec{k_a} \overset{k_h}{\rightarrow} \vec{k_r}, \fnn{region}, se, i 
    \vdash^{\np} k[k_c] :: st \Rightarrow (k \sqcup se(i)) :: \epsilon} \\\\
  
  \dfrac{P_m[i] = \ins{arraylength}\ \ \ k \in \mathcal{S}\ \ \ k_c \in \sext 
    \ \ \ \forall j \in \fn{region}{i, \np}, k \leq se(j) \ \ \ 
    \fn{Handler}{i, \np} \uparrow\ \ \ k \leq \vec{k_r}[\np]}
  {\Gamma, \fnn{ft}, \vec{k_a} \overset{k_h}{\rightarrow} \vec{k_r}, \fnn{region}, se, i 
    \vdash^{\np} k[k_c] :: st \Rightarrow} \\\\
  \dfrac{P_m[i] = \ins{arrayload}\ \ \ \ k_1, k_2 \in \mathcal{S}\ \ \ \ k_c \in \sext
	  \ \ \ \forall j \in \fn{region}{i, \norm}, k_2 \leq se(j)}
  {\Gamma, \fnn{ft}, \vec{k_a} \overset{k_h}{\rightarrow} \vec{k_r}, \fnn{region}, se, i 
    \vdash^{\norm} k_1 :: k_2[k_c]  :: st \Rightarrow 
    \fn{lift_{k_2}}{\big( (k_1 \sqcup k_2) \sqcupext k_c \big) :: st}} \\\\
	
	\dfrac{P_m[i] = \ins{arrayload}\ \ \ \ k_1, k_2 \in \mathcal{S}\ \ \ \ k_c \in \sext
	  \ \ \ \forall j \in \fn{region}{i, \np}, k_2 \leq se(j)\ \ \ \ 
	  \fn{Handler}{i, \np} = t}
  {\Gamma, \fnn{ft}, \vec{k_a} \overset{k_h}{\rightarrow} \vec{k_r}, \fnn{region}, se, i 
    \vdash^{\np} k_1 :: k_2[k_c]  :: st \Rightarrow (k_2 \sqcup se(i)) :: \epsilon} \\\\
    
  \dfrac{P_m[i] = \ins{arrayload}\ \ \ \ k_1, k_2 \in \mathcal{S}\ \ \ \ k_c \in \sext
    \ \ \ \forall j \in \fn{region}{i, \np}, k_2 \leq se(j)\ \ \ \ 
	  \fn{Handler}{i, \np} \uparrow\ \ \ \ k_2 \leq \vec{k_r}[\np]}
  {\Gamma, \fnn{ft}, \vec{k_a} \overset{k_h}{\rightarrow} \vec{k_r}, \fnn{region}, se, i 
    \vdash^{\np} k_1 :: k_2[k_c]  :: st \Rightarrow } \\\\  
		
  \dfrac{\begin{gathered}P_m[i] = \ins{arraystore}\ \ \ \ 
    ((k_2 \sqcup k_3) \sqcupext k_1) \leqext k_c\ \ \ \ 
    k_2, k_3 \in \mathcal{S}\ \ \ \ k_1, k_c \in \sext \\
    \forall j \in \fn{region}{i, \norm}, k_2 \leq se(j) \end{gathered}}
  {\Gamma, \fnn{ft}, \vec{k_a} \overset{k_h}{\rightarrow} \vec{k_r}, \fnn{region}, se, i 
    \vdash^{\norm} k_1 :: k_2 :: k_3[k_c] :: st \Rightarrow \fn{lift_{k_2}}{st}}\\\\	
		
	\dfrac{\begin{gathered}P_m[i] = \ins{arraystore}\ \ \ \ 
    ((k_2 \sqcup k_3) \sqcupext k_1) \leqext k_c\ \ \ \ 
    k_2, k_3 \in \mathcal{S}\ \ \ \ k_1, k_c \in \sext \\
    \forall j \in \fn{region}{i, \np}, k_2 \leq se(j) \ \ \ \ \fn{Handler}{i, \np} = t
    \end{gathered}}
  {\Gamma, \fnn{ft}, \vec{k_a} \overset{k_h}{\rightarrow} \vec{k_r}, \fnn{region}, se, i 
    \vdash^{\np} k_1 :: k_2 :: k_3[k_c] :: st \Rightarrow 
    (k_2 \sqcup se(i)) :: \epsilon}\\\\		
	
	\dfrac{\begin{gathered}P_m[i] = \ins{arraystore}\ \ \ \ 
    ((k_2 \sqcup k_3) \sqcupext k_1) \leqext k_c\ \ \ \ 
    k_2, k_3 \in \mathcal{S}\ \ \ \ k_1, k_c \in \sext \\
    \forall j \in \fn{region}{i, \np}, k_2 \leq se(j) \ \ \ \ 
    \fn{Handler}{i, \np} \uparrow\ \ \ \ k_2 \leq \vec{k_r}[\np]
    \end{gathered}}
  {\Gamma, \fnn{ft}, \vec{k_a} \overset{k_h}{\rightarrow} \vec{k_r}, \fnn{region}, se, i 
    \vdash^{\np} k_1 :: k_2 :: k_3[k_c] :: st \Rightarrow }\\\\	
    
  \dfrac{\begin{gathered} P_m[i] = \ins{invoke}\ m_\mathrm{ID}\ \ \ \ 
    \fn{length}{st_1} = \fn{nbArguments}{m_\mathrm{ID}} \ \ \ \ 
    \Gamma_{m_\mathrm{ID}}[k] = \vec{k'_a} \overset{k'_h}{\rightarrow}\vec{k'_r} \\ 
    \forall i \in [0, \fn{length}{st_1} - 1]. st_1[i] \leq \vec{k'_a}[i + 1] \ \ \ \ 
    k \leq \vec{k'_a}[0] \ \ \ \ k \sqcup k_h \sqcup se(i) \leq k'_h\ \\ 
    k_e=\bigsqcup \big\{ \vec{k'_r}[e]\ |\ e \in \fn{excAnalysis}{m_\mathrm{ID}}\big\} 
    \ \ \ \ \forall j \in \fn{region}{i, \norm}, k \sqcup k_e \leq se(j)
  \end{gathered}}
  {\Gamma, \fnn{ft}, \vec{k_a} \overset{k_h}{\rightarrow} \vec{k_r}, \fnn{region}, se, i 
    \vdash^{\norm} st_1 :: k :: st_2 \Rightarrow 
    \text{lift}_{k \sqcup k_e} \big( (\vec{k'_r}[n] \sqcup se(i)) :: st_2)\big)} \\\\  
    
\end{array}
\end{array}}\]
\end{figure*}
\newpage \null \newpage
\begin{figure*}[!htbp]
\[\boxed{\begin{array}{c}\hspace{0.9\textwidth} \\					

  \dfrac{
    \begin{gathered} P_m[i] = \ins{invoke}\ m_\mathrm{ID}\ \ \ \ 
    \fn{length}{st_1} = \fn{nbArguments}{m_\mathrm{ID}} \ \ \ \ 
    \Gamma_{m_\mathrm{ID}}[k] = \vec{k'_a} \overset{k'_h}{\rightarrow}\vec{k'_r}\\ 
    \forall i \in [0, \fn{length}{st_1} - 1]. st_1[i] \leq \vec{k'_a}[i + 1] \ \ \ \ 
    k \leq \vec{k'_a}[0] \ \ \ \ k \sqcup k_h \sqcup se(i) \leq k'_h\\
    e \in \fn{excAnalysis}{m_\mathrm{ID}} \cup \{\np\} \ \ \ 
    \fn{Handler}{i, e} = t\ \ \ \ 
    \forall j \in \fn{region}{i, e}, k \sqcup k'_r[e] \leq se(j)
  \end{gathered}}
  {\Gamma, \fnn{ft}, \vec{k_a} \overset{k_h}{\rightarrow} \vec{k_r}, \fnn{region}, se, i 
    \vdash^{e} st_1::k::st_2\Rightarrow (k\sqcup\vec{k'_r}[e])::\epsilon} \\\\	

  \dfrac{
    \begin{gathered} P_m[i] = \ins{invoke}\ m_\mathrm{ID}\ \ \ \ 
    \fn{length}{st_1} = \fn{nbArguments}{m_\mathrm{ID}} \ \ \ \ 
    \Gamma_{m_\mathrm{ID}}[k] = \vec{k'_a} \overset{k'_h}{\rightarrow}\vec{k'_r}\\ 
    \forall i \in [0, \fn{length}{st_1} - 1]. st_1[i] \leq \vec{k'_a}[i + 1] \ \ \ \ 
    k \leq \vec{k'_a}[0] \ \ \ \ k \sqcup k_h \sqcup se(i) \leq k'_h  \ \ \ \ 
    k \sqcup se(i) \sqcup \vec{k'_r}[e] \leq \vec{k_r}[e]\\
    e \in \fn{excAnalysis}{m_\mathrm{ID}} \cup \{\np\}\ \ \ \ 
    \fn{Handler}{i, e} \uparrow\ \ \ \ 
    \forall j \in \fn{region}{i, e}, k \sqcup k'_r[e] \leq se(j)
  \end{gathered}}
  {\Gamma, \fnn{ft}, \vec{k_a} \overset{k_h}{\rightarrow} \vec{k_r}, \fnn{region}, se, i 
    \vdash^{e} st_1 :: k :: st_2 \Rightarrow } \\\\	
		
	\dfrac{
	  \begin{gathered} P_m[i] = \ins{throw}\ \ \ \ 
	  e \in \fn{classAnalysis}{i} \cup \{\np\}\ \ \ \  
		\forall j \in \fn{region}{i,e}, k \leq se(j)\ \ \ \ \fn{Handler}{i, e} = t
  \end{gathered}}
	{\Gamma, \fnn{ft}, \vec{k_a} \overset{k_h}{\rightarrow} \vec{k_r}, \fnn{region}, se, i 
	  \vdash^{e} k :: st \Rightarrow (k \sqcup se(i)) :: \epsilon} \\\\
		
	\dfrac{
	  \begin{gathered} P_m[i] = \ins{throw}\ \ \ \ 
	  e \in \fn{classAnalysis}{i} \cup \{\np\}  \ \ \ \ 
		k \leq \vec{k_r}[e]\ \ \ \ \forall j \in \fn{region}{i,e}, k \leq se(j)\ \ \ \ 
		\fn{Handler}{i, e} \uparrow
  \end{gathered}}
	{\Gamma, \fnn{ft}, \vec{k_a} \overset{k_h}{\rightarrow} \vec{k_r}, \fnn{region}, se, i 
	  \vdash^{e} k :: st \Rightarrow } \\\\	
\end{array} }
\]
\caption{JVM Transfer Rule}
\label{figure:javaTypeRuleFull}
\end{figure*}

\section{Full DEX Operational Semantics and Transfer Rules}
\label{section:appendix_dex}

\newcommand{\eqcond}[4]{\mathbf{eqcond}(#1, #2, #3, #4)}

The following figure~\ref{figure:dexOperationalSemanticFull} is the full operational 
semantics for DEX in section~\ref{dexTypeSystem}. It is similar to that of JVM, with
several differences, e.g. the state in DEX does not have operand stack but its 
functionality is covered by the registers (local variables) $\rho$.
The function $\fnn{fresh} : \text{Heap} \rightarrow \mathcal{L}$ is an 
allocator function that given a heap returns the location for that object. The function 
$\fnn{default} : \mathcal{C} \rightarrow \mathcal{O}$ returns for each class a default 
object of that class. For every field of that default object, the value will be $0$ if 
the field is numeric type, and $null$ if the field is of object type. Similarly 
$\fnn{defaultArray} : \mathbb{N} \times \mathcal{T}_D \rightarrow (\mathbb{N} 
\rightharpoonup \mathcal{V})$. The $\rightsquigarrow$ relation which defines transition 
between state is $\rightsquigarrow\subseteq \object{State} \times (\object{State} + 
\mathcal{V \times \object{Heap}})$.

The operator $\oplus$ denotes the function 
where $\rho \oplus \{r \mapsto v\}$ means a new function $\rho'$ such that 
$\forall i \in \fn{dom}{\rho} \backslash \{r\}. 
\rho'(i) = \rho(i)$ and $\rho'(r) = v$. The operator $\oplus$ is overloaded to also 
mean the update of a field on an object, or update on a heap. 

To handle exception, program will also comes equipped with two parameters 
$\fnn{classAnalysis}$ and $\fnn{excAnalysis}$. $\fnn{classAnalysis}$ contains 
information on possible classes of exception of a program point, and 
$\fnn{excAnalysis}$ contains possible escaping exception of a method.

There is also additional partial function for method $m$ $\fnn{Handler_m} : 
\mathcal{PP} \times \mathcal{C} \rightharpoonup \mathcal{PP}$ which gives the handler 
address for a given program point and exception. Given a program point $i$ and an 
exception thrown $c$, if $\fn{Handler_m}{i, c} = t$ then the control will be 
transferred to program point $t$, if the handler is undefined (noted 
$\fn{Handler_m}{i, c} \uparrow$) then the exception is uncaught in method $m$. 


The next figure~\ref{figure:dexTypeRuleFull} is the full version of 
figure~\ref{figure:dexTypeRuleSelected} in section~\ref{dexTypeSystem}.
The full typing judgement takes the form of $\Gamma, \fnn{ft}, \fnn{region}, 
sgn, se, i \vdash^{\tau} rt \Rightarrow rt'$
where $\Gamma$ is the table of method policies, $\fnn{ft}$ is the global policy for
fields, $\fnn{region}$ is the CDR information for the current method, 
$sgn$ is the policy for the current method taking the form of 
$\vec{k_a}\overset{k_h}{\rightarrow}\vec{k_r}$, $se$ is the security environment,
$i$ is the current program point, $rt$ is the register typing for the current 
instruction, $rt'$ is the register typing after the instruction is executed. 


As in the main paper, we may not write the full notation whenever it is clear from the 
context. In the table of operational semantics, we may drop the subscript
$m, \norm$ from $\rightsquigarrow$, e.g. we may write $\rightsquigarrow$ instead of
$\rightsquigarrow_{m, \norm}$ to mean the same thing. In the table of transfer rules,
we may drop the superscript of tag from $\vdash^{\tau}$ and write $\vdash$ instead.
The same case applies to the typing judgement, we may write $i \vdash^{\tau} rt 
\Rightarrow rt'$ 
instead of $\Gamma, \fnn{ft}, \fnn{region}, 
\vec{k_a}\overset{k_h}{\rightarrow}\vec{k_r}, se, i \vdash^{\tau} rt \Rightarrow rt'$.


\begin{figure*}[!htbp]
\[\setlength{\arraycolsep}{0pt}
\boxed{\begin{array}{c}
\begin{array}{ccc} \hspace{0.4\textwidth} & \hspace{0.4\textwidth} & 
    \hspace{0.2\textwidth}\\
  \hfill \dfrac{P_m[i] = \dexins{const}{r, v}\ \ r \in \fn{dom}{\rho}}
	{\langle i, \rho, h \rangle \rightsquigarrow_{m, \norm} \langle i+1, \rho \oplus 
	  \{r \mapsto v\}, h \rangle} \hfill &
	  
	\hfill \dfrac{P_m[i] = \dexins{move}{r, r_s}\ \ r \in \fn{dom}{\rho}}
	{\langle i, \rho, h \rangle \rightsquigarrow_{m, \norm} \langle i+1, \rho \oplus 
	  \{r \mapsto \rho(r_s)\}, h \rangle} \hfill &
	  
	\hfill \dfrac{P_m[i] = \dexins{goto}{t}}
	{\langle i, \rho, h \rangle\ \rightsquigarrow\ \langle t, \rho, h \rangle} \hfill \\
\end{array}\\
\begin{array}{ccc} \hspace{0.3\textwidth} & \hspace{0.3\textwidth} & 
    \hspace{0.4\textwidth} \\
	\hfill \dfrac{P[i]_m = \dexins{ifeq}{r, j}\ \ \ \rho(r) = 0}
	{\langle i, \rho, h \rangle \rightsquigarrow_{m, \norm} 
	  \langle t, \rho, h \rangle} \hfill &
	
	\hfill \dfrac{P_m[i] = \dexins{ifeq}{r, t}\ \ \ \rho(r) \neq 0}
	{\langle i,\rho,h \rangle \rightsquigarrow_{m, \norm} 
	  \langle i+1, \rho, h \rangle} \hfill &
	
	\hfill \dfrac{P[i]_m = \dexins{return}{r_s}\ \ r_s\in \fn{dom}{\rho}}
	{\langle i, \rho, h \rangle \rightsquigarrow_{m, \norm} \rho(r_s), h} \hfill \\  
	
\end{array}\\
\begin{array}{c} \\
	
  \dfrac{P_m[i] = \dexins{binop}{op, r, r_a, r_b}\ \ \ \ 
    r, r_a, r_b \in \fn{dom}{\rho} \ \ \ \ n = \rho(r_a)\ \underline{op}\ \rho(r_b)}
	{\langle i, \rho, h \rangle \rightsquigarrow_{m, \norm} \langle i+1, \rho \oplus 
	  \{r \mapsto n\}, h \rangle} \\

\end{array}\\
\begin{array}{cc}\\
	
	\dfrac{P_m[i] = \dexins{iget}{r, r_o, f}\ \ \ \ 
	  \rho(r_o) \in \fn{dom}{h} \ \ \ \ f \in \fn{dom}{h(\rho(r_o))}}
	{\cstate{i, \rho, h} \rightsquigarrow_{m, \norm} \cstate{i + 1, \rho \oplus 
	  \{r \mapsto h(\rho(r_o)).f\}, h}} \hspace{0.5cm} &

  \dfrac{P_m[i] = \dexins{new}{r, c}\ \ \ \ l = \fn{fresh}{h}}
	{\cstate{i, \rho, h} \rightsquigarrow \cstate{i + 1, \rho \oplus \{r \mapsto l\}, 
	  h \oplus \{l \mapsto \fn{default}{c}\}}} \\
  
\end{array}\\
\begin{array}{cc} \hspace{0.5\textwidth} & \hspace{0.5\textwidth}\\

  \dfrac{P_m[i] = \dexins{iget}{r, r_o, f}\ \ \ \ \rho(r_o) = null\ \ \ \ 
    l' = \fn{fresh}{h}}
	{\cstate{i, \rho, h} \rightsquigarrow_{m, \np} 
	  \fn{RuntimeExcHandling}{h, l', \np, i, \rho}} &
	  
  \dfrac{P_m[i] = \dexins{iput}{r_s, r_o, f}\ \ \ \ 
	  \rho(r_o) = null \ \ \ \ l' = \fn{fresh}{h}}
	{\cstate{i, \rho, h} \rightsquigarrow_{n, \np} 
	  \fn{RuntimeExcHandling}{h, l', \np, i, \rho}} \\
	  
\end{array}\\
\begin{array}{c} \\
	\dfrac{P_m[i] = \dexins{iput}{r_s, r_o, f}\ \ \ \ 
	  \rho(r_o) \in \fn{dom}{h} \ \ \ \ f \in \fn{dom}{h(\rho(r_o))}}
	{\cstate{i, \rho, h} \rightsquigarrow_{n, \norm} 
	  \cstate{i + 1, \rho, os, h \oplus \{\rho(r_o) \mapsto h(\rho(r_o)) \oplus 
	  \{f \mapsto \rho(r_s)\}\}}} \\\\

  \dfrac{P_m[i] = \dexins{newarray}{r, r_l, t}\ \ \ \ l = \fn{fresh}{h}\ \ \ \ 
	  \rho(r_l) \geq 0}
	{\cstate{i, \rho, h} \rightsquigarrow \cstate{i + 1, \rho \oplus \{r \mapsto l\}, 
	  h \oplus \{l \mapsto (\rho(r_l), \fn{defaultArray}{\rho(r_l), t}, i)\}}} \\\\
	
	\dfrac{P_m[i] = \dexins{arraylength}{r, r_a}\ \ \ \ \rho(r_a) \in \fn{dom}{h}}
	{\cstate{i, \rho, h} \rightsquigarrow_{m, \norm}
	  \cstate{i + 1, \rho \oplus \{r \mapsto h(\rho(r_a)).\field{length}, h\}}}\\\\	
	  
	\dfrac{P_m[i] = \dexins{arraylength}{r, r_a}\ \ \ \ \rho(r_a) = null
	  \ \ \ \ l' = \fn{fresh}{h}}
	{\cstate{i, \rho, h} \rightsquigarrow_{m, \np} 
	  \fn{RuntimeExcHandling}{h, l', \np, i, \rho}}\\\\  
		
	\dfrac{P_m[i] = \dexins{aget}{r, r_a, r_i}\ \ \ \ \rho(r_a) \in \fn{dom}{h}\ \ \ \ 
		0 \leq \rho(r_i) < h(\rho(r_a)).\field{length}}
  {\cstate{i, \rho, h} \rightsquigarrow_{m, \norm} \cstate{i+1, \rho \oplus 
    \{r \mapsto h(\rho(r_a))[\rho(r_i)]\}, h}} \\\\
    
	\dfrac{P_m[i] = \dexins{aget}{r, r_a, r_i}\ \ \ \ \rho(r_a) = null\ \ \ \ 
		l' = \fn{fresh}{h}}
  {\cstate{i, \rho, h} \rightsquigarrow_{m, \np} 
    \fn{RuntimeExcHandling}{h, l', \np, i, \rho}} \\\\
		
	\dfrac{P_m[i] = \dexins{aput}{r_s, r_a, r_i}\ \ \ \ \rho(r_a) \in \fn{dom}{h}\ \ \ \
		0 \leq \rho(r_i) < h(\rho(r_a)).\field{length}}
  {\cstate{i, \rho, h} \rightsquigarrow_{m, \norm} 
    \cstate{i+1, \rho, h \oplus \{\rho(r_a) \mapsto 
    h(\rho(r_a)) \oplus \{\rho(r_i) \mapsto \rho(r_s)\}\}} } \\

\end{array}\\
\begin{array}{cc} \hspace{0.55\textwidth} & \hspace{0.45\textwidth} \\

  \dfrac{P_m[i] = \dexins{aput}{r_s, r_a, r_i}\ \ \ \ \rho(r_a) = null\ \ \ \
		l' = \fn{fresh}{h}}
  {\cstate{i, \rho, h} \rightsquigarrow_{m, \np} 
    \fn{RuntimeExcHandling}{h, l', \np, i, \rho}} &		
		
	\dfrac{P_m[i] = \dexins{moveresult}{r}\ \ \ \ r \in \fn{dom}{\rho}}
	{\cstate{i, \rho, h} \rightsquigarrow_{m, \norm} \cstate{i + 1, \rho \oplus
	  \{r \mapsto \rho(ret)\}, h}} \\\\

\end{array}	\\
\begin{array}{c}

  \dfrac{\begin{gathered}P_m[i] = \dexins{invoke}{n, m', \vec{p}}\ \ \ \ 
	  \vec{p} \in \fn{dom}{\rho}\ \ \ \ \langle 1, \{\vec{x} \mapsto \vec{p}\}, h \rangle 
	  \rightsquigarrow^{+}_{m'} v, h'
	\end{gathered}}
	{\langle i, \rho, h\rangle\ \rightsquigarrow_{m,\norm}\ \langle i+1, \rho \oplus 
	  \{ret \mapsto v\}, h'\rangle} \\\\	
	
	\dfrac{\begin{gathered}P_m[i] = \dexins{invoke}{n, m',\vec{p}}\ \ \ \ 
	  \vec{p} \in \fn{dom}{\rho}\ \ \ \ \langle 1, \{\vec{x} \mapsto \vec{p}\}, h \rangle 
	  \rightsquigarrow^{+}_{m'} \langle l' \rangle, h'\\
		e = \fn{class}{h'(l')}\ \ \ \ \fn{Handler_m}{i, e} = t\ \ \ \ 
		e \in \fn{excAnalysis}{m'}
	\end{gathered}}
	{\langle i, \rho, h\rangle\ \rightsquigarrow_{m, e}\ \langle t, \rho \oplus 
	  \{ex \mapsto l'\}, h' \rangle} \\\\		

  \dfrac{P_m[i] = \dexins{invoke}{n, m', \vec{p}}\ \ \ \ l' = \fn{fresh}{h}\ \ \ \ 
	  \rho(\vec{p}[0]) = null}
	{\cstate{i, \rho, h} \rightsquigarrow_{m, \np} 
	  \fn{RuntimeExcHandling}{h, l', \np, i, \rho}} \\\\

\end{array}\\
\end{array}}\]
\end{figure*}

\begin{figure*}
\[\boxed{
\begin{array}{c}
\begin{array}{c}

  \dfrac{\begin{gathered}P_m[i] = \dexins{invoke }{n, m', \vec{p}}\ \ \ \ 
    \vec{p} \in \fn{dom}{\rho}\ \ \ \ \langle 1, \{\vec{x} \mapsto \vec{p}\}, h \rangle 
    \rightsquigarrow^{+}_{m'} \langle l' \rangle, h'\\
		e = \fn{class}{h'(l')}\ \ \ \ \fn{Handler_m}{i, e} \uparrow\ \ \ \ 
		e \in \fn{excAnalysis}{m'}
	\end{gathered}}
	{\langle i, \rho, h\rangle\ \rightsquigarrow_{m, e} \langle l' \rangle, h'} \\\\
		
	\dfrac{\begin{gathered} P_m[i] = \dexins{throw}{r}\ \ \ \ \rho(r) \in \fn{dom}{h}
	  \ \ \ \ e = \fn{class}{h(\rho(r))}\ \ \ \ 
		\fn{Handler_m}{i, e} = t\ \ \ \ e \in \fn{classAnalysis}{m, i}\end{gathered}}			
	{\cstate{i, \rho, h} \rightsquigarrow_{m, e} \cstate{t, \rho \oplus 
	  \{ex \mapsto \rho(r)\}, h}} \\\\
		
	\dfrac{\begin{gathered} P_m[i] = \dexins{throw}{r}\ \ \ \ \rho(r) \in \fn{dom}{h}
	  \ \ \ \ e = \fn{class}{h(\rho(r))}\ \ \ \ 
    \fn{Handler_m}{i, e} \uparrow\ \ \ \ e \in \fn{classAnalysis}{m, i}
  \end{gathered}}			
	{\cstate{i, \rho, h} \rightsquigarrow_{m, e} \langle \rho(r) \rangle, h} \\\\	

\end{array}\\
\begin{array}{cc}

  \dfrac{P_m[i] = \dexins{throw}{r}\ \ \ \ l' = \fn{fresh}{h}\ \ \ \ \rho(r) = null}
	{\cstate{i, \rho, h} \rightsquigarrow_{m, \np} 
	  \fn{RuntimeExcHandling}{h, l', \np, i, \rho}} & 

	\dfrac{P_m[i] = \dexins{moveexception} r\ \ \ \ r \in \fn{dom}{\rho}}
	{\cstate{i, \rho, h} \rightsquigarrow_{m, \norm} \cstate{i+1, \rho \oplus 
	  \{r \mapsto \rho(ex)\}, h}}\\\\

\end{array}\\
\begin{array}{c}
			
	\multicolumn{1}{l}{\fnn{RuntimeExcHandling} : 
	\object{Heap} \times \mathcal{L} \times \mathcal{C} \times \mathcal{PP} \times 
	(\mathcal{R} \rightharpoonup \mathcal{V}) \rightarrow \object{State} + (\mathcal{L} 
	\times \object{Heap}) \text{ defined as} }\\\\
	
	\multicolumn{1}{l}{\fn{RuntimeExcHandling}{h, l', C, i, \rho} =
	\left\{ \begin{array}{ll} \cstate{t, \rho \oplus \{ex \mapsto l'\}, 
	h \oplus \{l' \mapsto \fn{default}{C}\}} & \text{if } \fn{Handler_m}{i, C} = t \\
	\langle l' \rangle, h \oplus \{l' \mapsto \fn{default}{C}\} & \text{if } 
	\fn{Handler_m}{i, C} \uparrow\end{array}\right. }\\\\

\end{array}

\end{array}} \] 
\caption{DEX Operational Semantic}
\label{figure:dexOperationalSemanticFull}
\end{figure*}



\begin{figure*}
\[\setlength{\arraycolsep}{0pt}
\boxed{
\begin{array}{c}
\begin{array}{ccc}
	\hspace{0.25\textwidth} & \hspace{0.35\textwidth} & \hspace{0.35\textwidth}\\

	\hfill \dfrac{P_m[i] = \dexins{const}{r, v}}
	{se, i \vdash rt \Rightarrow rt \oplus \{r \mapsto se(i)\}} \hfill &

  \dfrac{P_m[i] = \dexins{new}{r, c}}
	{se, i \vdash^{\norm} rt \Rightarrow rt \oplus \{r \mapsto se(i)\}} &

  \hfill \dfrac{P_m[i] = \dexins{move}{r, r_s}}
	{se, i \vdash rt \Rightarrow rt \oplus
	  \{r \mapsto \big(rt(r_s) \sqcup se(i)\big)\}} \hfill \\

%
	  
\end{array}\\
\begin{array}{cc}	\hspace{0.48\textwidth} & \hspace{0.47\textwidth} \\	

	
	\dfrac{P_m[i] = \dexins{return}{r_s}\ \ \ \ se(i) \sqcup rt(r_s) \leq \vec{k_r}[n]}
	{\vec{k_a} \overset{k_h}{\rightarrow} \vec{k_r}, se, i \vdash rt \Rightarrow} &  
  
  	\hfill \dfrac{P_m[i] = \dexins{binop}{op, r, r_a, r_b}}
	{se, i \vdash rt \Rightarrow rt \oplus 
	  \{r \mapsto	\big(rt(r_a) \sqcup rt(r_b) \sqcup se(i)\big)\} } \hfill \\
	  
%
  
\end{array}\\
%
%

\begin{array}{c}\hspace{0.9\textwidth}\\

  \dfrac{P_m[i] = \dexins{ifeq}{r, t}\ \ \ \ 
    \forall j' \in \fn{region}{i, \norm}, se(i) \sqcup rt(r) \leq se(j')}
	{\Gamma, \fnn{ft}, \vec{k_a} \overset{k_h}{\rightarrow} \vec{k_r}, 
	  \fnn{region}, se, i \vdash rt \Rightarrow rt} \\\\
	  
  \dfrac{P_m[i] = \dexins{ifneq}{r, t}\ \ \ \ 
    \forall j' \in \fn{region}{i, \norm}, se(i) \sqcup rt(r) \leq se(j')}
	{\Gamma, \fnn{ft}, \vec{k_a} \overset{k_h}{\rightarrow} \vec{k_r}, 
	  \fnn{region}, se, i \vdash rt \Rightarrow rt} \\




\end{array}\\


\begin{array}{c}\hspace{0.9\textwidth} \\	     
%


  \dfrac{\begin{gathered}P_m[i] = \dexins{iget}{r, r_o, f}\ \ \ \ 
    rt(r_o) \in \mathcal{S}\ \ \ \ 
	  \forall j \in \fn{region}{i, \norm}, rt(r_o) \leq se(j) \end{gathered}}
	{\Gamma, \fnn{ft}, \vec{k_a} \overset{k_h}{\rightarrow} \vec{k_r}, \fnn{region}, 
	  se, i \vdash^{\norm} rt \Rightarrow rt \oplus 
	  \{r \mapsto \big((rt(r_o) \sqcup se(i)) \sqcupext \fn{ft}{f}\big)\}} \\\\ 
\dfrac{\begin{gathered}P_m[i] = \dexins{iget}{r, r_o, f}\ \ \ \ \ 
	  rt(r_o) \in \mathcal{S}\ \ \ \ 
	  \forall j \in \fn{region}{i, \np}, rt(r_o) \leq se(j) \ \ \ \ 
	  \fn{Handler}{i, \np} = t \end{gathered}}
	{\Gamma, \fnn{ft}, \vec{k_a} \overset{k_h}{\rightarrow} \vec{k_r}, \fnn{region}, 
	  se, i \vdash^{\np} rt \Rightarrow \vec{k_a} \oplus 
	  \{ex \mapsto (rt(r_o) \sqcup se(i))\} }\\\\


	\dfrac{\begin{gathered}P_m[i] = \dexins{iget}{r, r_o, f}\ \ \ 
	  rt(r_o) \in \mathcal{S}\ \ \ 
	  \forall j \in \fn{region}{i, \np}, rt(r_o) \leq se(j) \ \ \ 
	  \fn{Handler}{i, \np} \uparrow\ \ \ se(i) \sqcup rt(r_o) \leq \vec{k_r}[\np] \end{gathered}}
	{\Gamma, \fnn{ft}, \vec{k_a} \overset{k_h}{\rightarrow} \vec{k_r}, \fnn{region}, 
	  se, i \vdash^{\np} rt \Rightarrow }\\\\


  \dfrac{\begin{gathered}
    P_m[i] = \dexins{iput}{r, r_o, f}\ \ \ \ rt(r) \in \sext\ \ \ \ 
	  rt(r_o) \in \mathcal{S}\ \ \ \ 
	  (rt(r_o) \sqcup se(i)) \sqcupext rt(r_0) \leq \fn{ft}{f}
	  \ \ \ \ k_h \leq \fn{ft}{f}\\
	  \forall j \in \fn{region}{i, \norm}, rt(r_o) \leq se(j)
	  \end{gathered}}
	{\Gamma, \fnn{ft}, \vec{k_a} \overset{k_h}{\rightarrow} \vec{k_r}, \fnn{region}, 
	  se, i \vdash^{\norm} rt \Rightarrow rt} \\

\end{array}\\

\begin{array}{c}\hspace{0.9\textwidth}\\

\end{array}\\

\end{array}}\]
\end{figure*}

\begin{figure*}
\[\boxed{\begin{array}{c}
\begin{array}{c}\hspace{0.9\textwidth}\\	
  \dfrac{\begin{gathered}
    P_m[i] = \dexins{iput}{r_s, r_o, f}\ \ \ \ rt(r_s) \in \sext\ \ \ \ 
	  rt(r_o) \in \mathcal{S}\ \ \ 
	  (rt(r_o) \sqcup se(i)) \sqcupext rt(r_s) \leq \fn{ft}{f}\\
	  \forall j \in \fn{region}{i, \np}, rt(r_o) \leq se(j) \ \ \ \ 
	  \fn{Handler}{i, \np} = t
	  \end{gathered}}
	{\Gamma, \fnn{ft}, \vec{k_a} \overset{k_h}{\rightarrow} \vec{k_r}, \fnn{region}, 
	  se, i \vdash^{\np} rt \Rightarrow 
	  \vec{k_a} \oplus \{ex \mapsto rt(r_o) \sqcup se(i) \}} \\\\

  \dfrac{\begin{gathered}
    P_m[i] = \dexins{iput}{r_s, r_o, f}\ \ \ \ rt(r_s) \in \sext\ \ \ \ 
	  rt(r_o) \in \mathcal{S}\ \ \ 
	  (rt(r_o) \sqcup se(i)) \sqcupext rt(r_s) \leq \fn{ft}{f}\\
	  \forall j \in \fn{region}{i, \np}, rt(r_o) \leq se(j) \ \ \ \ 
	  \fn{Handler}{i, \np} \uparrow\ \ \ \ se(i) \sqcup rt(r_o) \leq \vec{k_r}[\np]
	  \end{gathered}}
	{\Gamma, \fnn{ft}, \vec{k_a} \overset{k_h}{\rightarrow} \vec{k_r}, \fnn{region}, 
	  se, i \vdash^{\np} rt \Rightarrow} \\\\
  \dfrac{P_m[i] = \dexins{newarray}{r, r_l, t}\ \ \ \ rt(r_l) \in \mathcal{S}\ \ \ \ 
	  rt(r_l)[\fn{at}{i}] \leq \vec{k_a}(r)}
  {\Gamma, \fnn{ft}, \vec{k_a} \overset{k_h}{\rightarrow} \vec{k_r}, \fnn{region}, 
    se, i \vdash^{\norm} rt \Rightarrow 
    rt \oplus \{r \mapsto rt(r_l)[\fn{at}{i}]\}} \\\\	

  \dfrac{P_m[i] = \dexins{arraylength}{r, r_a}\ \ \ \ k[k_c] = rt(r_a)\ \ \ \ 
	  k \in \mathcal{S}\ \ \ k_c \in \sext\ \ \ \ 
	  \forall j \in \fn{region}{i, \norm}, k \leq se(j)}
	{\Gamma, \fnn{ft}, \vec{k_a} \overset{k_h}{\rightarrow} \vec{k_r}, \fnn{region}, 
	  se, i \vdash^{\norm} rt \Rightarrow rt \oplus \{r \mapsto k\}}\\\\	

  \dfrac{\begin{gathered}P_m[i] = \dexins{arraylength}{r, r_a}\ \ \ \ 
	  k[k_c] = rt(r_a)\ \ \ \ 
	  k \in \mathcal{S}\ \ \ k_c \in \sext\ \ \ \ k \leq \vec{k_a}(r)\\ 
	  \forall j \in \fn{region}{i, \np}, k \leq se(j)\ \ \ \ 
	  \fn{Handler}{i, \np} = t\end{gathered}}
	{\Gamma, \fnn{ft}, \vec{k_a} \overset{k_h}{\rightarrow} \vec{k_r}, \fnn{region}, 
	  se, i \vdash^{\np} rt \Rightarrow 
	  \vec{k_a} \oplus \{ex \mapsto (k \sqcup se(i))\}}\\\\	
	  
	\dfrac{\begin{gathered}P_m[i] = \dexins{arraylength}{r, r_a}\ \ \ \ 
	  k[k_c] = rt(r_a)\ \ \ \ 
	  k \in \mathcal{S}\ \ \ k_c \in \sext\ \ \ \ 
	  k \leq \vec{k_a}(r)\\ 
	  \forall j \in \fn{region}{i, \np}, k \leq se(j)\ \ \ \ 
	  \fn{Handler}{i, \np} \uparrow\ \ \ \ se(i) \sqcup k \leq \vec{k_a}[\np]\end{gathered}}
	{\Gamma, \fnn{ft}, \vec{k_a} \overset{k_h}{\rightarrow} \vec{k_r}, \fnn{region}, 
	  se, i \vdash^{\np} rt \Rightarrow }\\\\	

  \dfrac{P_m[i] = \dexins{aget}{r, r_a, r_i}\ \ \ \ k[k_c] = rt(r_a)\ \ \ \ 
    k, rt(r_i) \in \mathcal{S}\ \ \ k_c \in \sext\ \ \ \ 
    \forall j \in \fn{region}{i, \norm}, k \leq se(j)}
	{\Gamma, \fnn{ft}, \vec{k_a} \overset{k_h}{\rightarrow} \vec{k_r}, \fnn{region}, 
	  se, i \vdash^{\norm} rt \Rightarrow rt \oplus 
	  \{r \mapsto \big((se(i) \sqcup k \sqcup rt(r_i)\big) \sqcupext k_c)\}} \\\\	

  \dfrac{\begin{gathered}P_m[i] = \dexins{aget}{r, r_a, r_i}\ \ \ \ 
    k[k_c] = rt(r_a)\ \ \ \ k, rt(r_i) \in \mathcal{S}\ \ \ 
    k_c \in \sext \\
    \forall j \in \fn{region}{i, \np}, k \leq se(j)\ \ \ \ \fn{Handler}{i, \np} = t
    \end{gathered}}
	{\Gamma, \fnn{ft}, \vec{k_a} \overset{k_h}{\rightarrow} \vec{k_r}, \fnn{region}, 
	  se, i \vdash^{\np} rt \Rightarrow 
	  \vec{k_a} \oplus \{ex \mapsto (k \sqcup se(i))\}}\\

\end{array}\\



\begin{array}{c}\hspace{0.9\textwidth}\\

  \dfrac{\begin{gathered}P_m[i] = \dexins{aget}{r, r_a, r_i}\ \ \ \ 
    k[k_c] = rt(r_a)\ \ \ \ k, rt(r_i) \in \mathcal{S}\ \ \ \ 
    k_c \in \sext\\
    \forall j \in \fn{region}{i, \np}, k \leq se(j)\ \ \ \ 
    \fn{Handler}{i, \np} \uparrow \ \ \ \ se(i) \sqcup k \leq \vec{k_r}[\np] \end{gathered}}
	{\Gamma, \fnn{ft}, \vec{k_a} \overset{k_h}{\rightarrow} \vec{k_r}, \fnn{region}, 
	  se, i \vdash^{\np} rt \Rightarrow }\\\\	


  \dfrac{\begin{gathered}P_m[i] = \dexins{aput}{r_s, r_a, r_i}\ \ \ \ 
    k[k_c] = rt(r_a)\ \ \ \ k, rt(r_i) \in \mathcal{S}\ \ \ \ 
    k_c, rt(r_s) \in \sext\\
		((k \sqcup rt(r_i)) \sqcupext rt(r_s)) \leqext k_c\ \ \ \ 
		\forall j \in \fn{region}{i, \norm}, k \leq se(j) \end{gathered}}
	{\Gamma, \fnn{ft}, \vec{k_a} \overset{k_h}{\rightarrow} \vec{k_r}, \fnn{region}, 
	  se, i \vdash^{\norm} rt \Rightarrow rt}\\\\


  \dfrac{\begin{gathered}P_m[i] = \dexins{aput}{r_s, r_a, r_i}\ \ \ \ 
    k[k_c] = rt(r_a)\ \ \ \ 
    k, rt(r_i) \in \mathcal{S}\ \ \ \ k_c, rt(r_s) \in \sext\\ 
		((k \sqcup sec(r_i)) \sqcupext sec(r_s)) \leqext k_c\ \ \ \ 
		\forall j \in \fn{region}{i, \np}, k \leq se(j) \ \ \ \ \fn{Handler}{i, \np} = t
		\end{gathered}}
	{\Gamma, \fnn{ft}, \vec{k_a} \overset{k_h}{\rightarrow} \vec{k_r}, \fnn{region}, 
	  se, i \vdash^{\np} rt \Rightarrow
	  \vec{k_a} \oplus\{ex \mapsto (k \sqcup se(i)) \}}\\\\

  \dfrac{\begin{gathered}P_m[i] = \dexins{aput}{r_s, r_a, r_i}\ \ \ \ 
    k[k_c] = rt(r_a)\ \ \ \ k, rt(r_i) \in \mathcal{S}\ \ \ \ 
    k_c, rt(r_s) \in \sext\\ 
    ((k \sqcup rt(r_i)) \sqcupext rt(r_s)) \leqext k_c\ \ \ \ 
		\forall j \in \fn{region}{i, \np}, k \leq se(j) \ \ \ \ 
		\fn{Handler}{i, \np} \uparrow\ \ \ \ se(i) \sqcup k \leq \vec{k_r}[\np] \end{gathered}}
	{\Gamma, \fnn{ft}, \vec{k_a} \overset{k_h}{\rightarrow} \vec{k_r}, \fnn{region}, 
	  se, i \vdash^{\np} rt \Rightarrow }\\\\

\end{array}
\end{array} }\]
\end{figure*}

\begin{figure*}
\[\boxed{\begin{array}{c}
\begin{array}{c}
  \dfrac{P_m[i] = \dexins{moveresult}{r}}
	{\Gamma, \fnn{ft}, \vec{k_a} \overset{k_h}{\rightarrow} \vec{k_r}, \fnn{region}, 
	  se, i \vdash^{\norm} rt \Rightarrow	
	  rt \oplus \{r \mapsto se(i) \sqcup rt(ret)\} }\\\\


%

\dfrac{\begin{gathered} P_m[i] = \dexins{invoke}{n, m', \vec{p}}\ \ 
    \Gamma_{m'}[rt(\vec{p}[0])] = \vec{k'_a} \overset{k'_h}{\rightarrow}\vec{k'_r} 
    \ \ \ rt(\vec{p}[0]) \sqcup k_h \sqcup se(i) \leq k'_h\ \ \ \ 
    \forall 0 \leq i < n. rt(\vec{p}[i]) \leq \vec{k'_a}[i] \\
		k_e = \bigsqcup \big\{ \vec{k'_r}[e]\ |\ e \in \fn{excAnalysis}{m'}\big\} \ \ \ \ 
		\forall j \in \fn{region}{i, \norm}, rt(\vec{p}[0]) \sqcup k_e \leq se(j)
	\end{gathered}}
	{\Gamma, \fnn{ft}, \vec{k_a} \overset{k_h}{\rightarrow} \vec{k_r}, \fnn{region}, 
	  se, i \vdash^{\norm} rt \Rightarrow  
	  \big( rt \oplus \{ret \mapsto \vec{k'_r}[n] \sqcup se(i)\})\big)} \\

\end{array}\\



\begin{array}{c}\hspace{0.9\textwidth}\\

  \dfrac{\begin{gathered} P_m[i] = \fn{invoke}{n, m', \vec{p}}\ \ 
    \Gamma_{m'}[rt(\vec{p}[0])] = \vec{k'_a} \overset{k'_h}{\rightarrow}\vec{k'_r} 
    \ \ \ \ rt(\vec{p}[0]) \sqcup k_h \sqcup se(i) \leq k'_h\ \ \ \ 
    \forall 0 \leq i < n. rt(\vec{p}[i]) \leq \vec{k'_a}[i] \\ 
		\fn{Handler}{i, e} = t\ \ \ 
		e \in \fn{excAnalysis}{m'} \cup \{\np\}\ \ \ \ 
		\forall j \in \fn{region}{i, e}, rt(\vec{p}[0]) \sqcup k'_r[e] \leq se(j)
	\end{gathered}}
	{\Gamma, \fnn{ft}, \vec{k_a} \overset{k_h}{\rightarrow} \vec{k_r}, \fnn{region}, 
	  se, i \vdash^{e} rt \Rightarrow \vec{k_a} \oplus 
	  \{ex \mapsto (rt(\vec{p}[0]) \sqcup \vec{k'_r}[e])\} } \\\\

  \dfrac{\begin{gathered} P_m[i] = \dexins{invoke}{n, m', \vec{p}}\ \ 
    \Gamma_{m'}[rt(\vec{p}[0])] = \vec{k'_a} \overset{k'_h}{\rightarrow}\vec{k'_r} 
    \ \ \ rt(\vec{p}[0]) \sqcup k_h \sqcup se(i) \leq k'_h\ \ \ \ 
    \forall 0 \leq i < n. rt(\vec{p}[i]) \leq \vec{k'_a}[i]\\ 
		rt(\vec{p}[0]) \sqcup se(i) \sqcup \vec{k'_r}[e] \leq \vec{k_r}[e]\ \ \ \  
		e \in \fn{excAnalysis}{m'} \cup \{\np\}\\
		\fn{Handler}{i, e} \uparrow \ \ \ \ 
		\forall j \in \fn{region}{i, e}, rt(\vec{p}[0]) \sqcup k'_r[e] \leq se(j)
  \end{gathered}}
	{\Gamma, \fnn{ft}, \vec{k_a} \overset{k_h}{\rightarrow} \vec{k_r}, \fnn{region}, 
	  se, i \vdash^{e} rt \Rightarrow} \\\\

	\dfrac{\begin{gathered}P_m[i] = \dexins{throw}{r}\ \ \ \ 
	  e \in \fn{classAnalysis}{i} \cup \{\np\} \ \ \ \ 
	  \forall j \in \fn{region}{i,e}, rt(r) \leq se(j)\ \ \ \ 
	  \fn{Handler}{i, e} = t \end{gathered}}
  {\Gamma, \fnn{ft}, \vec{k_a} \overset{k_h}{\rightarrow} \vec{k_r}, \fnn{region}, se,
    i \vdash^{e} rt \Rightarrow rt \oplus \{ex \mapsto (rt(r) \sqcup se(i))\}}\\\\	
	  
	\dfrac{\begin{gathered}P_m[i] = \dexins{throw}{r}\ \ \ \ 
	  e \in \fn{classAnalysis}{i} \cup \{\np\}\ \ \ \ 
	  \forall j \in \fn{region}{i,e}, rt(r) \leq se(j)\\ 
	  \fn{Handler}{i, e} \uparrow\ \ \ \ se(i) \sqcup rt(r) \leq \vec{k_r}[e] \end{gathered}}
	{\Gamma, \fnn{ft}, \vec{k_a} \overset{k_h}{\rightarrow} \vec{k_r}, \fnn{region}, 
	  se, i \vdash^{e} rt \Rightarrow } \\\\

 	\dfrac{P_m[i] = \dexins{moveexception}{r}}
 	{\Gamma, \fnn{ft}, \vec{k_a} \overset{k_h}{\rightarrow} \vec{k_r}, \fnn{region}, 
 	  se, i \vdash^{\norm} rt \Rightarrow 
 	  rt \oplus \{r \mapsto (rt(ex) \sqcup se(i))\} } \\\\

%
\end{array}

\end{array}}\]
\caption{DEX Transfer Rule}
\label{figure:dexTypeRuleFull}
\end{figure*}
\newpage

\section{Auxilliary Lemmas}

These lemmas are useful to prove the soundness of DEX type system.

\begin{lemma}
\label{AuxHeap1}
  Let $k \in S$ a security level, for all heap $h \in \mathrm{Heap}$ and object~/~array 
  $o \in \mathcal{O}$ (or $o \in \mathcal{A}$), 
  $h \preceq_k h \oplus \{\fn{fresh}{h} \mapsto o\}$
\end{lemma}

\begin{lemma}
\label{AuxHeap2}
  For all heap $h, h_0 \in \mathrm{Heap}$, object $o \in \mathcal{O}$ and 
  $l = \fn{fresh}{h}$, $h \sim_{\beta} h_0$ implies 
  $h \oplus\{l \mapsto o\} \sim_{\beta} h_0$
\end{lemma}

\begin{lemma}
\label{AuxHeap3}
  For all heap $h, h_0 \in \mathrm{Heap}$ and $\fn{ft}{f} \nleq \kobs$, 
  $h \sim_{\beta} h_0$ implies 
  $h \oplus \{l \mapsto h(l) \oplus \{f \mapsto v\}\} \sim_{\beta} h_0$
\end{lemma}

\begin{lemma}
\label{AuxHeap4}
  For all heap $h, h_0 \in \mathrm{Heap}$, $r \in R$, $\rho \in R \rightharpoonup V$, 
  $rt \in \RTSpec$, $\rho(r) \in \fn{dom}{h}$, $\rho(r)$ is an array, 
  integer $0 \leq i < h(\rho(r)).\field{length}$, 
  $rt(\rho(r)) = k[k_c]$ and $k_c \nleqext \kobs$, $h \sim_{\beta} h_0$ implies 
  $h \oplus \{\rho(r) \mapsto h(\rho(r)) \oplus \{i \mapsto v\}\} \sim_{\beta} h_0$
\end{lemma}

\begin{lemma}
\label{AuxHeap5}
  For all heap $h, h', h_0 \in \mathrm{Heap}$, $k \nleq \kobs$, and 
  $h \preceq_{k} h'$, $h \sim_{\beta} h_0$ implies $h' \sim_{\beta} h_0$
\end{lemma}

\begin{lemma}
\label{AuxHeap6}
  If $h_1 \sim_{\beta} h_2$, if $l_1 = \fn{fresh}{h_1}$ and $l_2 = \fn{fresh}{h_2}$ 
  then the following properties hold
  \begin{itemize}
    \item $\forall C,$ 
    $h_1 \oplus \{l_1 \mapsto \fnn{default}_C\} \sim_{\beta} h_2$
    
    \item $\forall C,$ 
    $h_1 \sim_{\beta} h_2 \oplus \{l_2 \mapsto \fnn{default}_C\}$
    
    \item $\forall C,$ $h_1 \oplus \{l_1 \mapsto \fnn{default}_C\}$ 
      $\sim_{\beta} h_2 \oplus \{l_2 \mapsto \fnn{default}_C\}$
      
    \item $\forall l, t, i, $
    $h_1 \oplus \{l_1 \mapsto (l, \fn{defaultArray}{l, t}, i)\} \sim_{\beta} h_2$
    
    \item $\forall l, t, i, $ 
    $h_1 \sim_{\beta} h_2 \oplus \{l_2 \mapsto (l, \fn{defaultArray}{l, t}, i)\}$
    
    \item $\forall l, t, i, l', t', i'$ 
    $h_1 \oplus \{l_1 \mapsto (l, \fn{defaultArray}{l, t}, i)\}$ 
      $\sim_{\beta} h_2 \oplus \{l_2 \mapsto (l', \fn{defaultArray}{l', t'}, i')\}$
  \end{itemize}
\end{lemma}

\begin{lemma}
\label{AuxRegs1}
  $\rho_1 \sim_{rt_1, rt_2, \beta} \rho_2$ implies for any register $r \in \rho_1$ : 
  \begin{itemize}
    \item either $rt_1(r) = rt_2(r), rt_1(r) \leq \kobs$ and 
    $\rho_1(r) \sim_{\beta} \rho_2(r)$
    
    \item or $rt_1(r) \nleq \kobs$ and $rt_2(r) \nleq \kobs$
  \end{itemize}
\end{lemma}

\section{Proof that typable \dvm{I} implies non-interference}

In this appendix, we present the soundness of our type system for DEX program i.e. 
typable DEX program implies that the program is safe. We also base our proof 
construction on the work Barthe et. al., including the structuring of the submachine.
In the paper, we present the type system for the aggregate of the submachines. In the 
proof construction, we will have 4 submachines: standard instruction without modifying 
the heap (\dvm{I}), object and array instructions (\dvm{O}), method invocation 
(\dvm{C}), and exception mechanism (\dvm{G}).

There are actually more definitions on indistinguishability that would be required to 
establish that typability implies non-interference. Before we go to the definition of 
operand stack indistinguishability, there is a definition of high registers : let 
$\rho \in \RTVal$ be register mapping and $rt \in \RTSpec$ be a registers typing.

Several notes here in this submachine, since the execution is always expected to return
normally, the form of the policy for return value only takes the form of $k_r$ instead
of $\vec{k_r}$. There is also no need to involve the heap and $\beta$ mapping, 
therefore we will drop them from the proofs.

\begin{definition}[State indistinguishability]
	Two states $\cstate{i, \rho}$ and $\cstate{i', \rho'}$ are indistinguishable w.r.t. 
	$rt, rt' \in \RTSpec$, denoted $\cstate{i, \rho} \sim_{\vec{k_a}, rt, rt'} 
	\cstate{i', \rho'}$, iff $\rho \sim_{\vec{k_a}, rt, rt'} \rho'$
\end{definition}

\begin{lemma}[Locally Respects]
\label{lemma:LocallyRespectI}
  Let $(i, \rho_1), (i, \rho_2) \in \mathrm{State}_I$ be two \dvm{I} states at the same 
  program point $i$ and let two registers types $rt_1, rt_2 \in \RTSpec$ such that 
  $s_1 \sim_{\vec{k_a}, rt_1, rt_2} s_2$.
  \begin{itemize}
	  \item Let $s'_1, s'_2 \in \mathrm{State}_I$ and $rt'_1, rt'_2 \in \RTSpec$ such 
	  that $s_1 \rightsquigarrow s'_1$, $s_2 \rightsquigarrow s'_2$, 
	  $i \vdash rt_1 \Rightarrow rt'_1$, and $i \vdash rt'_2 \Rightarrow rt'_2$, then 
	  $s'_1 \sim_{\vec{k_a}, rt'_1, rt'_2} s'_2$.
		
	  \item Let $v_1, v_2 \in \mathcal{V}$ such that $s_1 \rightsquigarrow v_1$, 
	  $s_2 \rightsquigarrow v_2$, $i \vdash rt_1 \Rightarrow$, and 
	  $i \vdash rt'_2 \Rightarrow $, then $k_r \leq \kobs$ implies $v_1 \sim v_2$.
  \end{itemize}	
\end{lemma}
\begin{IEEEproof}
  By contradiction. Assume that all the precedent are true, but the conclusion is 
  false. That means, $s'_1$ is distinguishable from $s'_2$, which means that 
  $\rho_{s'_1} \nsim \rho_{s'_2}$, where $\rho_{s'}$ is part of $s'_1$ and 
  $\rho_{s'_2}$ are parts of $s'_2$. This can be the case only if the instruction at 
  $i$ is modifying some low values in $\rho_1$ and $\rho_2$ to have different 
  values. We will do this by case for possible instructions :
  \begin{itemize}
	  \item $\fn{move}{r, r_s}$.  
%
      This case is trivial, as the 
			distinguishability for $\rho_{s'_1}$ and $\rho_{s'_2}$ will depend only on the 
			source register. If the source register is low, then since we have that 
			$\rho_1 \sim \rho_2$, they have to have the same value 
			($\rho_1(r_s) = \rho_2(r_s)$), therefore the value put in $r$ will be the same as 
			well. If the source register is high, then the target register will have high 
			security level as well (the security of both values will be 
			$rt(r_s) \sqcup se(i)$, where $rt(r_s) \nleq \kobs$), thus preserving the 
			indistinguishability.
				
		\item $\fn{binop}{r, r_a, r_b}$.  Following the argument from $\ins{move}$,
%
      the 
			distinguishability for $\rho_{s'_1}$ and $\rho_{s'_2}$ will depend only on the 
			source registers. If source registers are low, then since we have that 
			$\rho_1 \sim \rho_2$, they have to have the same values 
			($\rho_1(r_a) = \rho_2(r_a)$ and $\rho_1(r_b) = \rho_2(r_b)$), therefore the 
			result of binary operation will be the same (no change in indistinguishability). 
			If any of the source register is high, then the target register will have high 
			security level as well (the security level of the resulting value will be 
			$rt(r_a) \sqcup rt(r_b) \sqcup se(i)$, where $rt(r_a) \nleq \kobs$ and/or
			$rt(r_b) \nleq \kobs$), thus preserving the indistinguishability.
		
		\item $\fn{const}{r, v}$. Nothing to prove here, the instruction will always give 
		the same value anyway, regardless whether the security level of the register to 
		store the value is high or low.
		
		\item $\fn{goto}{j}$. Nothing to prove here, as the instruction only modify the 
		program counter.
				
		\item $\fn{return}{r_s}$. This is a slightly different case here than before, where 
		we are comparing the results instead of the state ($v_1 \sim v_2$). Again, the 
		reasoning is that to have different result and they are distinguishable, we need 
		the register from which the value is returned to be high ($rt(r_s) \nleq \kobs$), 
		but the security level of the return value of the method is low ($k_r \leq \kobs$). 
		But this is already taken care of by the transfer rule which state 
		$rt(r_s) \leq k_r$. Therefore, a contradiction.
				
		\item $\fn{ifeq}{r, t}$. A special case where there might be a branching thus the 
		states compared are at two different program counters. If the register used in 
		comparison is low ($rt(r) \leq \kobs$), we know that the program counter will be 
		the same and there will be nothing left to prove ($\ins{ifeq}$ is just modifying 
		program counter). If the register is high ($rt(r) \nleq \kobs$), the operational
		semantics tells us that there is no modification to the registers.
    Therefore, register wise these two states are indistinguishable.
	\end{itemize}
\end{IEEEproof}

\begin{lemma}[High Branching]
\label{lemma:HighBranchingI}
  Let $s_1, s_2 \in \mathrm{State}_I$ be two \dvm{I} states at the same program point 
  $i$ and let two registers types $rt_1, rt_2 \in \RTSpec$ such that 
  $s_1 \sim_{\vec{k_a}, rt_1, rt_2} s_2$. If two states 
  $\langle i_1, \rho'_1 \rangle, \langle i_2, \rho'_2 \rangle \in \mathrm{State}_I$ and 
  two registers type $rt'_1, rt'_2 \in \RTSpec$ s.t. $i_1 \neq i_2$, 
  $s_1 \rightsquigarrow \langle i_1, \rho'_1 \rangle$, 
  $s_2 \rightsquigarrow \langle i_2, \rho'_2 \rangle$, 
  $i \vdash rt_1 \Rightarrow rt'_1$, $i \vdash rt_2 \Rightarrow rt'_2$ then 
  $\forall j \in \fn{region}{i}, se(j) \nleq \kobs$.
\end{lemma}
\begin{IEEEproof}
  This is already by definition of the branching instruction ($\ins{ifeq}$ and 
  $\ins{ifneq}$). 
  $se(i)$ will be high because $r$ will by definition be high. This level can not 
  be low, because if the level is low, then the register $r$ is low and by the 
  definition of indistinguishability will have to have the same values, and therefore 
  will take the same program counter. 
  Since $se$ is high for scope of the region, we have 
  $\forall j \in \fn{region}{i}, se(j) \nleq \kobs$.
\end{IEEEproof}



\begin{lemma}[indistinguishablility double monotony]
\label{lemma:IndistinguishabilityDoubleMonotony}
	if $s \sim_{\vec{k_a}, S, T} t, S \sqsubseteq U$, and $T \sqsubseteq U$ then 
	$s \sim_{\vec{k_a}, U, U} t$
\end{lemma}

\begin{lemma}[indistinguishablility single monotony]
\label{lemma:IndistinguishabilitySingleMonotony}
	if $s \sim_{\kobs, S, T} t, S \sqsubseteq S'$ and $S$ is high then 
	$s \sim_{\kobs, S', T} t$
\end{lemma}

\section{Proof that typable \dvm{O} implies non-interference}

Indistinguishability between states can be defined with the additional definition 
of heap indistinguishability, so we do not need additional indistinguishability 
definition. In the \dvm{O} part, we only need to appropriate the lemmas used to 
establish the proof.

\begin{definition}[State indistinguishability]
	Two states $\cstate{i, \rho, h}$ and $\cstate{i', \rho', h'}$ are indistinguishable 
	w.r.t. a partial function $\beta \in \mathcal{L} \rightharpoonup \mathcal{L}$, and 
	two registers typing $rt, rt' \in \RTSpec$, denoted 
	$\cstate{i, \rho, h} \sim_{\vec{k_a}, rt, rt', \beta} \cstate{i', \rho', h'}$, iff 
	$\rho \sim_{\vec{k_a}, rt, rt'} \rho'$ and $h \sim_{\beta} h'$ hold.
\end{definition}

\begin{lemma}[Locally Respects]
  Let $\beta$ a partial function $\beta \in \mathcal{L} \rightharpoonup \mathcal{L}$, 
  $s_1, s_2 \in \mathrm{State}_O$ be two \dvm{O} states at the same program point $i$ 
  and let two registers types $rt_1, rt_2 \in \RTSpec$ such that 
  $s_1 \sim_{\vec{k_a}, rt_1, rt_2, \beta} s_2$.
  \begin{itemize}
		\item Let $s'_1, s'_2 \in \mathrm{State}_O$ and $rt'_1, rt'_2 \in \RTSpec$ such 
		that $s_1 \rightsquigarrow s'_1, s_2 \rightsquigarrow s'_2$, 
		$i \vdash rt_1 \Rightarrow rt'_1$, and $i \vdash rt'_2 \Rightarrow rt'_2$, then 
		there exists $\beta' \in \mathcal{L} \rightharpoonup \mathcal{L}$ such that 
		$s'_2 \sim_{\vec{k_a}, rt'_1, rt'_2, \beta'} s'_2$ and $\beta \subseteq \beta'$.
		
		\item Let $v_1, v_2 \in \mathcal{V}$ such that $s_1 \rightsquigarrow v_1$, 
		$s_2 \rightsquigarrow v_2$, $i \vdash rt_1 \Rightarrow $, and 
		$i \vdash rt'_2 \Rightarrow $, then $k_r \leq \kobs$ implies $v_1 \sim_{\beta} v_2$.
\end{itemize}	
\end{lemma}
\begin{IEEEproof}
  By contradiction. Assume that all the precedent are true, but the conclusion is 
  false. That means, $s'_1$ is distinguishable from $s'_2$, which means either 
  $\rho'_{1} \nsim \rho'_{2}$ or $h'_{1} \nsim h'_{2}$, where $\rho'_{1}, h'_{1}$ are 
  parts of $s'_1$ and $\rho'_{2}, h'_{2}$ are parts of $s'_2$.
	\begin{itemize}
		\item assume $h'_{1} \nsim h'_{2}$. This can be the case only if the instruction at 
		$i$ are $\ins{iput}, \ins{newarray}$, and $\ins{aput}$. 
		\begin{itemize}
			\item $\dexins{iput}{r_s, r_o, f}$ can only cause the difference by putting 
			different values ($\rho_{1}(r_s) \neq \rho_{2}(r_s)$) with 
			$rt_1(r_s)$ $\nleq \kobs$ and $rt_2(r_s) \nleq \kobs$ on a field where 
			$\fn{ft}{f} \leq \kobs$. But the transfer rule for $\ins{iput}$ states that the 
			security level of the field has to be at least as high as $r_s$, i.e. 
			$sec \leq \fn{ft}{f}$ where $sec = rt_1(r_s) = rt_2(r_s)$. A plain 
			contradiction.
			
			\item $\dexins{aput}{r_s, r_a, r_i}$ can only cause the difference by putting 
			different values ($\rho_{1}(r_s) \neq \rho_{2}(r_s)$) with $rt_1(r_s)$ 
			$\nleq \kobs$ and $rt_2(r_s) \nleq \kobs$ on an array whose content is low 
			($k_c \leq \kobs, k[k_c]$ is the security level of the array). But the typing 
			rule for $\ins{aput}$ states that the security level of the array content has to 
			be at least as high as $r_s$, i.e. $sec \leq k_c$ where 
			$sec = rt_1(r_s) = rt_2(r_s)$. A plain contradiction.
			
			\item $\dexins{newarray}{r_a, r_l, t}$ can only cause the difference by creating 
			array of different lengths ($\rho_{1}(r_l) \neq \rho_{2}(r_l)$) with 
			$rt_1(r_s) \nleq \kobs$ and $rt_2(r_s) \nleq \kobs$. But if that's the case, 
			then that means this new array does not have to be included in the mapping 
			$\beta'$ and therefore the heap will stay indistinguishable. A contradiction.
		\end{itemize}
		
		\item assume $\rho'_{1} \nsim \rho'_{2}$. This can be the case only if the 
		instruction at $i$ is modifying some low values in $\rho'_{1}$ and $\rho'_{2}$ to 
		have different values. There are only three possible instructions in this extended 
		submachine which can cause $\rho'_{1} \nsim \rho'_{2}$: $\ins{iget}$, $\ins{aget}$, 
		and $\ins{arraylength}$. We already have by the assumption that the original state 
		is indistinguishable, which means that the heaps are indistinguishable as well 
		($h_1 \sim h_2$). 
%
      Based on the transfer rule we
	    have that the security level of the value put in the target register will at least
	    be as high as the source. If the security is low, we know from the assumption of
	    indistinguishability that the value is the same, thus it will maintain 
	    registers indistinguishability. If the security is high, then the value put in
	    the target register will also have high security level, maintaining registers
	    indistinguishability. 
		\end{itemize}
\end{IEEEproof}

\begin{lemma}[High Branching]
\label{lemma:HighBranchingO}
  Let $\beta$ a partial function $\beta \in \mathcal{L} \rightharpoonup \mathcal{L}$, 
  $s_1, s_2 \in \mathrm{State}_O$ be two \dvm{O} states at the same program point $i$ 
  and let two registers types $rt_1, rt_2 \in \RTSpec$ such that 
  $s_1 \sim_{\vec{k_a}, rt_1, rt_2, \beta} s_2$. Let two states 
  $\langle i_1, \rho'_1, h'_1 \rangle, \langle i_2, \rho'_2, h'_2 \rangle \in 
  \mathrm{State}_O$ and two registers type $rt'_1, rt'_2 \in \RTSpec$ s.t. 
  $i_1 \neq i_2, s_1 \rightsquigarrow \langle i_1, \rho'_1, h'_1 \rangle$, 
  $s_2 \rightsquigarrow \langle i_2, \rho'_2, h'_2 \rangle$. If 
  $i \vdash rt_1 \Rightarrow rt'_1, i \vdash rt_2 \Rightarrow rt'_2$ then 
  $\forall j \in \fn{region}{i}, se(j) \nleq \kobs$
\end{lemma}


\section{Proof that typable \dvm{C} implies security}

Since now the notion of secure program also defined with \emph{side-effect-safety} due 
to method invocation, we also need to establish that typable program implies that it is 
\emph{side-effect-safe}. We show this by showing the property that all instruction step 
transforms a heap $h$ into a heap $h'$ s.t. $h \preceq_{k_h} h'$.

\begin{lemma}
\label{lemma:SideEffectSafetyC_sequential}
	Let $\langle i, \rho, h \rangle, \langle i', \rho', h' \rangle \in \mathrm{State}_C$ 
	be two states s.t. $\state{} \rangle \rightsquigarrow_m \state{'}$. Let two registers 
	types $rt, rt' \in \RTSpec$ s.t. 
	$\fnn{region}, se, \vec{k_a} \overset{k_h}{\rightarrow} k_r, i \vdash^{\norm} rt 
	\Rightarrow rt'$ and $P[i] \neq \ins{invoke}$, then $h \preceq_{k_h} h'$
\end{lemma}
\begin{IEEEproof}
	The only instruction that can cause this difference is $\ins{newarray}$, $\ins{new}$, 
	$\fnn{iput}$, and $\fnn{aput}$. For creating new objects or arrays, 
	Lemma~\ref{AuxHeap1} shows that they still preserve the side-effect-safety. For 
	$\ins{iput}$, the transfer rule implies $k_h \leq \fn{ft}{f}$. Since there will be no 
	update such that $k_h \nleq \fn{ft}{f}$, $h \preceq_{k_h} h'$ holds.
\end{IEEEproof}

\begin{lemma}
\label{lemma:SideEffectSafetyC_return}
	Let $\langle i, \rho, h \rangle \in \mathrm{State}_C$ be a state, 
	$h' \in \object{Heap}$, and $v \in V$ s.t. 
	$\langle i, \rho, h \rangle \rightsquigarrow v, h'$. Let $rt \in \RTSpec$ 
	s.t. $\fnn{region}, se, \vec{k_a} \overset{k_h}{\rightarrow} k_r, i \vdash^{\norm}
	rt \Rightarrow$, then $h \preceq_{k_h} h'$
\end{lemma}
\begin{IEEEproof}
	This only concerns with $\ins{return}$ instruction at the moment. And it's clear 
	that return instruction will not modify the heap therefore $h \preceq_{k_h} h'$ holds
\end{IEEEproof}
		
\begin{lemma}
\label{lemma:SideEffectSafetyC_invoke}
	For all method $m$ in $P$, let ($\fnn{region_m}, \fnn{jun_m}$) be a safe CDR for $m$. 
	Suppose all methods $m$ in $P$ are typable with respect to $\fnn{region_m}$ and to 
	all signatures in $\fn{Policies_{\Gamma}}{m}$.
	Let $\state{}, \state{'} \in \mathrm{State}_C$ be two states s.t. 
	$\state{} \rightsquigarrow_m \state{'}$. Let two registers types 
	$rt, rt' \in \RTSpec$ s.t. 
	$\fnn{region}$, $se$, $\vec{k_a} \overset{k_h}{\rightarrow} k_r$, $i \vdash^{\norm} 
	rt \Rightarrow rt'$ and $P[i] = \ins{invoke}$, then $h \preceq_{k_h} h'$
\end{lemma}
\begin{IEEEproof}
  Assume that the method called by $\ins{invoke}$ is $m_0$. The instructions contained 
  in $m'$ can be any of the instructions in DEX, including another $\ins{invoke}$ to 
  another method. Since we are not dealing with termination~/~non-termination, we can 
  assume that for any instruction $\ins{invoke}$ called, it will either return normally 
  or throws an exception. Therefore, for any method $m_0$ called by $\ins{invoke}$, 
  there can be one or more chain of call 
  \[ m_0 \rightsquigarrow m_1 \rightsquigarrow ... \rightsquigarrow m_n\] 
  where $m \rightsquigarrow m'$ signify that an instruction in method $m$ calls $m'$.
  Since the existence of such call chain is assumed, we can use induction on the length
  of the longest call chain. The base case would be the length of the chain is $0$, 
  which means we can just invoke Lemma~\ref{lemma:SideEffectSafetyC_sequential} and 
  Lemma~\ref{lemma:SideEffectSafetyC_return} because all the instructions contained in 
  this method $m_0$ will fall to either one of the two above case.
	
	The induction step is when we have a chain with length $1$ or more and we want
	to establish that assuming the property holds when the length of call chain is $n$, 
	then the property also holds when the length of call chain is $n+1$. In this case, we 
	just examine possible instructions in $m_0$, and proceed like the base case except 
	that there is also a possibility that the instruction is $\ins{invoke}$ on $m_1$. 
	Since the call chain is necessarily shorter now $m_0 \rightsquigarrow m_1$ is dropped 
	from the call chain, we know that $\ins{invoke}$ on $m_1$ will fulfill 
	\emph{side-effect-safety}. Since all possible instructions are maintaining 
	\emph{side-effect-safety}, we know that this lemma holds.
\end{IEEEproof}

Since all typable instructions implies \emph{side-effect-safety}, then we can state
the lemma saying that typable program will be \emph{side-effect-safe}.

\begin{lemma} 
	For all method $m$ in $P$, let ($\fnn{region_m}, \fnn{jun_m}$) be a safe CDR for $m$. 
	Suppose all methods $m$ in $P$ are typable with respect to $\fnn{region_m}$ and to 
	all signatures in $\fn{Policies_{\Gamma}}{m}$. Then all method $m$ is 
	\emph{side-effect-safe} w.r.t. the heap effect level of all the policies in 
	$\fn{Policies_{\Gamma}}{m}$.
\end{lemma}

	Then, like the previous machine, we need to appropriate the unwinding lemmas. The 
	unwinding lemmas for \dvm{O} stay the same, and the one for instruction 
	$\ins{moveresult}$ is straightforward. Fortunately, $\fnn{Invoke}$ is not a branching 
	source, so we don't need to appropriate the high branching lemma for this instruction 
	(it will be for exception throwing one in the subsequent machine).

\begin{lemma}[Locally Respect Lemma]
	Let $P$ a program and a table of signature $\Gamma$ s.t. all of its method $m'$ are 
	\emph{non-interferent} w.r.t. all the policies in $\fn{Policies_{\Gamma}}{m'}$ and 
	\emph{side-effect-safe} w.r.t. the heap effect level of all the policies in 
	$\fn{Policies_{\Gamma}}{m'}$. Let $m$ be a method in $P$, 
	$\beta \in L \rightharpoonup L$ a partial function, $s_1, s_2 \in \mathrm{State}_C$ 
	two \dvm{C} states at the same program point $i$ and two registers types 
	$rt_1, rt_2 \in \RTSpec$ s.t. $s_1 \sim_{\kobs, rt_1, rt_2, \beta} s_2$. If there 
	exist two states $s'_1. s'_2 \in \mathrm{State}_C$ and two registers types 
	$rt'_1, rt'_2 \in \RTSpec$ s.t. 
	\[\begin{array}{ccc}
	  s_1 \rightsquigarrow_m s'_1 & \text{and} &
	  \Gamma, \mathrm{region}, se, \vec{k_a} \overset{k_h}{\rightarrow} k_r, i \vdash 
	  rt_1 \Rightarrow rt'_1 
	\end{array}\]
  and
  \[\begin{array}{ccc}
	  s_2 \rightsquigarrow_{m'} s'_2 & \text{and} &
	  \Gamma, \mathrm{region}, se, \vec{k_a} \overset{k_h}{\rightarrow} k_r, i \vdash 
	  rt_2 \Rightarrow rt'_2
	\end{array}\]
  then there exists $\beta' \in L \rightharpoonup L$ s.t. 
	$s'_1 \sim_{\kobs, rt'_1, rt'_2, \beta'} s'_2$ and $\beta \subseteq \beta'$.
\end{lemma}
\begin{IEEEproof}
	By contradiction. Assume that all the precedent are true, but the conclusion is 
	false. That means, $s'_1$ is distinguishable from $s'_2$, which means either 
	$\rho'_{1} \nsim \rho'_{2}$ or $h'_{1} \nsim h'_{2}$, where $\rho'_{1}, h'_{1}$ are 
	parts of $s'_1$ and $\rho'_{2}, h'_{2}$ are parts of $s'_2$.
  \begin{itemize}
    \item
    Assume $h'_{1} \nsim h'_{2}$. $\fn{invoke}{m, n, \vec{p}}$ can only cause the state 
    to be distinguishable if the arguments passed to the function have some difference. 
    And since the registers are indistinguishable in the initial state, this means that 
    those registers with different values are registers with security level higher than 
    $\kobs$ (let's say this register $x$). By the transfer rules of $\fnn{invoke}$, 
    this will imply that the $\vec{k'_a}[x] \nleq \kobs$ (where $0 \leq x \leq n$). Now 
    assume that there is an instruction in $m$ using the value in $x$ to modify the 
    heap, which can not be the case because in \dvm{O} we already proved that the 
    transfer rules prohibit any object~/~array manipulation instruction to update the 
    low field~/~content with high value.

    \item
    Assume $\rho_{s'} \nsim \rho_{t'}$. $\fnn{invoke}$ only modifies the pseudo-
    register $ret$ with the values that will be dependent on the security of the return 
    value. Because we know that the method invoked is non-interferent and the arguments 
    are indistinguishable, therefore we can conclude that the result will be 
    indistinguishable as well (which will make $ret$ also indistinguishable). As for 
    $\ins{moveresult}$, we can follow the arguments in $\fnn{move}$ (in \dvm{I}), 
    except that the source is now the pseudo-register $ret$.
  \end{itemize}
\end{IEEEproof}

\section{Proof that typable \dvm{G} implies security}

Like the one in \dvm{C}, we also need to firstly prove the \emph{side-effect-safety} of 
a program if it's typable. Fortunately, this proof extends almost directly from the one 
in \dvm{C}. The only difference is that there is a possibility for invoking a function 
which throws an exception and the addition of $\ins{throw}$ instruction. The proof for 
invoking a function which throws an exception is the same as the usual $\ins{invoke}$, 
because we do not concern whether the returned value $r$ is in $L$ or in $V$. The one 
case for $\ins{throws}$ use the same lemma \ref{AuxHeap1} as it differs only in the 
allocation of exception in the heap. The complete definition :

\begin{lemma}
\label{lemma:SideEffectSafetyG_sequential}
	Let $\langle i, \rho, h \rangle, \langle i', \rho', h' \rangle \in \mathrm{State}_G$ 
	be two states s.t. $\state{} \rightsquigarrow_m \state{'}$. Let two registers types 
	$rt, rt' \in \RTSpec$ s.t. 
	$\fnn{region}, se, \vec{k_a} \overset{k_h}{\rightarrow} k_r, i \vdash 
	rt \Rightarrow rt'$ and $P[i] \neq \ins{invoke}$, then $h \preceq_{k_h} h'$
\end{lemma}
\begin{IEEEproof}
	The only instruction that can cause this difference are
	array~/~object manipulation instructions that throws a null pointer exception.
	For creating new objects or arrays and allocating the space for exception, 
	Lemma~\ref{AuxHeap1} shows that they still preserve the \emph{side-effect-safety}.
	$\ins{throw}$ instruction itself does not allocate space for exception, so no 
	modification to the heap.
\end{IEEEproof}

\begin{lemma}
\label{lemma:SideEffectSafetyG_return}
	Let $\state{} \in \mathrm{State}_G$ be a state, $h' \in \object{Heap}$, and $v \in V$ 
	s.t. $\state{} \rightsquigarrow v, h'$. Let $rt \in \RTSpec$ s.t. 
	$\fnn{region}, se,\vec{k_a} \overset{k_h}{\rightarrow} k_r, i \vdash rt \Rightarrow$, 
	then $h \preceq_{k_h} h'$
\end{lemma}
\begin{IEEEproof}
	This can only be one of two cases, either it is $\ins{return}$ instruction or 
	uncaught exception. For return instruction, it's clear that it will not modify the 
	heap therefore $h \preceq_{k_h} h'$ holds. For uncaught exception, the only 
	difference is that we first need to allocate the space on the heap for the exception, 
	and again we use lemma \ref{AuxHeap1} to conclude that it will still make 
	$h \preceq_{k_h} h'$ holds
\end{IEEEproof}

\begin{lemma}
\label{lemma:SideEffectSafetyG_invoke}
  Let for all method $m \in P$, ($\fnn{region}_m$, $\fnn{jun}_m$) a safe CDR for $m$. 
	Suppose all methods $m \in P$ are typable w.r.t. $\fnn{region}_m$ and to all 
	signatures in $\fnn{Policies_{\Gamma}}(m)$.
	Let $\state{}, \state{'} \in \mathrm{State}_G$ be two states s.t. 
	$\state{} \rightsquigarrow_m \state{'}$. Let two registers types 
	$rt, rt' \in \RTSpec$ s.t. 
	$\fnn{region}$, $se$, $\vec{k_a} \overset{k_h}{\rightarrow} k_r$, $i \vdash 
	rt \Rightarrow rt'$ and $P[i] = \ins{invoke}$, then $h \preceq_{k_h} h'$
\end{lemma}
\begin{IEEEproof} 
	In the case of $\ins{invoke}$ executing normally, we can refer to the proof in 
	Lemma~\ref{lemma:SideEffectSafetyC_invoke}. In the case of caught exception, if it is 
	caught then we can follow the same reasoning in 
	Lemma~\ref{lemma:SideEffectSafetyG_sequential}). In the case of uncaught exception it 
	will fall to Lemma~\ref{lemma:SideEffectSafetyG_return}.
\end{IEEEproof}

\begin{lemma}
\label{lemma:SafeDEXG}
	Let for all method $m \in P$, ($\fnn{region}_m$, $\fnn{jun}_m$) a safe CDR for $m$. 
	Suppose all methods $m \in P$ are typable w.r.t. $\fnn{region}_m$ and to all 
	signatures in $\fnn{Policies_{\Gamma}}(m)$. Then all method $m$ are 
	\emph{side-effect-safe} w.r.t. the heap effect level of all the policies in 
	$\fnn{Policies_{\Gamma}}(m)$. 
\end{lemma}
\begin{IEEEproof}
	We use the definition of typable method and 
	Lemma~\ref{lemma:SideEffectSafetyG_sequential}, 
	Lemma~\ref{lemma:SideEffectSafetyG_return}, and 
	Lemma~\ref{lemma:SideEffectSafetyG_invoke}. Given typable method, for a derivation
	\[\state{_0} \rightsquigarrow_{m, \tau_0} \state{_1} \dots 
		\rightsquigarrow_{m, \tau_n} (r, h)\]
	there exists $RT \in \mathcal{PP} \rightarrow \RTSpec$ and 
	$rt_1, \dots rt_n \in$ $\RTSpec$ s.t.
	\[i_0 \vdash^{\tau_0} RT_{i_0} \Rightarrow rt_1\ \ i_1 \vdash^{\tau_1} RT_{i_1} 
	  \Rightarrow rt_2, \dots i_n \vdash^{\tau_n} RT_{i_n} \Rightarrow \]
	Using the lemmas, then we will get
	\[ h_0 \preceq_{k_h} h_1 \preceq_{k_h} \dots \preceq h_n \preceq h \]
	which we can use the transitivity of $\preceq_{k_h}$ to 
	conclude that $h_0 \preceq_{k_h} h$ (the definition of \emph{side-effect-safety}).
\end{IEEEproof}

\begin{definition}[High Result]
	Given $(r, h) \in (V + L ) \times \object{Heap}$ and an output level $\vec{k_r}$, the 
	predicate $\fn{highResult_{\vec{k_r}}}{r, h}$ is defined as :
	\[\setlength{\arraycolsep}{0pt}\begin{array}{cc}
		\dfrac{\vec{k_r}[n] \nleq \kobs\ \ \ \ v \in V}{\fn{highResult_{k_r}}{v, h}} & 
		\hspace{9pt}
		\dfrac{\vec{k_r}[\fn{class}{h(l)}] \nleq \kobs\ \ \ 
		  l \in \fn{dom}{h}}{\fn{highResult_{k_r}}{\langle l \rangle, h}} 
	\end{array}\]
\end{definition}

\begin{definition}[Typable Execution]$\\$
\begin{itemize}
	\item An execution step $\state{} \rightsquigarrow_{m, \tau} \state{'}$ is typable 
	w.r.t. $RT \in PP \rightarrow \RTSpec$ if there exists $rt'$ s.t. 
	$i \vdash^{\tau} RT_i \Rightarrow rt'$ and $rt' \sqsubseteq RT_{i'}$
	
	\item An execution step $\state{} \rightsquigarrow_{m, \tau} (r, h')$ is typable 
	w.r.t. $RT \in PP \rightarrow \RTSpec$ if $i \vdash^{\tau} RT_i \Rightarrow$
	
	\item An execution sequence $s_0 \rightsquigarrow_{m, \tau_0} s_1 
	\rightsquigarrow_{m, \tau_1} ... s_k \rightsquigarrow_{m, \tau_k} (r, h')$ is typable 
	w.r.t. $RT \in PP \rightarrow \RTSpec$ if :
	\begin{itemize}
		\item $\forall i, 0 \leq i < k, s_i \rightsquigarrow_{m, \tau_i} s_{i+1}$ is 
		typable w.r.t. $RT$;
		
		\item $s_n \rightsquigarrow_{m, \tau_n} (r, h')$ is typable w.r.t. $RT$.
	\end{itemize}
\end{itemize}
\end{definition}

\begin{lemma} [High Security Environment High Result]
\label{lemma:HighSEHighResult}
  Let $\state{}, \state{'} \in \text{State}_G$, $se(i)$ is high, 
  $\state{} \sim_{\beta} \state{'}$
  $i \mapsto$ and $\state{} \rightsquigarrow (r, h_r)$, then 
  $\fn{highResult}{r, h_r}$ and $h_r \sim_{\beta} h'$.
\end{lemma}
\begin{IEEEproof}
  We do a structural induction on the instruction in $i$. This lemma
  is only applicable if the instruction at $i$ is either a return instruction,
  or a possibly throwing instruction with uncaught exception.
  \begin{itemize}
    \item Return : the transfer rule has a constraint that $\vec{k_r}[n]$ will
      be at least as high as $se$ which is high. So by definition the lemma holds.
      Since return does not modify heaps, we know that $h_r \sim_{\beta} h'$ 
    
    \item Invoke : the transfer rule where the instruction is throwing an uncaught
      exception $e$ has constraint saying that $\vec{k_r}[e]$ will be at least as
      high as $se$, the level of exception thrown by the method invoked, and the object
      level on which the method is invoked. We know $se$ is high, so by definition the lemma holds.
      Heap wise, we know that the exception is newly generated so we can use 
      Lemma~\ref{AuxHeap6} to say that $h_r \sim_{\beta} h'$.
      
    \item Iget : the transfer rule where the instruction is throwing an uncaught
      exception $\np$ has constraint saying that $\vec{k_r}[\np]$ will be at least as
      high as $se$ and the security level of the object. 
      We know $se$ is high, so by definition the lemma holds.
      Heap wise, we know that the exception is newly generated so we can use 
      Lemma~\ref{AuxHeap6} to say that $h_r \sim_{\beta} h'$.
      
    \item Iput : Same as Iget.
    
    \item Aget : the transfer rule where the instruction is throwing an uncaught
      exception $\np$ has constraint saying that $\vec{k_r}[\np]$ will be at least as
      high as $se$ and the security level of the array. 
      We know $se$ is high, so by definition the lemma holds.
      Heap wise, we know that the exception is newly generated so we can use 
      Lemma~\ref{AuxHeap6} to say that $h_r \sim_{\beta} h'$.
      
    \item Aput : Same as Aget.
    \item Arraylength : Same as Aget.
    
    \item Throw : the transfer rule has a constraint that for any uncaught exception $e$, 
      $\vec{k_r}[e]$ will be at least as high as $se$ which is high. 
      So by definition the lemma holds. Throw does not modify heap as well, so we have
      $h_r \sim_{\beta} h'$.
  \end{itemize}
\end{IEEEproof}

\begin{lemma} [High Region High Result]
\label{lemma:HighRegionHighResult}
  Let $se$ be high in $\fn{region}{s, \tau}$, $\fn{jun}{s,\tau}$ is never defined,
  $\state{_0}, \state{'} \in \text{State}_G$,
  $\state{_0} \sim_{\beta} \state{'}$ and there is an execution trace 
  \[\state{_0} \rightsquigarrow_{m, \tau_0} \dots \state{_k} 
      \rightsquigarrow_{m, \tau_k} (r, h_r)\]
  where $\state{_0} \in \fn{region}{s, \tau}$. 
  Then $\fn{highResult}{r, h_r}$ and $h_r \sim_{\beta} h'$.
\end{lemma}
\begin{IEEEproof}
  We do induction on the length of the execution. 
  For the base case where $k$ is $0$ we can use Lemma~\ref{lemma:HighSEHighResult}.
  In the induction step, we know that $\state{_k}$ is in $\fn{region}{s, \tau}$ using
  SOAP2 and eliminating the case where it is a junction point by the assumption that
  $\fn{jun}{s, \tau}$ is never defined. Since we now have shorter execution length, we
  can apply induction hypothesis. Heap wise, we know from the transfer rules that
  field / array update will be bounded by $se$, thus we have $h_r \sim_{\beta} h'$
  by Lemma~\ref{AuxHeap3} and Lemma~\ref{AuxHeap4}.
\end{IEEEproof}

\begin{lemma}[High Register Stays]
\label{lemma:HighRegisterStays}
  Let $\state{_0} \in \text{State}_G$ and there is an execution step such that
  \[\state{_0} \rightsquigarrow_{m, \tau_0} \dots \state{_k}\] 
  and $se$ is high in $i_0, \dots, i_k$, if $RT_0(r)$ is high then $RT_k(r)$ is high. 
\end{lemma}
\begin{IEEEproof}
  We do induction on the length of execution, and we do case analysis on
  do possible instructions. If the length of execution is $0$ we have
  $\state{_0} \rightsquigarrow_{m, \tau_0} \state{_k}$. 
  The instruction is not a return point, since it contradicts the assumption.
  If the instruction is an instruction that modify $r$, we know that  
  these instructions will update the register $r$
  with security level at least as high as $se$, so the base case holds.
  In the induction step, we show using the same argument as the base case that
  $RT_1(r)$ is high, therefore now we can invoke induction hypothesis on the
  trace $\state{_1} \rightsquigarrow_{m, \tau_1} \dots \state{_k}$
\end{IEEEproof}

\begin{lemma}[Changed Register High]
\label{lemma:ChangedRegisterHigh}
  Let $\state{_0} \in \text{State}_G$ and there is an execution step such that
  \[\state{_0} \rightsquigarrow_{m, \tau_0} \dots \state{_k}\] 
  and $se$ is high in $i_0, \dots, i_k$, and the value of $r$ is changed by one
  or more instruction in the execution trace, then $RT_k(r)$ is high. 
\end{lemma}
\begin{IEEEproof}
  We do case analysis on where the first change might happen [1] then we
  do case analysis on all of the register modifying instructions change register
  $r$ to high [2]
  and invoke Lemma~\ref{lemma:HighRegisterStays} to claim that they stay high
  until it reaches $\state{_k}$. 
  All these instructions which modify register $r$ will update the register $r$
  with security level at least as high as $se$ so we already have [2].
  Since we assume that there is a change, [1] trivially holds.
\end{IEEEproof}

\begin{lemma}[Unchanged Register Stays]
\label{lemma:UnchangedRegisterStays}
  Let $\state{_0} \in \text{State}_G$ and there is an execution step such that
  \[\state{_0} \rightsquigarrow_{m, \tau_0} \dots \state{_k}\] 
  and the value of $r$ is not changed during the execution trace, 
  then $RT_0(r) = RT_k(r)$ and $\rho_0(r) = \rho_k(r)$. 
\end{lemma}
\begin{IEEEproof}
  We do induction on the length of execution, and we do case analysis on
  do possible instructions. If the length of execution is $0$ we have
  $\state{_0} \rightsquigarrow_{m, \tau_0} \state{_k}$. 
  The instruction is not a return point, since it contradicts the assumption.
  If the instruction is an instruction that modify $r$, we know that it contradicts
  our assumption. If the instruction does not modify $r$ then the base case holds by definition.
  In the induction step, we show using the same argument as the base case that
  $RT_0(r) = RT_1(r)$ and $\rho_0(r) = \rho_1(r)$, 
  therefore now we can invoke induction hypothesis on the
  trace $\state{_1} \rightsquigarrow_{m, \tau_1} \dots \state{_k}$.
\end{IEEEproof}

\begin{lemma}[Junction Point Indistinguishable]
\label{lemma:JunctionIndist}
	Let $\beta$ a partial function $\beta \in L \rightharpoonup L$ and 
	$\state{_0}, \state{'_0} \in \mathrm{State}_G$ two \dvm{G} states such that
  \[\begin{array}{cc}
  \multicolumn{2}{c}{\state{_0} \sim_{RT_{i_0}, RT_{i'_0}, \beta} \state{'_0}}
  \end{array}\]
  and $i_0 = i'_0$.
  
  Suppose that $se$ is high in region $\fn{region_m}{i_0, \tau_0}$ and also in region 
  $\fn{region_m}{i'_0, \tau'_0}$. Suppose we have a derivation
  \[\state{_0} \rightsquigarrow_{m, \tau_0} \dots \state{_k} 
    \rightsquigarrow_{m, \tau_k} (r, h)\] 
  and suppose this derivation is typable w.r.t. $RT$. Suppose we have a derivation
  \[\state{'_0} \rightsquigarrow_{m, \tau'_0} \dots \state{'_k} 
    \rightsquigarrow_{m, \tau'_k} (r', h')\]
  and suppose this derivation is typable w.r.t. $RT$. Then one of the following case 
  holds:
  \begin{enumerate}
	  \item there exists $j, j'$ with $0 \leq j \leq k$ and $0 \leq j' \leq k'$ s.t.
	  \[i_j = i'_j\hspace{0.5cm} \text{and} 
	    \hspace{0.5cm}\state{_j} \sim_{RT_{i_j}, RT_{i'_{j'}}, \beta} \state{'_{j'}}\]
	  
	  \item $(r, h) \sim_{\vec{k_r}, \beta} (r', h')$
  \end{enumerate}
\end{lemma}
\begin{IEEEproof} 
  We do a case analysis on whether a junction point is defined for both of the execution traces.
  There are 3 possible cases :
  \begin{enumerate}
    \item junction point is defined for both of the execution. We trace any changed registers
      during the execution. If the register is changed, then we can invoke 
      Lemma~\ref{lemma:ChangedRegisterHigh} to claim that the register is high and we know that
      high register does not affect indistinguishability. If the register
      is not changed, then we can invoke Lemma~\ref{lemma:UnchangedRegisterStays} to obtain 
      $RT_{i_j}(r) = RT_0(r)$ and $\rho_{i_j}(r) = \rho_0(r)$ and $\rho_{i'_{j'}}(r) = \rho'_0(r)$.
      If $RT_0(r)$ is low, then we know that $\rho_0(r) = \rho'_0(r)$, thus we can obtain that
      $\rho_{i_j}(r) = \rho_{i'_{j'}}(r)$. Otherwise, we know that $RT_{i_j}(r) = RT_{i'_{j'}}(r)$ is high.
      Whatever the case it does not affect indistinguishability. 
      Since for all register in $RT_{i_j}$ ($RT_{i'_{j'}}$) they are either changed or unchanged, 
      we can obtain $\state{_j} \sim_{RT_{i_j}, RT_{i'_{j'}}, \beta} \state{'_{j'}}$ and we are
      in Case 1.
    
    \item only one execution has junction point. For the part where junction point is not
      defined (assume it is the execution ending in $(r,h)$), 
      we can invoke lemma~\ref{lemma:HighRegionHighResult} to obtain $\fn{highResult}{r,h}$. 
      On the other execution path, we know from SOAP5 that the junction point is in the region 
      ($\fn{jun_m}{i'_0, \tau'_0} \in \fn{region}{i_0, \tau_0}$).
      Hence we can invoke lemma~\ref{lemma:HighRegionHighResult} again to obtain $\fn{highResult}{r',h'}$
      since $se$ is high in $\fn{region}{i_0, \tau_0}$, and prove that we are in Case 2. 
    
    \item both of the execution traces have no junction point. In this case since we know
      that $se$ is high in the region, we can just invoke lemma~\ref{lemma:HighRegionHighResult} on both executions 
      to obtain $\fn{highResult}{r,h}$ and $\fn{highResult}{r',h'}$, hence we are in Case 2.
  \end{enumerate}
\end{IEEEproof}

\begin{lemma}[High Branching] 
\label{lemma:HighBranchingG}
	Let all method $m'$ in $P$ are non-interferent w.r.t. all the policies in 
	$\fn{Policies_{\Gamma}}{m'}$. Let $m$ be a method in $P$, 
	$\beta \in L \rightharpoonup L$ a partial function, $s_1, s_2 \in \mathrm{State}_G$ 
	and two registers types $rt_1, rt_2 \in \RTSpec$ s.t.
	\[s_1 \sim_{rt_1, rt_2, \beta} s_2\]
\begin{enumerate}
	\item If there exists two states $\state{'_1}, \state{'_2} \in \mathrm{State}_G$ and  
	two registers types $rt'_1, rt'_2 \in \RTSpec$ s.t. $i'_1 \neq i'_2$
  \[\begin{array}{cc}
    s_1 \rightsquigarrow_{m, \tau_1} \state{'_1} & 
    s_2 \rightsquigarrow_{m, \tau_2} \state{'_2}\\
    i \vdash^{\tau_1} rt_1 \Rightarrow rt'_1 & 
    i \vdash^{\tau_2} rt_2 \Rightarrow rt'_2
  \end{array}\]
  then
  \[\begin{array}{c}
    se$ is high in $\fn{region}{i, \tau_1}\\
    se$ is high in $\fn{region}{i, \tau_2}
  \end{array}\]
	
	\item If there exists a state $\state{'_1} \in \mathrm{State}_G$, a final result 
	$(v_2, h'_2) \in (V + L) \times \mathrm{Heap}$ and a registers types 
	$rt'_1 \in \RTSpec$ s.t.
  \[\begin{array}{cc}
    s_1 \rightsquigarrow_{m, \tau_1} \state{'_1} &
    s_2 \rightsquigarrow_{m, \tau_2} (r_2, h'_2)\\
    i \vdash^{\tau_1} rt_1 \Rightarrow rt'_1 & 
    i \vdash^{\tau_2} rt_2 \Rightarrow
	\end{array}\]
  then
  \[\begin{array}{c}
    se$ is high in $\fn{region}{i, \tau_1}
  \end{array}\]
  
\end{enumerate}
\end{lemma} 
\begin{IEEEproof}
  By case analysis on the instruction executed.
  \begin{itemize}
    \item $\ins{ifeq}$ and $\ins{ifneq}$: the proof's outline follows from before as 
    there is no possibility for exception here.

    \item $\ins{invoke}$ : there are several cases to consider for this instruction to 
    be a branching instruction:
    
      \textbf{1)} both are executing normally. Since the method we are invoking are 
      non-interferent, and we have that $\rho_1 \sim_{rt_1, rt_2, \beta} \rho_2$, we 
      will also have indistinguishable results. Since they throw no 
      exceptions there is no branching there.
        
      \textbf{2)} one of them is normal, the other throws an exception $e'$. Assume 
      that the policy for the method called is 
      $\vec{k'_a} \overset{k'_h}{\rightarrow} \vec{k'_r}$.
      This will imply that  $\vec{k'_r}[e'] \nleq \kobs$ otherwise the output will be 
      distinguishable. By the transfer rules we have that for all the regions $se$ is 
      to be at least as high as $\vec{k'_r}[e']$ (normal execution one is lub-ed with 
      $k_e = \bigsqcup \big\{ \vec{k'_r}[e]\ |\ e \in \fn{excAnalysis}{m'}\big\}$ where
      $e' \in \fn{excAnalysis}{m'}$, and $\vec{k'_r}[e']$ by itself for the exception 
      throwing one), thus we will have $se \nleq \kobs$ throughout the regions. 
      For the exception throwing one, if the exception is 
      caught, then we know we will be in the first case. 
      If the exception is uncaught, then we are in the second case.
        
      \textbf{3)} the method throws different exceptions (let's say $e_1$ and $e_2$). 
      Assume that the policy for the method called is 
      $\vec{k'_a} \overset{k'_h}{\rightarrow} \vec{k'_r}$. Again, since the outputs are 
      indistinguishable, this means that $\vec{k'_r}[e_1] \nleq \kobs$ and 
      $\vec{k'_r}[e_2] \nleq \kobs$. By the transfer rules, as before we will have $se$ 
      high in all the regions required. If the exception both are uncaught, then this 
      lemma does not apply. Assume that the $e_1$ is caught. We follow the argument
      from before that we will have 
      $se$ is high in $\fn{region}{i, \tau_1}$.
      If $e_2$ is caught,
      using the same argument we will have 
      $se$ is high in $\fn{region}{i, \tau_2}$ and we are in 
      the first case. If $e_2$ is uncaught, then we know that we are in the second case.
      The rest of the cases will be dealt with by first assuming that $e_2$ is caught.
    
	  \item object~/~array manipulation instructions that may throw a null pointer 
	  exception. This can only be a problem if one is null and the other is non-null.
	  From this, we can infer that register pointing to object~/~array reference will have
	  high security level (otherwise they have to have the same value). If this is the 
	  case, then from the transfer rules for handling null pointer we have that $se$ is 
	  high in region $\fn{region}{i_1, \np}$. 
	  
    Now, regarding the part which is not null, we also can deduce 
    that it is 
    from the transfer rules that we have $se$ will be high 
    in that region, which implies that $se$ will be high in $\fn{region}{i_2, \norm}$ 
	
    \item $\fnn{throw}$. Actually the reasoning for this instruction is closely similar 
    to the case of object~/~array manipulation instruction that may throw a null 
    pointer exception, except with additional possibility of the instruction throwing 
    different exception. Fortunately for us, this can only be the case if the security 
    level to the register pointing to the object to throw is high, therefore the 
    previous reasoning follows.
  \end{itemize} 
\end{IEEEproof}

\begin{lemma}[Locally Respect (Specialized)]
\label{lemma:LocallyRespectG}
	Let all method $m'$ in $P$ are non-interferent w.r.t. all the policies in 
	$\fn{Policies_{\Gamma}}{m'}$. Let $m$ a method in $P$, 
	$\beta \in L \rightharpoonup L$ a partial function, $s_1, s_2 \in \mathrm{State}_G$ 
	two \dvm{G} states at the same program point $i$ and two registers types 
	$rt_1, rt_2 \in \RTSpec$ s.t. $s_1 \sim_{rt_1, rt_2, \beta} s_2$.
  \begin{enumerate}
	  \item If there exists two states $s'_1, s'_2 \in \mathrm{State}_G$ 
	  and the program point of $s'_1$ is the same as $s'_2$ 
	  and two registers types $rt'_1, rt'_2 \in \RTSpec$ s.t.
		\[\begin{array}{cc}
			s_1 \rightsquigarrow_{m, \tau_1} s'_1 &
			s_2 \rightsquigarrow_{m, \tau_2} s'_2\\
			i \vdash^{\tau_1} rt_1 \Rightarrow rt'_1 & 
			i \vdash^{\tau_2} rt_2 \Rightarrow rt'_2
		\end{array}\]
		then there exists $\beta' \in L \rightharpoonup L$ s.t.
		\[s'_1 \sim_{rt'_1, rt'_2, \beta'} s'_2\ \ \ \ \beta \subseteq \beta'\]
	
	  \item If there exists a state $\state{'_1} \in \mathrm{State}_G$, a final result 
	  $(r_2, h'_2) \in (V + L) \times \mathrm{Heap}$ and a registers types 
	  $rt'_1 \in \RTSpec$ s.t.
		\[\begin{array}{cc}
			s_1 \rightsquigarrow_{m, \tau_1} \state{'_1} & 
			s_2 \rightsquigarrow_{m, \tau_2} (r_2, h'_2)\\
			i \vdash^{\tau_1} rt_1 \Rightarrow rt'_1 & i \vdash^{\tau_2} rt_2 \Rightarrow
		\end{array}\]
		then there exists $\beta' \in L \rightharpoonup L$ s.t.
		\[h'_1 \sim_{\beta'} h'_2,\ \ \ \ \fn{highResult_{\vec{k_r}}}{r_2, h'_2}\ \ \ \ 
		  \beta \subseteq \beta'\]
	
	  \item If there exists two final results $(r_1, h'_1), 
	  (r_2, h'_2) \in (V + L) \times \mathrm{Heap}$ s.t.
		\[\begin{array}{cc}
			s_1 \rightsquigarrow_{m, \tau_1} (r_1, h'_1) &
			s_2 \rightsquigarrow_{m, \tau_2} (r_2, h'_2) \\
			i \vdash^{\tau_1} rt_1 \Rightarrow & i \vdash^{\tau_2} rt_2 \Rightarrow
		\end{array}\]
		then there exists $\beta' \in L \rightharpoonup L$ s.t.
		\[(r_1, h'_1) \sim_{\beta'} (r_2, h'_2)\ \ \ \ \beta \subseteq \beta'\]
	\end{enumerate}
\end{lemma}
\begin{IEEEproof}
  Since we have already proved this lemma for all the instructions apart from the 
  exception cases, we only deal with the exception here. Moreover, we already proved 
  for the heap for instructions without exception to be indistinguishable. Therefore, 
  only instructions which may cause an exception are considered here, and only consider 
  the case where the registers may actually be distinguishable. Note that for exception
  case, the lemma is specialized to only affect those that have the same successor's
  program point.
  \begin{itemize}
    \item $\ins{invoke}$. There are 6 possible successors here, but we only consider 
    the 4 exception related one (since one of them can be subsumed by the other) :

      \textbf{1)} One normal and one has caught exception. 

      In this case we know that the lemma is not applicable since the successors have
      different program points.
		
      \textbf{2)} One normal and one has uncaught exception (the case where one throw 
      caught exception and one throw uncaught exception is proved using similar arguments). 
      In this case, we have one successor state while the other will return value or 
      location (case 2). So in this case we only need to prove that 
      $\fn{highResult_{k_r}}{r_2, h'_2}$ (the part about heap indistinguishability is 
      already proved). We can easily again appeal to the output distinguishability 
      since we already assumed that the method $m'$ is non-interferent. Since we have 
      the exception $e$ returned by the method $m'$ as high ($\vec{k'_r}[e]$), we can 
      now use the transfer rule which states that $\vec{k'_r}[e] \leq \vec{k_r}[e]$ and 
      establish that $\vec{k_r}[e] \nleq \kobs$ which in turns implies that 
      $\fn{highResult_{k_r}}{r_2, h'_2}$, thus we are in Case 2.

      \textbf{3)} Both has caught exception
      If they have the same exception, then we that the content of $ex$ register will be
      the same thus the register will be indistinguishable. The case where the exceptions
      are different is not covered in this lemma since they will lead to different handlers,
      thus different program point.
      
      \textbf{4)} Both has uncaught exception (and different exception on top of that). 
      Let's say the two exceptions are $e_1$ and $e_2$. For the beginning, we use the 
      output indistinguishability of the method to establish that 
      $\vec{k'_r}[e_1]\nleq \kobs$ and $\vec{k'_r}[e_2] \nleq \kobs$. Then, using the 
      transfer rules for uncaught exceptions whichs states 
      $\vec{k'_r}[e_1] \leq \vec{k_r}[e_1]$ and $\vec{k'_r} \leq \vec{k_r}[e_2]$ to 
      establish that $\vec{k_r}[e_1]$ and $\vec{k_r}[e_2]$ are high as well. Now, since 
      they are both high, we can claim that they are indistinguishable (output-wise), 
      therefore concluding the proof that we are in Case 3. 

    \item $\ins{iget}$. There are four cases to consider here:
    
      \textbf{1)} One is normal execution and one has caught null exception. 
      In this case we know that the lemma is not applicable since the successors have
      different program points.

      \textbf{2)} One is normal execution and one has uncaught null exception. The only 
      difference with the previous case is that there will be one execution  
      returning a location for exception instead. In this case, we only need to prove 
      that this return of value is high ($\fn{highResult_{k_r}}{r_2, h'_2}$). We know 
      that $r_o$ (the register containing the object) is high ($sec(r_o) \nleq \kobs$), 
      otherwise $s_1 \nsim_{rt_1, rt_2 \beta} s_2$. The transfer rule for 
      $\ins{iget}$ with uncaught exception states that $sec(r_o) \leq \vec{k_r}[\np]$, 
      which will give us $\vec{k_r}[\np] \nleq \kobs$, which will implies that 
      $\fn{highResult_{k_r}}{r_2, h'_2}$, thus we are in Case 2.

      \textbf{3)} Both has caught null exception. In this case, there are two things 
      that needs consideration: the new objects in the heap, and the pseudo-register 
      $ex$ containing the new null exception. Since we have $h_1 \sim_{\beta} h_2$ and 
      the exception is created fresh ($l_1 = \fn{fresh}{h_1}$, 
      $l_1 \mapsto \fnn{default_{\np}}$, $l_2 = \fn{fresh}{h_2}$, 
      $l_2 \mapsto \fnn{default_{\np}}$), by lemma \ref{AuxHeap6} we have that 
      $h'_1 \sim_{\beta} h'_2$ as well. Now for the pseudo-register $ex$, we take the 
      mapping $\beta'$ to be $\beta \oplus \{l_1 \mapsto l_2\}$, where 
      $l_1 = \fn{fresh}{h_1}$ and $l_2 = \fn{fresh}{h_2}$, both are used to store the 
      new exception. Under this mapping, we know that $l_1 \sim_{\beta'} l_2$ and this 
      will give us $\rho'_1 \sim_{\beta'} \rho'_2$ since $\rho'_1 = \{ex \mapsto l_1\}$ 
      and $\rho'_2 = \{ex \mapsto l_2\}$, thus we are in Case 1.
      
      \textbf{4)} Both has uncaught null exception. Following the previous arguments, 
      we have $h'_1 \sim_{\beta'} h'_2$, and $l_1 \sim_{\beta'} l_2$, which will give 
      us $(\langle l_1 \rangle, h'_1) \sim_{\beta'} (\langle l_2 \rangle, h'_2)$, thus 
      we are in Case 3.

    \item $\ins{iput}$, $\ins{aget}$, and $\ins{aput}$. The arguments closely follows 
    that of $\ins{iget}$

    \item $\ins{throw}$. There are four cases to consider here :

      \textbf{1)} Two same exception. In this case, we know that the exception will be 
      the same, therefore the value for $ex$ will be the same 
      ($ex = \rho_1(r_e) = \rho_2(r_e)$), thus giving us 
      $\rho'_1 \sim_{\beta'} \rho'_2$ if the exception is caught (Case 1).
      In the case where the exception is uncaught, we know that the value will be the 
      same, that is $\rho_1(r_e)$, therefore the output will be indistinguishable as 
      well (Case 3).
      
      \textbf{2)} Two different exceptions, both are caught. 
      In this case we know that the lemma is not applicable since the successors have
      different handlers (thus program points). 
      
      \textbf{3)} Two different exceptions, both are uncaught. The transfer rules 
      states that $rt'_1(r_e) \leq \vec{k_r}[e_1]$ and $rt'_2(r_e) \leq \vec{k_r}[e_2]$ 
      (where $r_e$ is the register containing the exception). Since $r_e$ must be high 
      to have different value, therefore $\vec{k_r}[e_1]$ and $\vec{k_r}[e_2]$ must be 
      high as well. With this, we will have that 
      $(r_1, h'_1) \sim_{\beta'} (r_2, h'_2)$ since both are high outputs (Case 3).
      
      \textbf{4)} Two different exceptions, one is caught one is uncaught. Similar to 
      the previous argument: we know that $\vec{k_r}[e]$ will be high, therefore we 
      will have $\fn{highResult_{\vec{k_r}}}{r_2, h'_2}$, $h'_1 \sim_{\beta'} h'_2$ 
      (throw instruction does not modify the heap), and $\beta \subseteq \beta'$ (Case 2).
  \end{itemize}
\end{IEEEproof}

\begin{lemma} [Typable DEX Program is Non-Interferent]
\label{lemma:DEXgNI}
  Suppose we have $\beta$ a partial function 
  $\beta \in \mathcal{L} \rightharpoonup \mathcal{L}$ and 
  $\state{_0}, \state{'_0} \in \mathrm{State}_G$ two \dvm{G} states s.t. $i_0 = i'_0$
  and $\state{_0} \sim_{RT_{i_0}, RT_{i'_0}, \beta} \state{'_0}$. Suppose we have a 
  derivation
    \[\state{_0} \rightsquigarrow_{m, \tau_0} \dots \state{_k} 
      \rightsquigarrow{m, \tau_k} (r, h)\]
  and suppose this derivation is typable w.r.t. RT. Suppose we also have another 
  derivation
    \[\state{'_0} \rightsquigarrow_{m, \tau'_0} \dots \state{'_k} 
      \rightsquigarrow{m, \tau'_k} (r', h')\]
  and suppose this derivation is typable w.r.t. RT. 
  Then what we want to prove is that there exsts 
  $\beta' \in \mathcal{L} \rightharpoonup \mathcal{L}$ s.t.
    \[(r, h) \sim_{\vec{k_a}, \beta'} (r', h') \text{ and } \beta \subseteq \beta'\]
\end{lemma}
\begin{IEEEproof}
  Following the proof in the side effect safety, we use induction on the length of 
  method call chain. 
  For the base case, there is no $\ins{invoke}$ instruction involved (method call chain
  with length $0$). A note about this setting is that we can use lemmas which assume 
  that all the methods are non-interferent since we are not going to call another
  method. To start the proof in the base case of induction on method call 
  chain length, we use induction on the length of $k$ and $k'$. 
  The base case is when $k = k' = 0$. In this case, we can use case 3 of 
  Lemma~\ref{lemma:LocallyRespectG}. There are several possible cases for the 
  induction step:
  \begin{enumerate}
    \item 
    $k > 0$ and $k' = 0$: then we can use case 2 of Lemma~\ref{lemma:LocallyRespectG}
    to get existence of $\beta' \in \mathcal{L} \rightharpoonup \mathcal{L}$ s.t.
      \[\begin{array}{ccccr}
        h_1 \sim_{\beta'} h', & \fn{highResult_{\vec{k_r}}}{r', h'} & \text{and}
        & \beta \subseteq \beta' & [1]
      \end{array}\]
    Using case 2 of Lemma~\ref{lemma:HighBranchingG} 
    we get
      \[se \text{ is high in } \fn{region}{i_0, \tau_1}\]
    where $\tau_1$ is the tag s.t. $i_0 \mapsto^{\tau_1} i_1$. SOAP2 gives us that 
    either$i_1 \in \fn{region}{i_0, \tau_1}$ or $i_1 = \fn{jun}{i_0, \tau_1}$ but the 
    later case is rendered impossible due to SOAP3. Applying 
    Lemma~\ref{lemma:HighRegionHighResult} we get
      \[\begin{array}{cccr}
        \fn{highResult_{\vec{k_r}}}{r, h'_1} & \text{and} & h_1 \sim_{\beta} h'_1 & [2]
      \end{array}\]
    Combining [1] and [2] we get
      \[\begin{array}{ccc}
        h_1 \sim_{\beta'} h'_1, & h_1 \sim_{\beta'} h', 
        (r, h) \sim (r', h') 
      \end{array}\]
    to conclude.
  
    \item $k = 0$ and $k' > 0$ is symmetric to the previous case.
  
    \item $k > 0$ and $k' > 0$. If the next instruction is at the same program point 
    ($i_1 = i'_1$) we can conclude using the case 1 of Lemma~\ref{lemma:LocallyRespectG}
    and induction hypothesis. Otherwise we will have registers typing $rt_1$ and 
    $rt'_1$ s.t.
      \[\begin{array}{cc}
        i_0 \vdash^{\tau_0} RT_{i_0} \Rightarrow rt_1, & rt_1 \sqsubseteq RT_{i_1} \\
        i'_0 \vdash^{\tau'_0} RT_{i'_0} \Rightarrow rt'_1, & rt'_1 \sqsubseteq RT_{i'_1}
      \end{array}\]
    Then using case 1 of Lemma~\ref{lemma:HighBranchingG} 
    we have  
      \[\begin{array}{c}
        se \text{ is high in } \fn{region}{i_0, \tau} \\
        se \text{ is high in } \fn{region}{i'_0, \tau'}
      \end{array}\]
    where $\tau, \tau'$ are tags s.t. $i_0 \mapsto^{\tau} i_1$ and 
    $i'_0 \mapsto^{\tau'} i'_1$. Using case 1 of Lemma~\ref{lemma:LocallyRespectG} 
    there exists $\beta', \beta \subseteq \beta'$ s.t. (with the help of 
    Lemma~\ref{lemma:IndistinguishabilitySingleMonotony}
    \[\state{_1} \sim_{RT_{i_1}, RT_{i'_1}, \beta'} \state{'_1}\]
    Invoking Lemma~\ref{lemma:JunctionIndist} will give us two
    cases:
    \begin{itemize}
      \item There exists $j, j'$ with $1 \leq j \leq k$ and $1 \leq j' \leq k'$ s.t.
      $i_j = i'_j$, and $\state{_j} \sim_{RT_{i_j}, RT_{i'_j}, \beta} \state{'_j}$.
      We can then use induction hypothesis on the rest of executions to conclude.
      
      \item $(r, h) \sim_{\vec{k_a}, \beta'} (r', h')$ and we can directly conclude.
    \end{itemize}
  \end{enumerate}
  
  After we established the base case, we can then continue to prove the induction on
  method call chain. In the case where an instruction calls another method, we will
  have the method non-interferent since they necessarily have shorter call chain length
  (induction hypothesis).
  
\end{IEEEproof}

Proof of Theorem~\ref{thm:DEXSoundness} is direct application of 
Lemma~\ref{lemma:DEXgNI} and Lemma~\ref{lemma:SafeDEXG}.

\end{document}